\documentclass[a4paper,11pt]{article}
\usepackage{jheppub}
\usepackage[T1]{fontenc} 
\usepackage{amsmath}
\usepackage[caption=false]{subfig}

\usepackage{breqn}

\def\eps{\epsilon}

\def\cO{  {\cal O}  }
\newcommand{\bea}{\begin{eqnarray}}
\newcommand{\eea}{\end{eqnarray}}
\newcommand{\dr}{{\rm d}}
\newcommand{\nn}{\nonumber}

\allowdisplaybreaks[4]

\preprint{TTP13-021,
SFB/CPP-13-39}
\title{Analytic results for planar three-loop four-point integrals
from a Knizhnik-Zamolodchikov equation}

\author[a]{Johannes M.\ Henn}
\author[b,d]{Alexander V.\ Smirnov}
\author[c]{Vladimir A.\ Smirnov}
\affiliation[a]{Institute for Advanced Study, Princeton, NJ 08540, USA}
\affiliation[b]{Scientific Research Computing Center, Moscow State University, 
119992 Moscow, Russia}
\affiliation[c]{Skobeltsyn Institute of Nuclear Physics of Moscow State 
University, 119992 Moscow, Russia}
\affiliation[d]{Institut f\"{u}r Theoretische Teilchenphysik, KIT, 76128 
Karlsruhe, Germany}
\emailAdd {jmhenn@ias.edu}
\emailAdd {asmirnov80@gmail.com}
\emailAdd {smirnov@theory.sinp.msu.ru}

\abstract{
We apply a recently suggested new strategy to solve differential equations
for master integrals for families of Feynman integrals.
After a set of master integrals has been found using the integration-by-parts method,
the crucial point of this strategy is to introduce a new basis where all master integrals 
are pure functions of uniform transcendentality.
In this paper, we apply this method to all planar three-loop four-point massless on-shell master integrals.
We explicitly find such a basis, and show that the differential equations are of the
Knizhnik-Zamolodchikov type. We explain how to solve the latter to all orders in 
the dimensional regularization parameter $\epsilon$, including all boundary constants,
in a purely algebraic way.
The solution is expressed in terms of harmonic polylogarithms.
We explicitly write out the Laurent expansion in $\epsilon$ for all master integrals up to weight six.
}


\begin{document}

\maketitle
\flushbottom

\section{Introduction}

The method of differential equations (DE) suggested in \cite{Kotikov:1990kg,Kotikov:1991pm}
is one of the
most powerful modern methods of evaluating multiloop Feynman integrals.
It was presented in a systematic form in \cite{Remiddi:1997ny,Gehrmann:1999as,Gehrmann:2000zt,Gehrmann:2001ck} where it was
successfully applied to the evaluation of four-point two-loop massless Feynman integrals 
with one leg off shell. In this formulation, DE are applied to the evaluation of
{\em master integrals} whose number is always finite ~\cite{Smirnov:2010hn}. 
This approach supposes that one has a solution of
integration by parts (IBP) relations \cite{Chetyrkin:1981qh} at hand, i.e. 
an algorithm which expresses any Feynman integral of 
a given family as a linear combination of the master integrals.\footnote{We use the term {\em family} of Feynman integrals to 
refer to a set of integrals sharing the same denominator factors, and possibly having numerators. In this terminology, 
an integral with all propagators present can be thought of as the parent integral,
and integrals with missing propagators as descendants.}
There are several public codes to solve IBP relations
\cite{Anastasiou:2004vj,Smirnov:2008iw,Smirnov:2013dia,Studerus:2009ye,vonManteuffel:2012np,Lee:2012cn} and many private codes.
In the present work, we applied the {\tt c++} version of {\tt FIRE} \cite{Smirnov:2008iw,Smirnov:2013dia}.

The idea of the method is to take derivatives of a given
master integral with respect to kinematical invariants and masses.
Then the result of this differentiation is written in terms of
Feynman integrals of the given family and, according to the known IBP
reduction, in terms of the master integrals.
In this way, one obtains a system of first-order differential equations
for the master integrals, and can then try to solve this system with appropriate boundary 
conditions. The method of DE was successfully applied in many calculations.
For reviews, see \cite{Argeri:2007up,Smirnov:2012gma}, and \cite{Czakon:2008ii,vonManteuffel:2012je} for some recent examples.

Despite its power and generality, one can encounter practical problems when using
this method for complicated families of Feynman integrals.
One difficulty can lie in the fact that the class of integral functions appropriate to
describe the solution is complicated, and it only becomes
apparent in the course of the calculation which class of functions is needed.
Another difficulty can arise when there are several master integrals that
satisfy coupled differential equations. These can turn out rather cumbersome to solve in
practice. Also, the results for the master integrals are often rather lengthy and their
structure is not particularly transparent.
 
Quite recently a new strategy of solving DE for master integrals
was suggested \cite{Henn:2013pwa} by one the authors of the present paper. 
When applicable, it overcomes the problems indicated above.
The key ingredient of this strategy is to choose a convenient basis of master integrals having 
desirable properties.
The goal is to choose all master integrals such that they are {\em pure} functions of uniform {\em weight}, 
i.e uniform degree of transcendentality. For generalized polylogarithms 
\cite{Chen1997,Goncharov:1998kja} that are defined through iterated integrals over logarithmic differential forms, the weight of a function is defined as the number of integrations needed to define it. A linear combination of such functions has uniform (i.e. homogeneous) weight if all its summands have the same weight. Finally, a function is called pure if the weight of its differential is lowered by one unit. This last property is motivated by the fact that such functions satisfy simple differential equations. This will be important in the following.
In the remainder of this paper, we will use the terms weight and (degree of) transcendentality without distinction.

The fact that certain loop integrals have uniform transcendentality was observed
in many calculations, especially in supersymmetric theories, see e.g. 
\cite{Smirnov:1999gc,Smirnov:2003vi,Bern:2005iz,Heinrich:2009be,Gehrmann:2011xn}, and more recently
in \cite{Drummond:2010mb,ArkaniHamed:2010gh}.\footnote{The concept of transcendentality also played an important role in a different context, 
at the level of anomalous dimensions of composite operators,
where the anomalous dimensions in $N=4$ SUSY Yang--Mills theory
may be obtained from the leading-transcendentality contributions in QCD \cite{Kotikov:2004er}. }
Certainly results for generic scattering amplitudes in QCD do not appear to have simple
transcendentality properties, at least in the way they are conventionally presented.
One may ask, however, whether such results can be written in terms of a finite number of building
blocks that have the properties discussed above.
Reference \cite{Henn:2013pwa} suggests that all master integrals can indeed be
chosen to be pure functions of uniform transcendentality, including the integrals needed for QCD, 
and provides criteria for finding such a basis.

Suppose that for a given family the set of master integrals has
already been identified,  using IBP relations.
The main point of the strategy of \cite{Henn:2013pwa} is then to turn to a 
new basis of the master integrals which all have uniform transcendentality.
This transition is given by a linear transformation in the space
of master integrals and the corresponding matrix is rational with
respect to dimension and usually algebraic w.r.t. kinematic invariants.

As explained in  \cite{Henn:2013pwa} one can use various strategies to
reveal uniformly transcendental master integrals. One efficient
method is to replace propagators by delta functions and analyze
whether the resulting expression is uniformly transcendental.
In other cases, explicit integral representations can be derived, using Feynman parameters
or other means \cite{ArkaniHamed:2012nw}, to make the transcendental properties of 
the answer manifest. We also wish to mention related work in the mathematical 
literature \cite{Brown:2009rc}.

Let us denote the kinematical variables by $x=(x_1,\ldots,x_{n})$, the set of $N$ basis integrals
by $f=(f_1,\ldots,f_{N})$, 
and let us work in $D=4-2\eps$ dimensions. The general set of differential equations takes the form
\begin{align}\label{diffeq_general}
 \partial_i
 f(\eps, x ) =  A_i(\eps, x) f(\eps, x) \,,
\end{align}
where $\partial_{i} = \frac{\partial}{\partial x_i }$, and
each $A_{i}$ is an $N \times N$ matrix.

The existence of a basis of master integrals with the above properties is 
closely related to the possibility to obtain a much simpler system of differential equations, 
as conjectured in \cite{Henn:2013pwa},
\begin{align}\label{diffeq_special}
\partial_{i} f(\eps, x) = \eps \, A_{i}(x) f(\eps, x) \,.
\end{align}
The essential difference w.r.t. (\ref{diffeq_general}) is that the matrix in
the equation is just proportional to $\eps$.
As a result such a system of equations can be solved in a very easy and natural way.
There is no general proof that, for any family of Feynman integrals, one
can turn from (\ref{diffeq_general}) to (\ref{diffeq_special}). However,
we are going to provide non-trivial examples of Feynman integrals where this
is possible and thereby arrive at new results.

In \cite{Henn:2013pwa}  it was shown that this strategy can successfully be applied
to all the on-shell massless two-loop Feynman integrals, and previous results, 
in particular, for the two double box integrals of this family \cite{Smirnov:1999gc,Anastasiou:2000kp}, 
can be reproduced.

\begin{figure}[t] 
\captionsetup[subfigure]{labelformat=empty}
\begin{center}
\subfloat[(A)]{\includegraphics[width=0.3\textwidth]{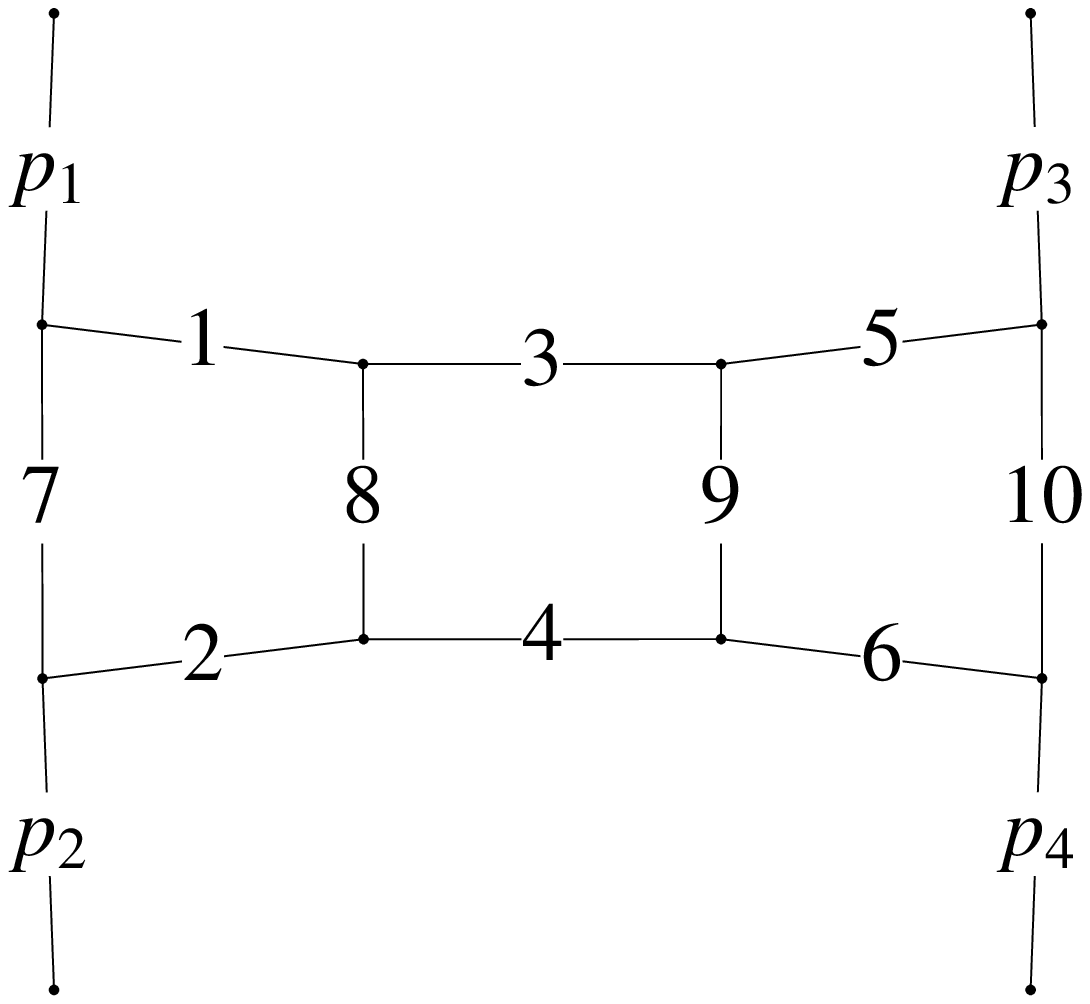}}
\hspace{2cm}
\subfloat[(E)]{\includegraphics[width=0.3\textwidth]{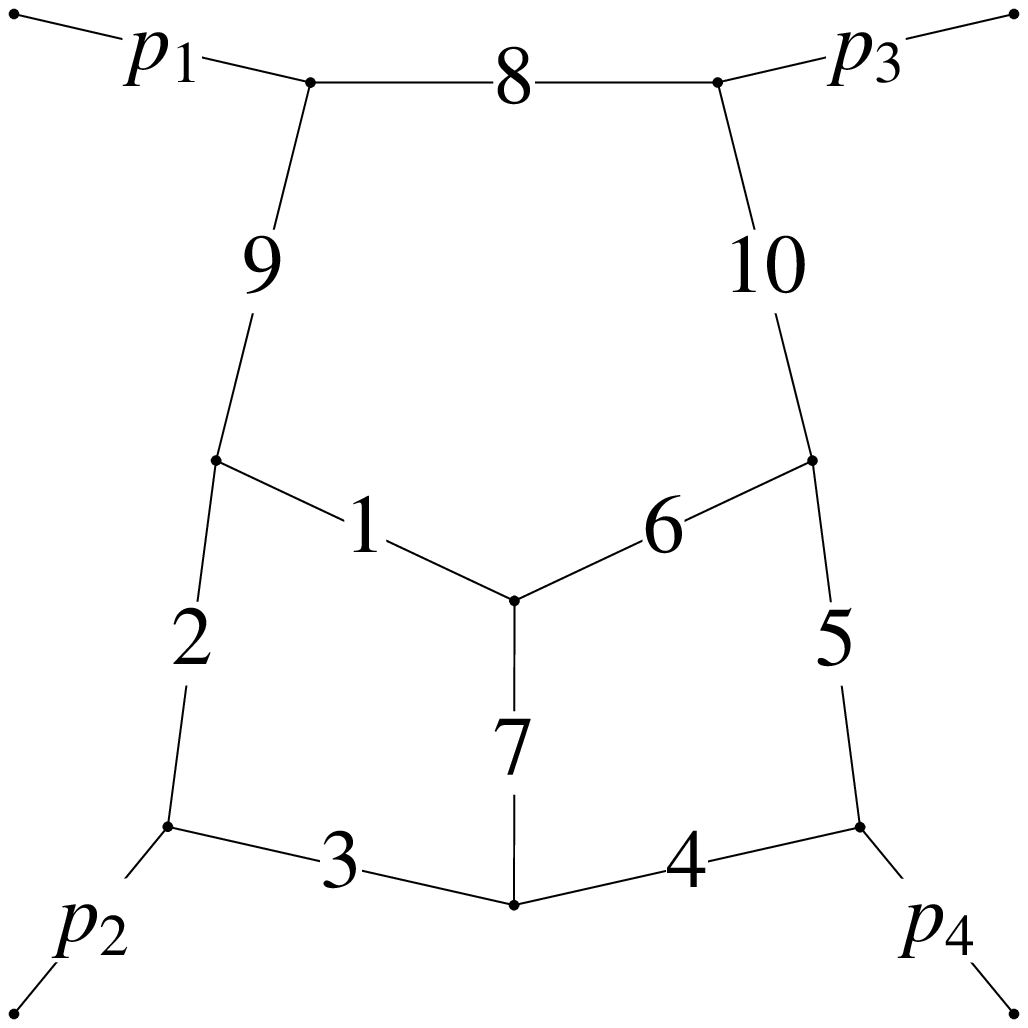}}
\caption{The triple box (A) and tennis court diagrams (E). Latin numbers refer to propagators associated to line parameters $a_{i}$, cf. eqs. (1.3) and (1.4). Lines associated to possible numerators are not shown in the figures.}
\label{3btk}
\end{center}
\end{figure}

The goal of the present paper is to derive new results with the strategy of 
\cite{Henn:2013pwa}. We will consider the two families of planar three-loop massless on-shell
integrals corresponding to the ladder (i.e. triple box) and the tennis court graph
shown in Fig.~\ref{3btk}. (The notation A and E for the families of master integrals follows that of \cite{Bern:2007hh}. Other letters stand for non-planar integrals.)
These integrals have fifteen indices: we associate the first ten of them to the edges of
these graphs, as shown in Fig.~\ref{3btk}, and the last five to numerators. 
Explicitly, we have
\bea
F^A_{a_1,\ldots,a_{15}}(s,t;D) &=&
\int\int\int \frac{\dr^Dk_1 \, \dr^Dk_2 \, \dr^Dk_3}{(-k_1^2)^{a_1}
[-(p_1 + p_2 + k_1)^2]^{a_2}
(-k_2^2)^{a_3}}
\nonumber \\ && \hspace*{-15mm}
\times \frac{[-(k_1 - p_3)^2]^{-a_{11}}[-(p_1 + k_2)^2]^{-a_{12}} [-(k_2 - p_3)^2]^{-a_{13}} }{
[-(p_1 + p_2 + k_2)^2]^{a_4}(-k_3^2)^{a_5}
[-(p_1 + p_2 + k_3)^2]^{a_6} [-(p_1 + k_1)^2]^{a_7} }
\nonumber \\ && \hspace*{-15mm}
\times \frac{ [-(p_1 + k_3)^2]^{-a_{14}} [-(k_1 - k_3)^2]^{-a_{15}}}{
[-(k_1 - k_2)^2]^{a_8} [-(k_2 - k_3)^2]^{a_9}
[-(k_3 - p_3)^2]^{a_{10}} }
\; ,
\label{diA}
\eea
and
\bea
F^E_{a_1,\ldots,a_{15}}(s,t;D) &=&
\int\int\int \frac{\dr^Dk_1 \, \dr^Dk_2 \, \dr^Dk_3}{[-(k_1 - k_3)^2]^{a_1}
[-(p_1 + k_1)^2 ]^{a_2}
[-(p_1 + p_2 + k_1)^2]^{a_3}}
\nonumber \\ && \hspace*{-15mm}
\times \frac{[-(p_1 + p_2 + k_3)^2]^{-a_{11}}[-(p_1 + k_2)^2 ]^{-a_{12}} [ -(k_1 - p_3)^2]^{-a_{13}} }{
[-(p_1 + p_2 + k_2)^2 ]^{a_4}[-(k_2 - p_3)^2]^{a_5}
[ -(k_2 - k_3)^2]^{a_6} [-(k_1 - k_2)^2 ]^{a_7} }
\nonumber \\ && \hspace*{-15mm}
\times \frac{(-k_1^2)^{-a_{14}} (-k_2^2)^{-a_{15}}}{
(-k_3^2)^{a_8} [-(p_1 + k_3)^2 ]^{a_9}
[ -(k_3 - p_3)^2]^{a_{10}} }
\; .
\label{diE}
\eea
Here $s=(p_1+p_2)^2$ and $t=(p_1 + p_3)^2$ denote the Mandelstam invariants.
For later use, we note that $u=(p_2+p_3)^2 =-s-t$.

As we explain presently, the master integrals for these two families represent all master integrals needed to evaluate any
massless planar on-shell three-loop four-point scattering amplitude.
We explicitly find a basis where all master integrals have uniform transcendentality, 
and show that the differential equations are of the
Knizhnik-Zamolodchikov type \cite{Knizhnik:1984nr}. 
We explain how to solve the latter to all orders in the dimensional regularization 
parameter $\epsilon$, including all boundary constants, in a purely algebraic way,
for {\em all} master integrals.
The solution is expressed in terms of harmonic polylogarithms.
We explicitly write out the Laurent expansion in $\epsilon$ for all master integrals up to weight six.
Up to now, two analytical results for integrals
of this family were known: for the triple box without numerator~\cite{Smirnov:2003vi}
and for the tennis court diagram with a special numerator~\cite{Bern:2005iz}.

We would also like mention a perhaps surprising outcome of our analysis.
As a by-product of our calculation, we also obtained analytic results for single-scale 
integrals appearing in form factors.
Na{\"i}vely, the DE method cannot be applied to these cases, since
their scale dependence is trivially fixed by their engineering dimension. However,
they are a part of the system of differential equations for the more general four-point
integrals, where they enter as boundary values. The latter, however, are greatly
constrained by the finiteness of planar integrals in the $u$-channel as $u \to 0$.
As we will discuss in more detail below, these consistency conditions fix all
boundary constants, up to trivial propagator-type integrals.
In this way, one obtains results for non-trivial single-scale integrals, 
to any order in $\eps$. One may verify agreement with the planar form factor integrals computed in 
references \cite{Gehrmann:2006wg,Heinrich:2007at,Baikov:2009bg,Heinrich:2009be,Gehrmann:2010ue,Lee:2010ik}. 
We find this way of computing these integrals rather elegant.

Let us now explain why the master integrals computed above are sufficient to describe
{\em all} the families of three-loop four-point planar on-shell massless diagrams
(which have fifteen indices, with the number of positive indices being lower or equal to ten.)
To see this, let us first observe that we can construct integrals with
the maximal number of positive indices by building graphs with trivalent vertices.
A quartic vertex can always be obtained as a special case, with one index being zero.
Let us then observe than the triple box and the tennis court
are the only graphs composed of cubic vertices with no triangles as subgraphs.
So, any other graph has at least one triangle subgraph. 
In this case, one can use the presence of such a triangle
and apply IBP relations to reduce an index, either internal
or external, of this triangle to zero starting from positive values~\cite{Chetyrkin:1981qh}.
In graph-theoretical language, this means shrinking the corresponding line to a point.
By analyzing various graphs obtained by this procedure
we can see that the resulting reduced graphs can be also obtained, in some way,
either from the triple box or/and from the tennis court.
 
This paper is organized as follows. 
In section \ref{sec:basis}, we explain the strategy we use for finding integrals that
give rise to pure functions of uniform transcendentality, providing several examples.
We then present our basis choice for the master integrals. In section \ref{sec:diffeq},
we present the differential equations satisfied by the latter, and explain how to solve
them in the $\eps$ expansion. We also discuss physical boundary conditions.
We analyze the structure of the solution.
Explicit results for the ten-propagator integrals are relegated to Appendix~B, and
for all integrals to the ancillary files {\tt resultA.m} and {\tt resultE.m}.
For convenience, we also present in these files the corresponding matrices appearing in the differential
equations.
We conclude in section \ref{sec:discussion}.

\section{Choice of integral basis}
\label{sec:basis}

An important part of the result of this paper is to provide a basis of master integrals
for the families of Feynman integrals A and E where each basis element is
a pure function of uniform weight.
Ideas for how to construct such a basis where outlined in ref. \cite{Henn:2013pwa}.
In practice, these lead to very useful criteria for choosing master integrals.
Here we wish to explain the criteria that we found most useful in the present context.

When constructing good candidate integrals at $(L+1)$ loops, it is very convenient
to have a solution of the problem at $L$ loops at hand, as one can often infer from this
which integrals to choose at the next loop order. We will see this in more detail in the following examples.
In the present case, the solution at two loops was presented in \cite{Henn:2013pwa}.

\subsection{Example 1: massless bubble subintegrals}
\begin{figure}[t] 
\captionsetup[subfigure]{labelformat=empty}
\begin{center}
\subfloat[(1)]{\includegraphics[width=0.17\textwidth]{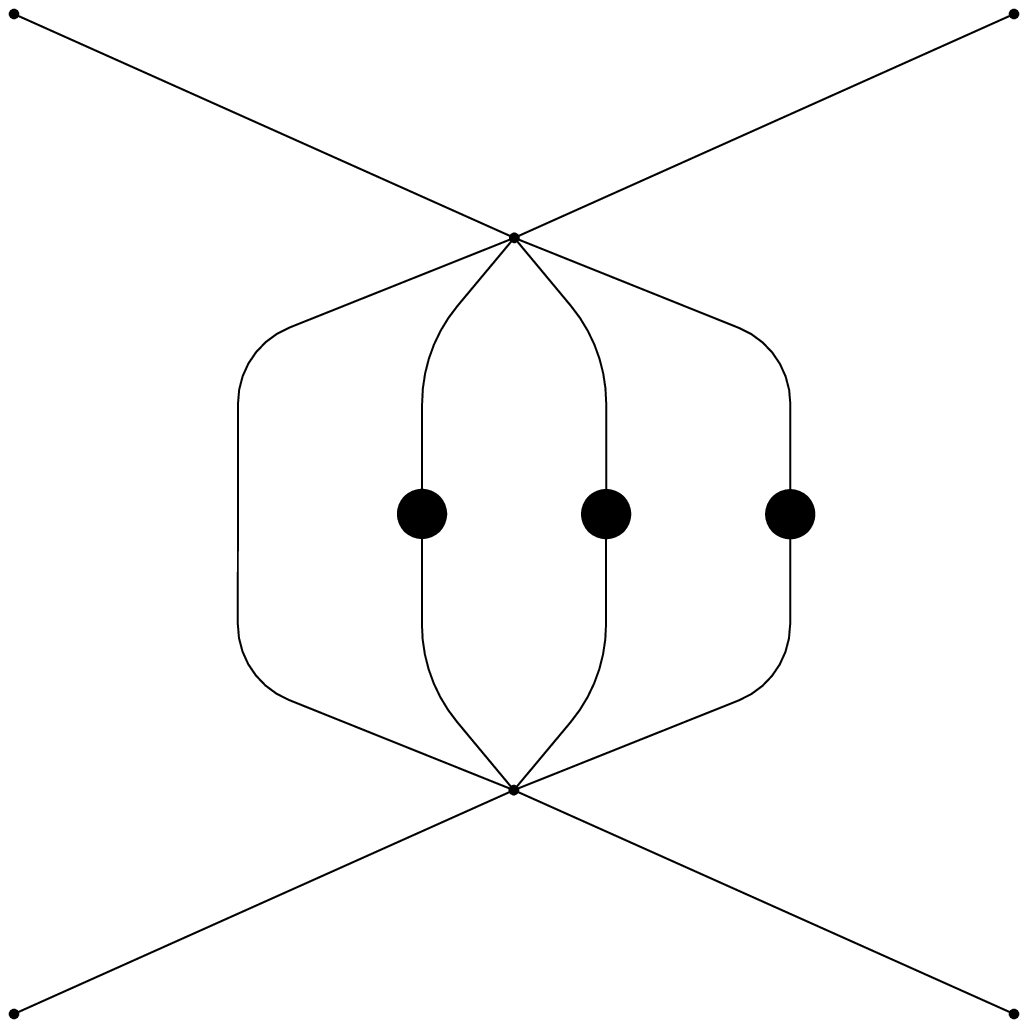}}
\subfloat[(2)]{\includegraphics[width=0.17\textwidth]{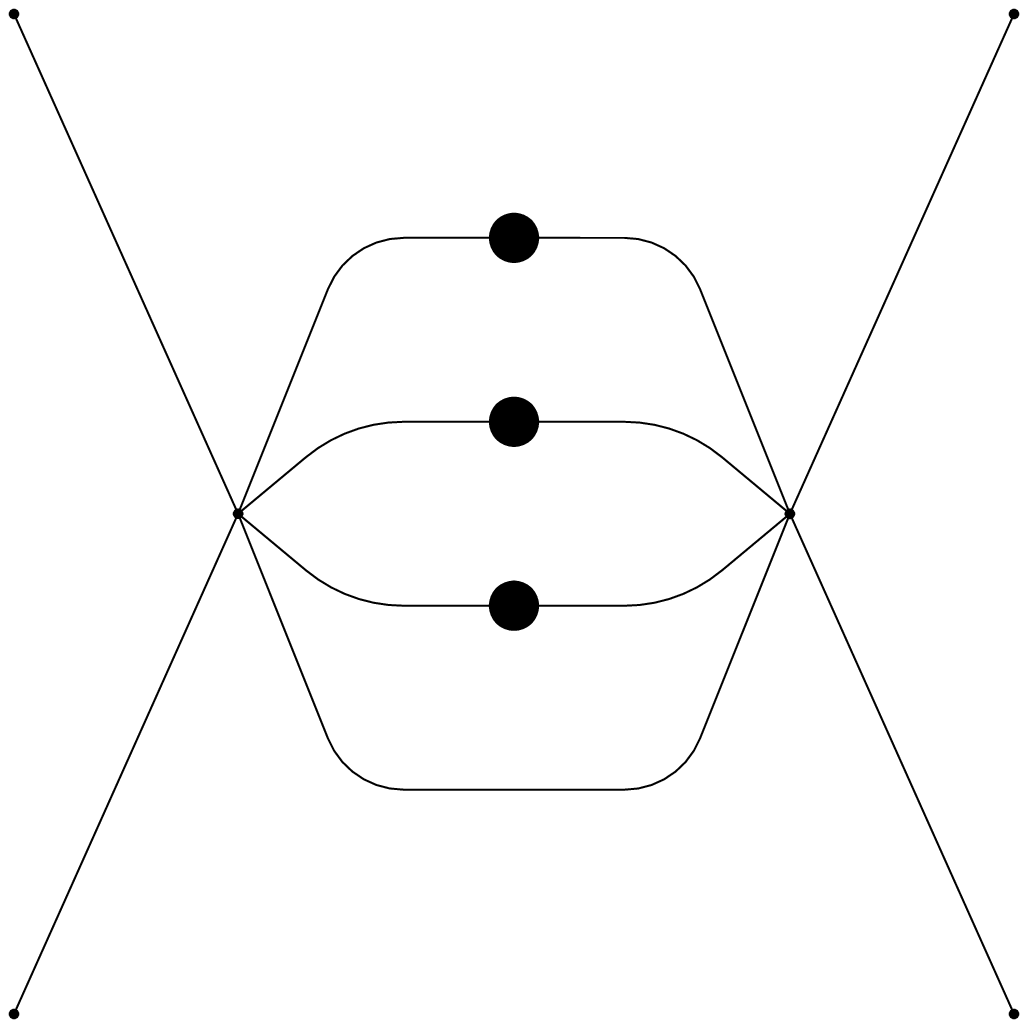}}
\subfloat[(3)]{\includegraphics[width=0.17\textwidth]{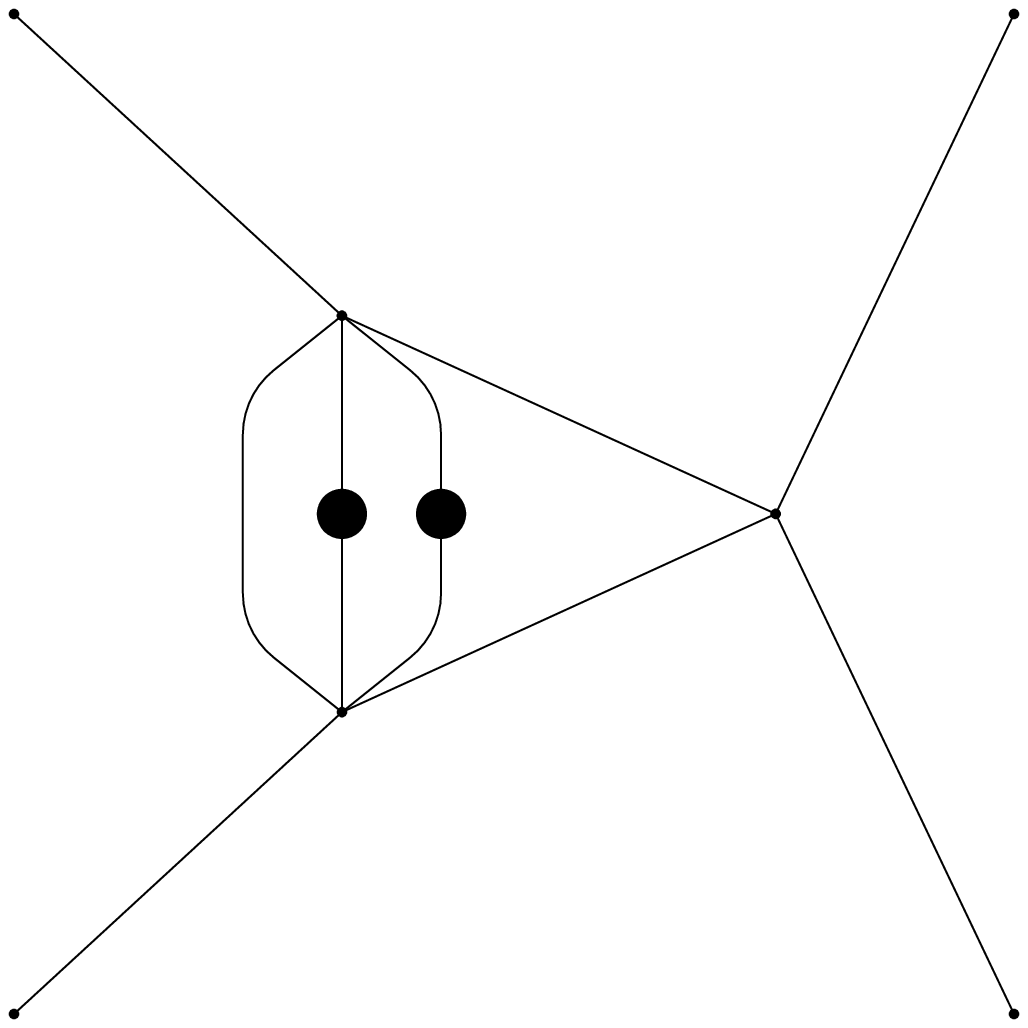}}
\subfloat[(4)]{\includegraphics[width=0.17\textwidth]{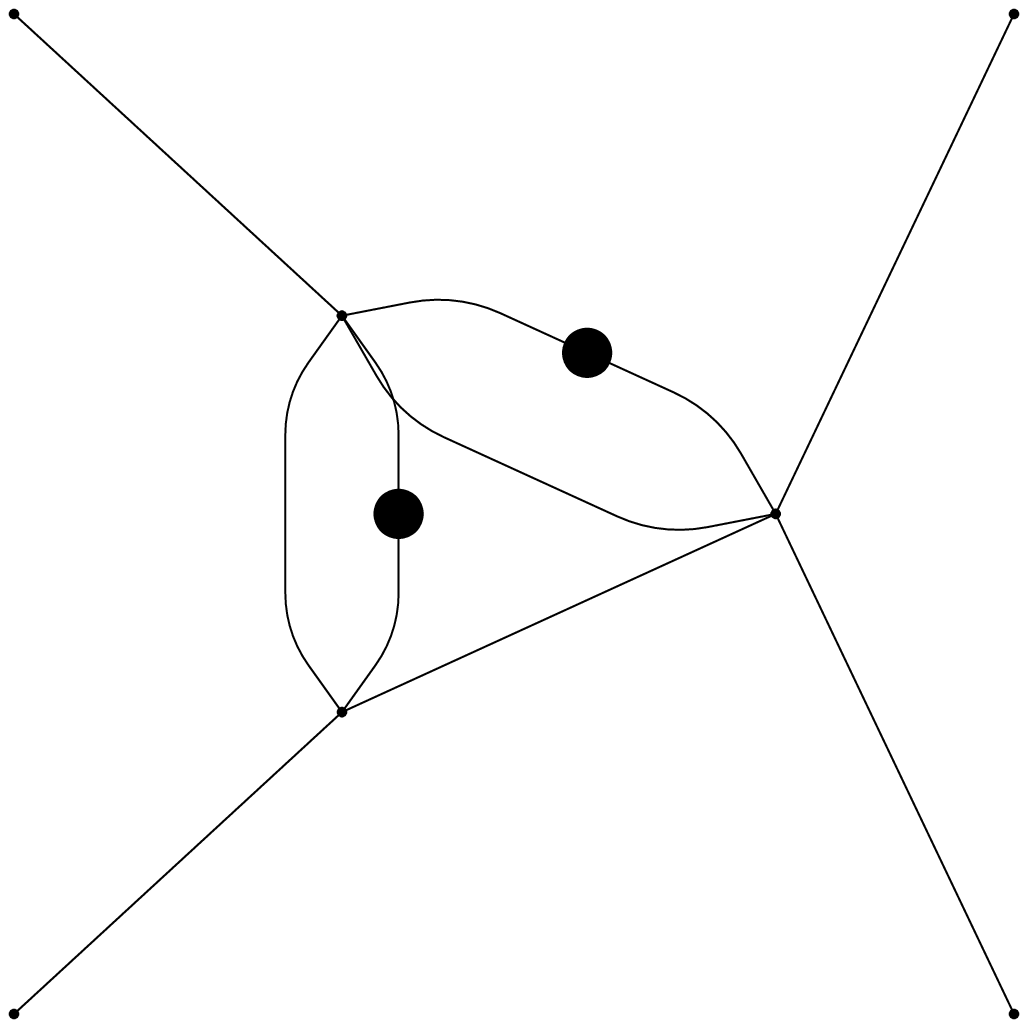}}
\subfloat[(5)*]{\includegraphics[width=0.17\textwidth]{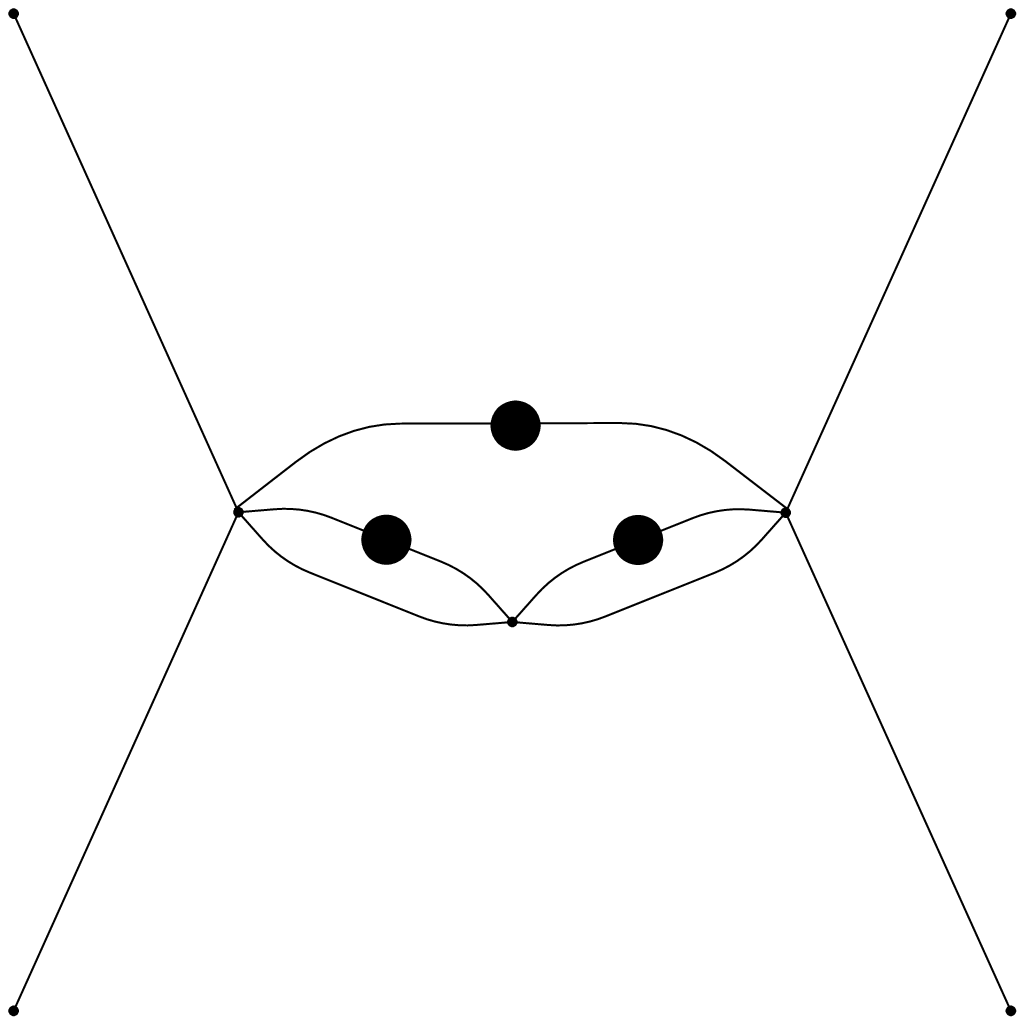}}
\newline
\subfloat[(6)]{\includegraphics[width=0.17\textwidth]{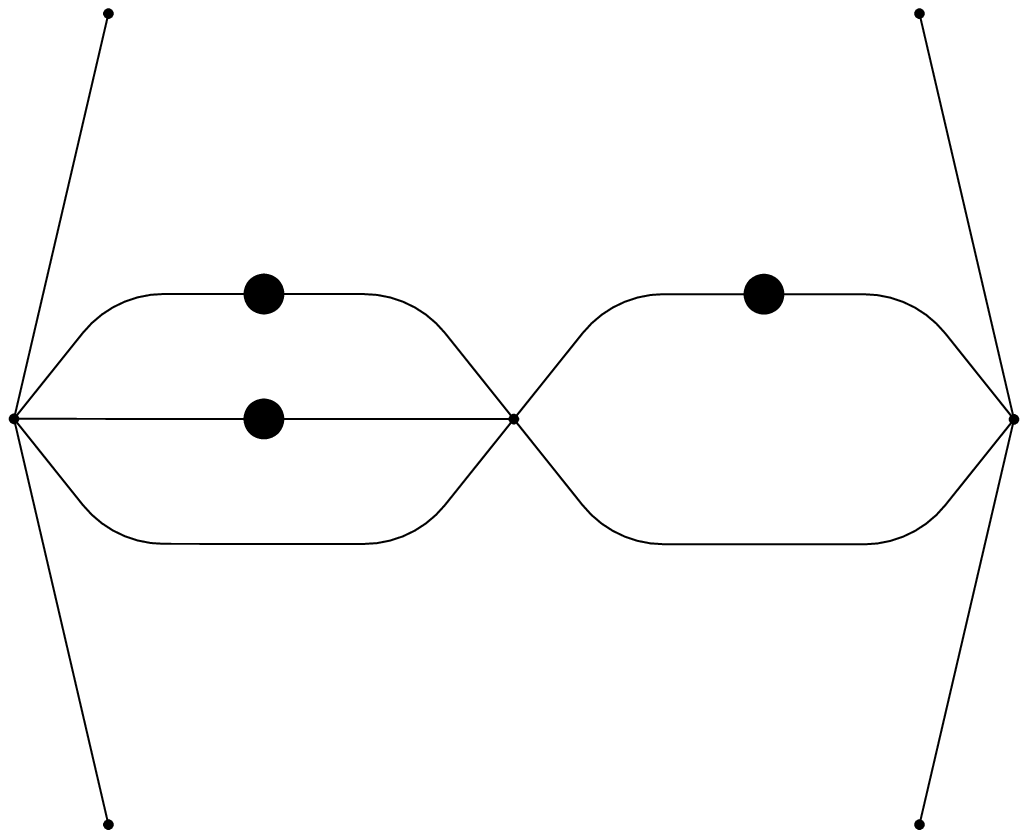}}
\subfloat[(7)]{\includegraphics[width=0.17\textwidth]{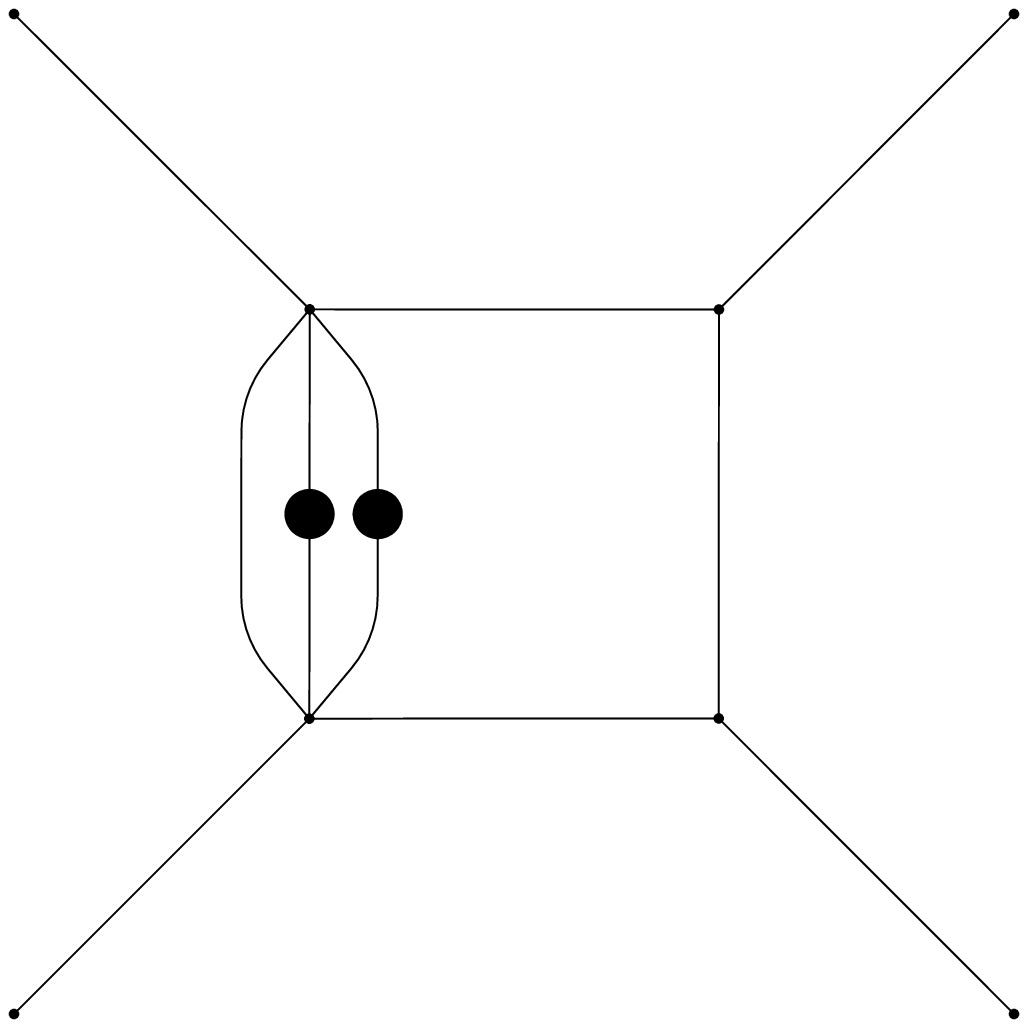}}
\subfloat[(8)]{\includegraphics[width=0.17\textwidth]{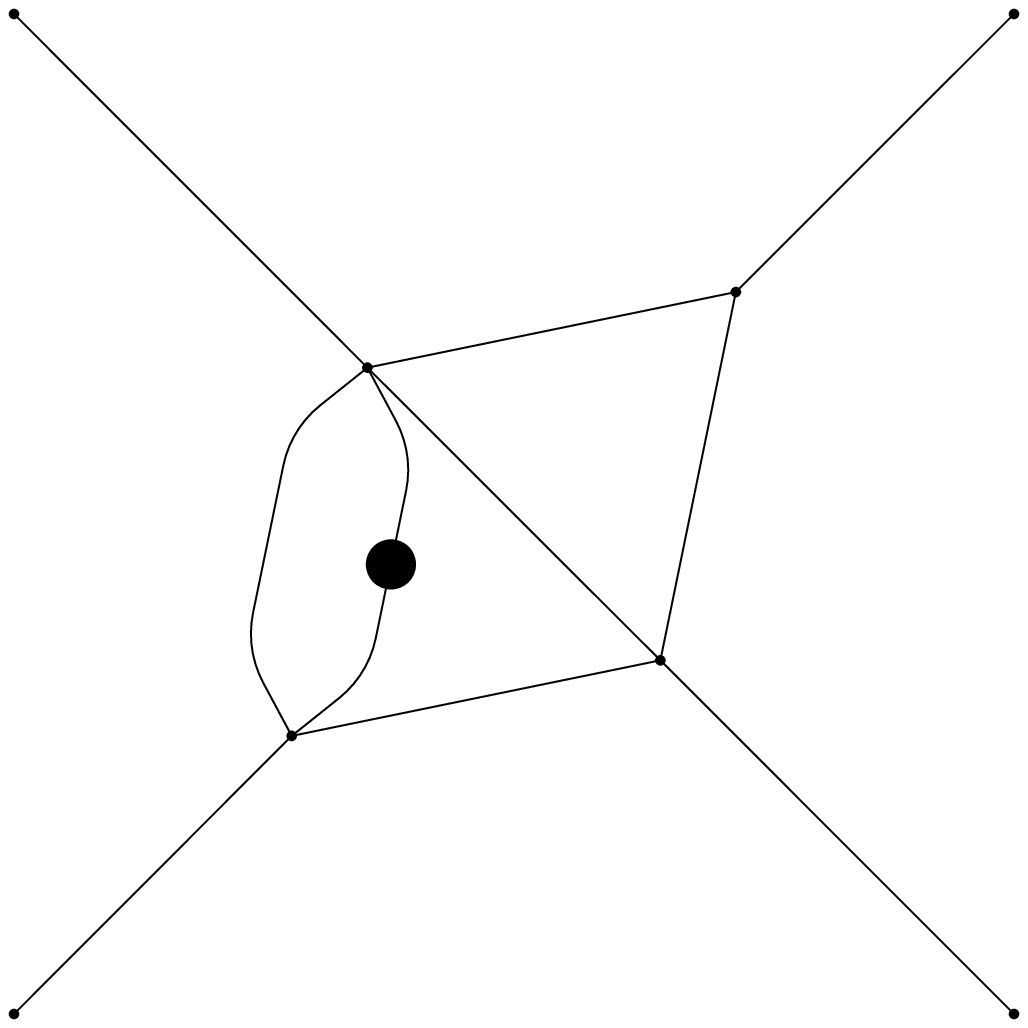}}
\subfloat[(9), (14)*]{\includegraphics[width=0.17\textwidth]{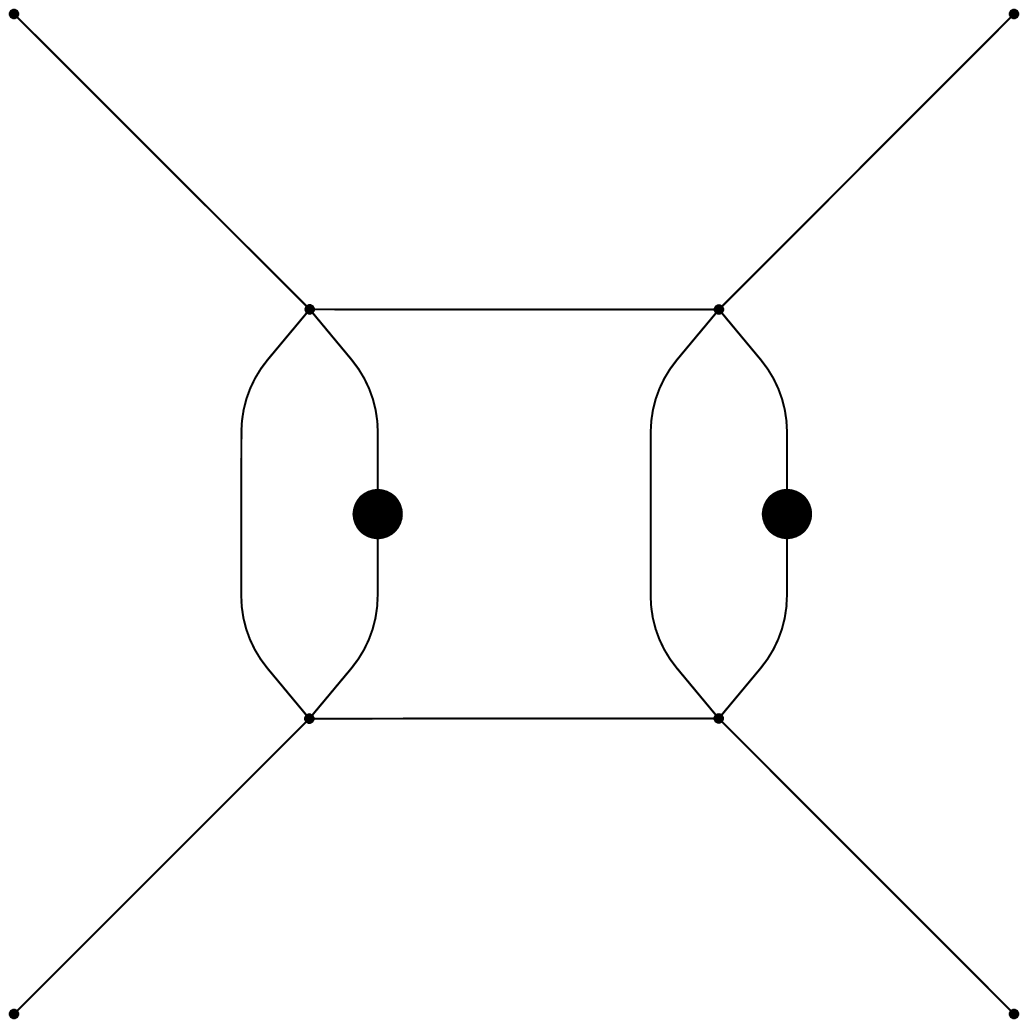}}
\subfloat[(10)]{\includegraphics[width=0.17\textwidth]{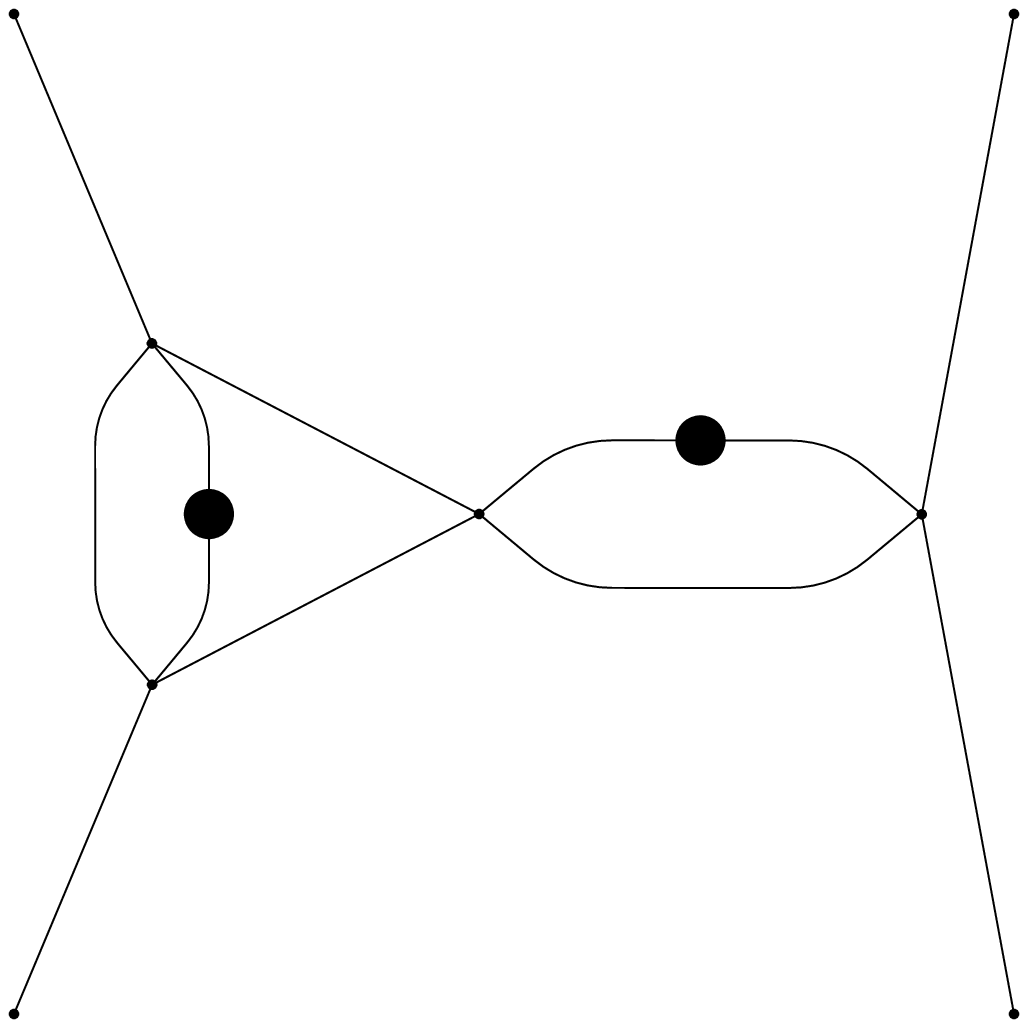}}
\newline
\subfloat[(11)]{\includegraphics[width=0.17\textwidth]{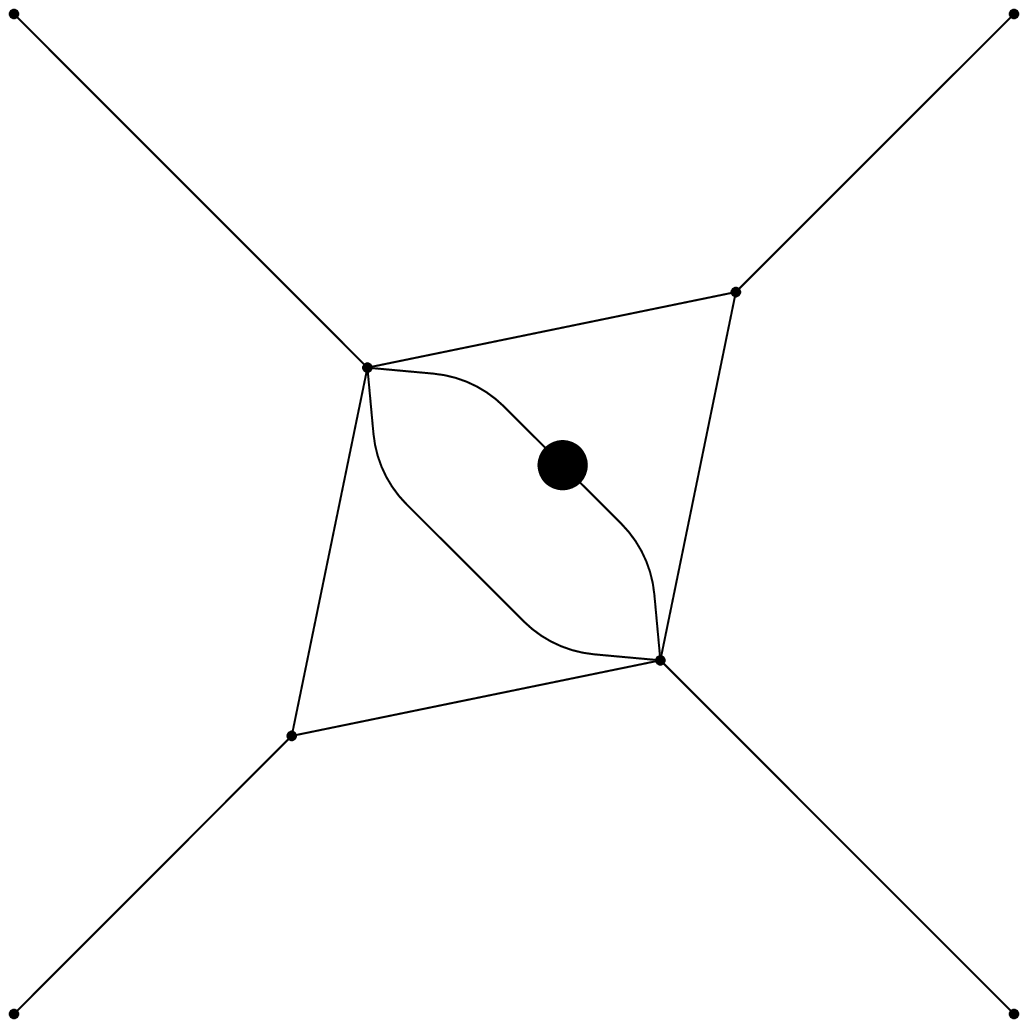}}
\subfloat[(12)]{\includegraphics[width=0.17\textwidth]{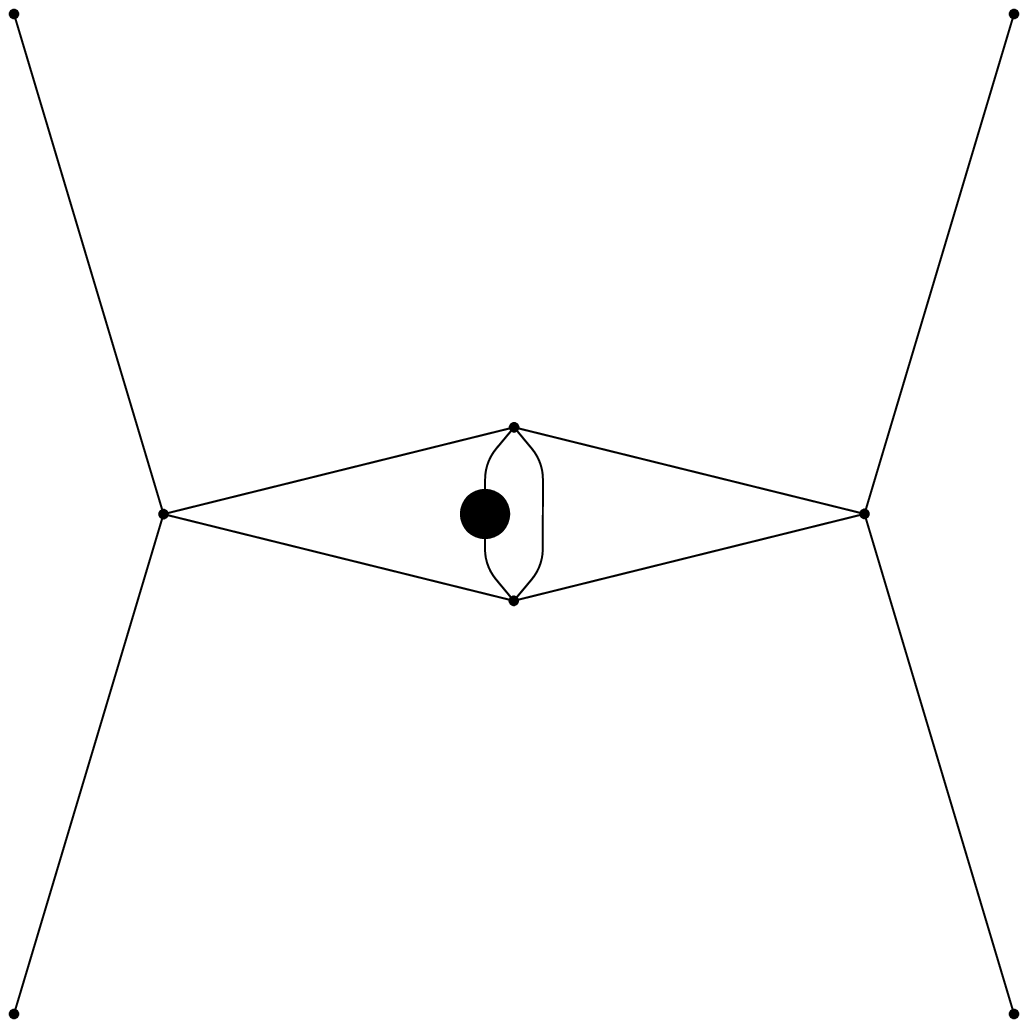}}
\subfloat[(13)]{\includegraphics[width=0.17\textwidth]{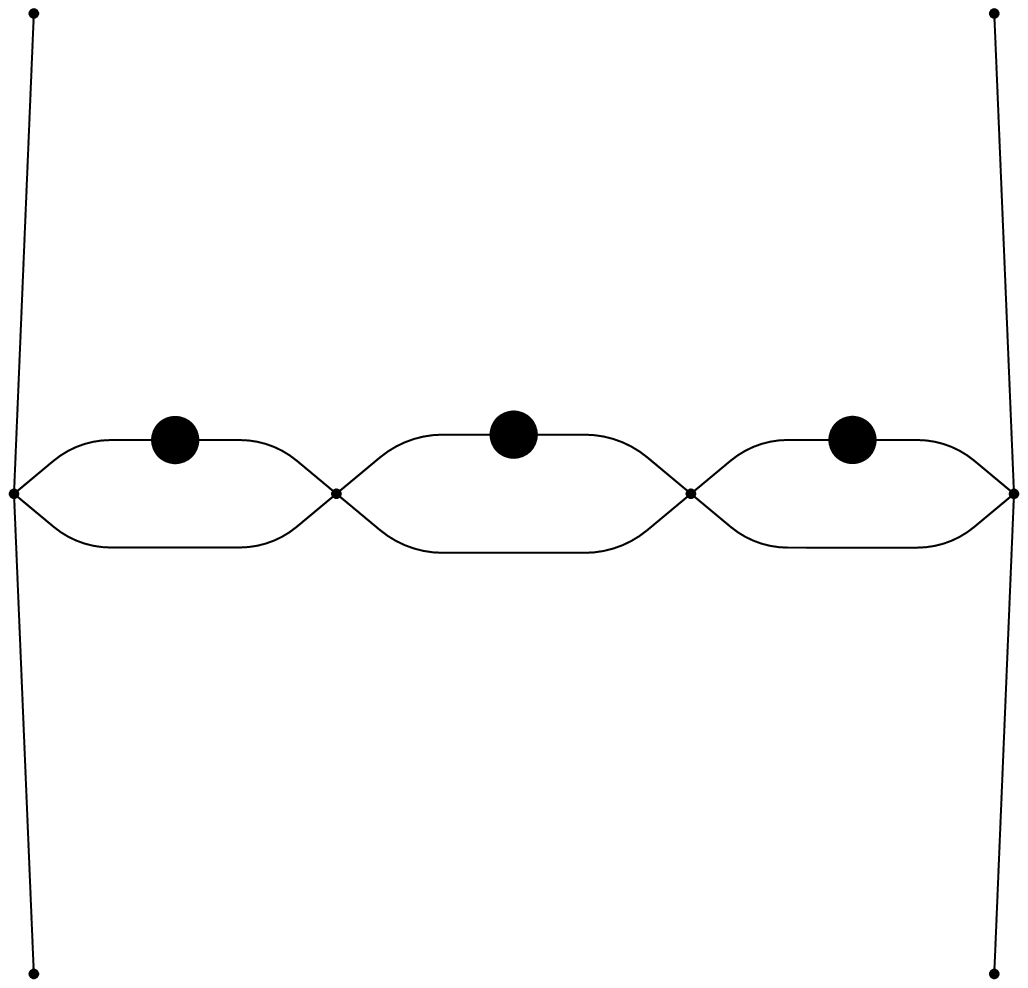}}
\subfloat[(18)*, (19)]{\includegraphics[width=0.17\textwidth]{ladder18.eps}}
\subfloat[(22), (23)*]{\includegraphics[width=0.17\textwidth]{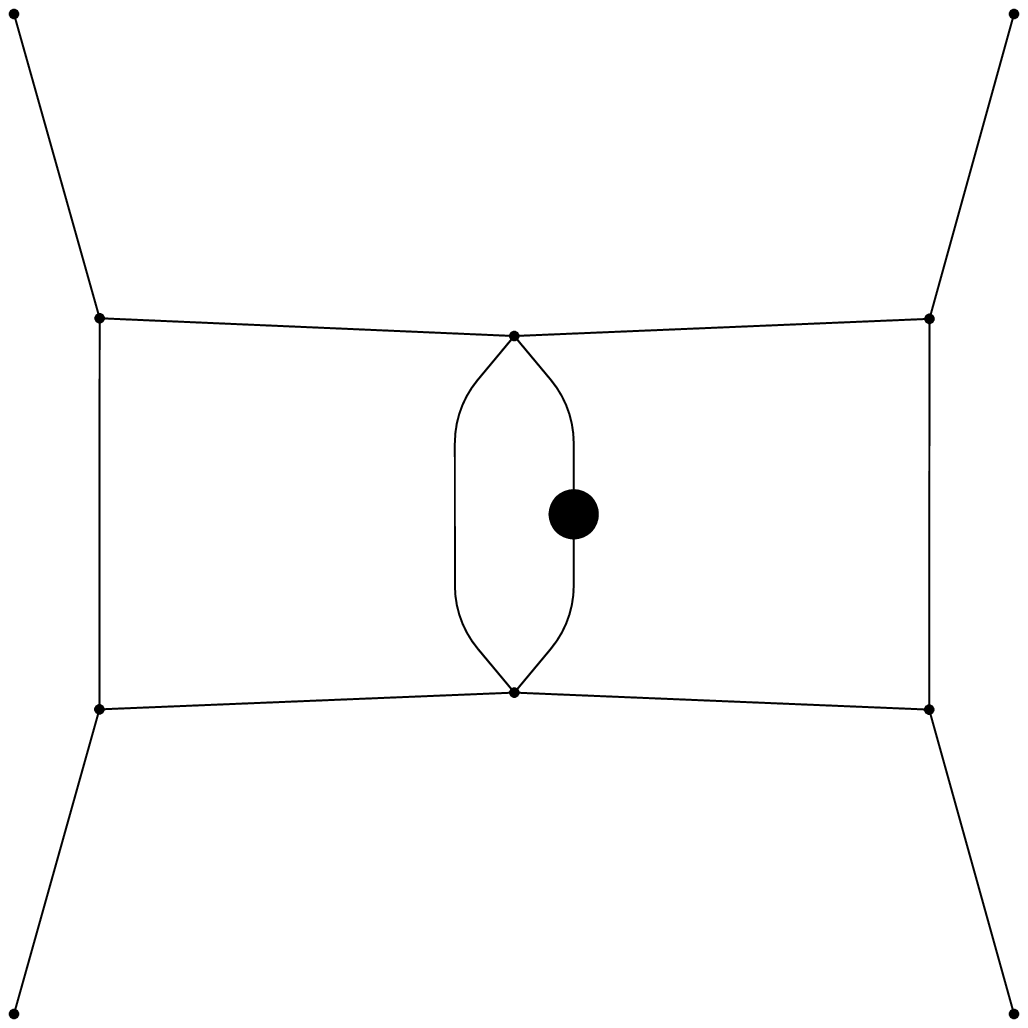}}
\caption{Master integrals for integral family A that have bubble subintegrals. 
Dots denote doubled propagators. 
An asterisk indicates that there are numerator factors not shown in the figure.
}
\label{fig:basisladders1}
\end{center}
\end{figure}

\begin{figure}[h] 
\captionsetup[subfigure]{labelformat=empty}
\begin{center}
\subfloat[(1)]{\includegraphics[width=0.17\textwidth]{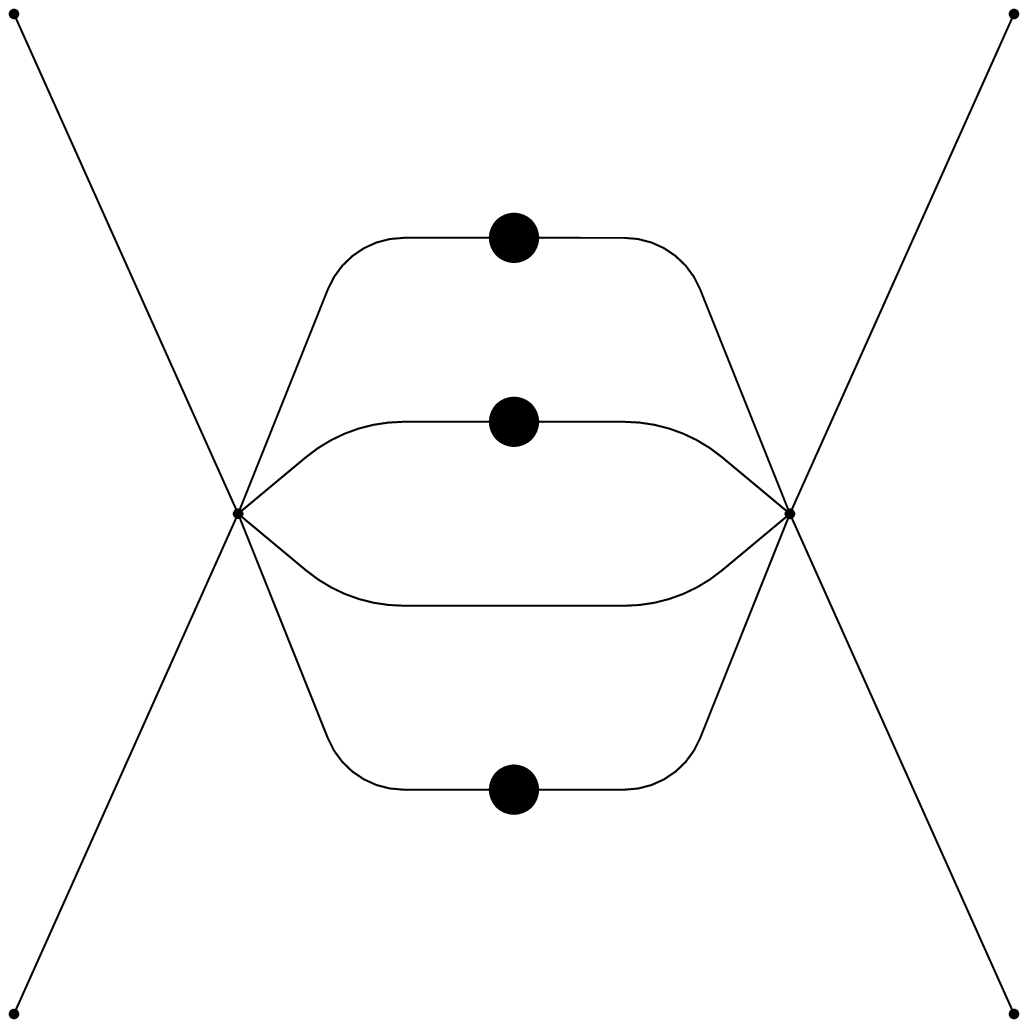}}
\subfloat[(2)]{\includegraphics[width=0.17\textwidth]{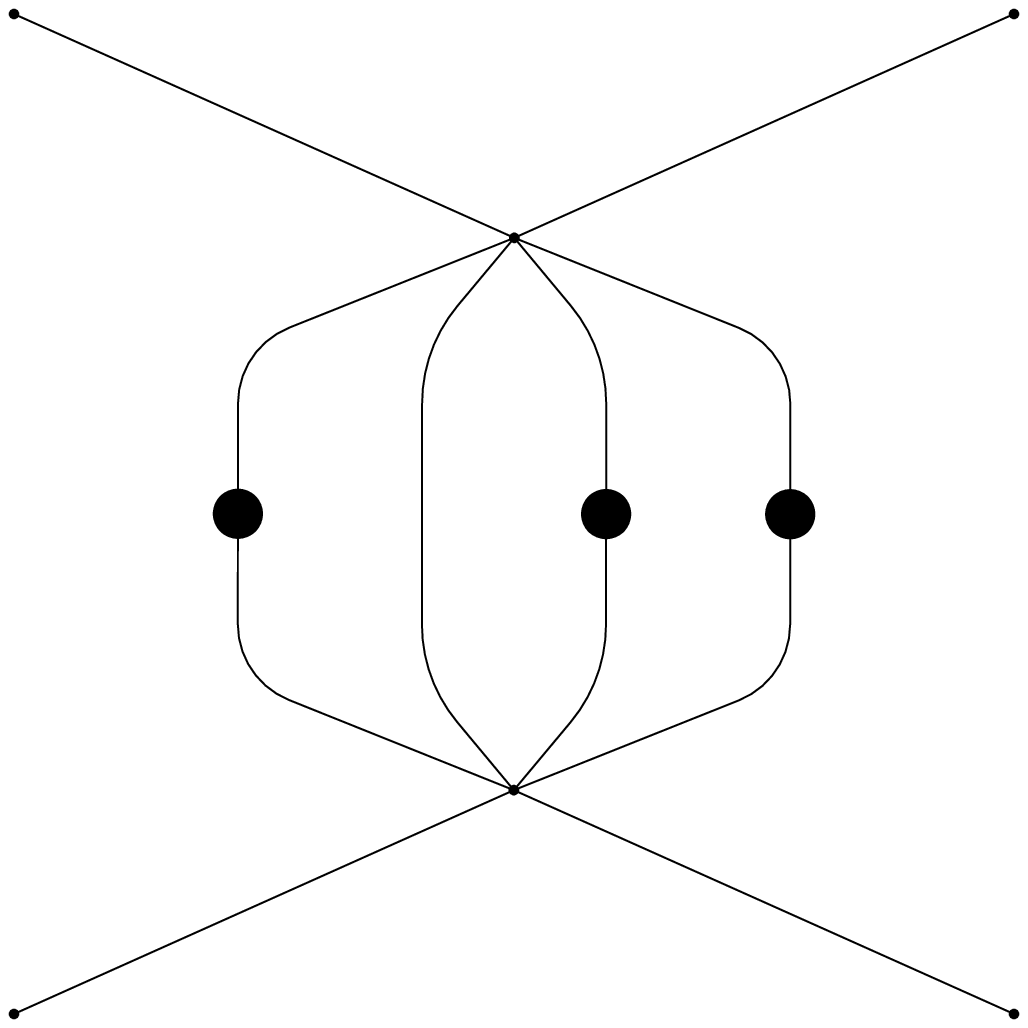}}
\subfloat[(3)]{\includegraphics[width=0.17\textwidth]{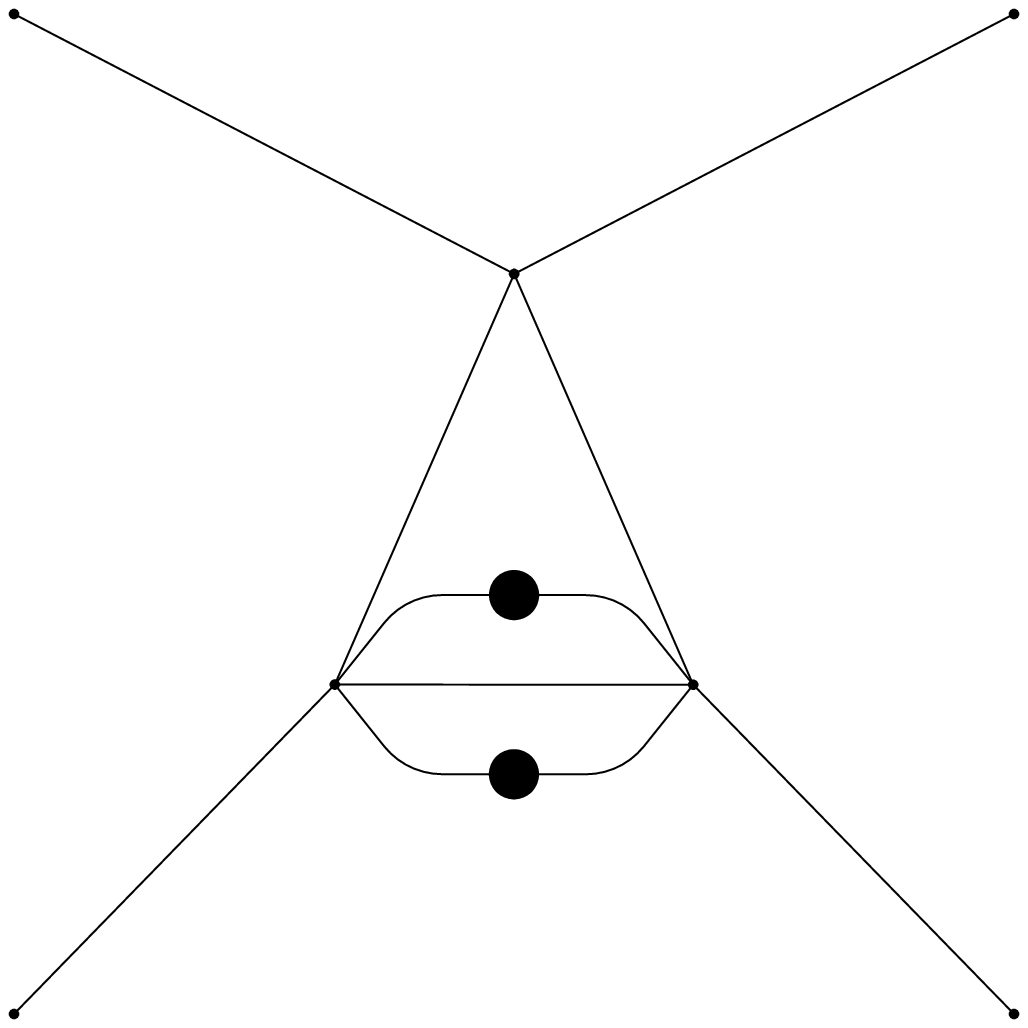}}
\subfloat[(4)]{\includegraphics[width=0.17\textwidth]{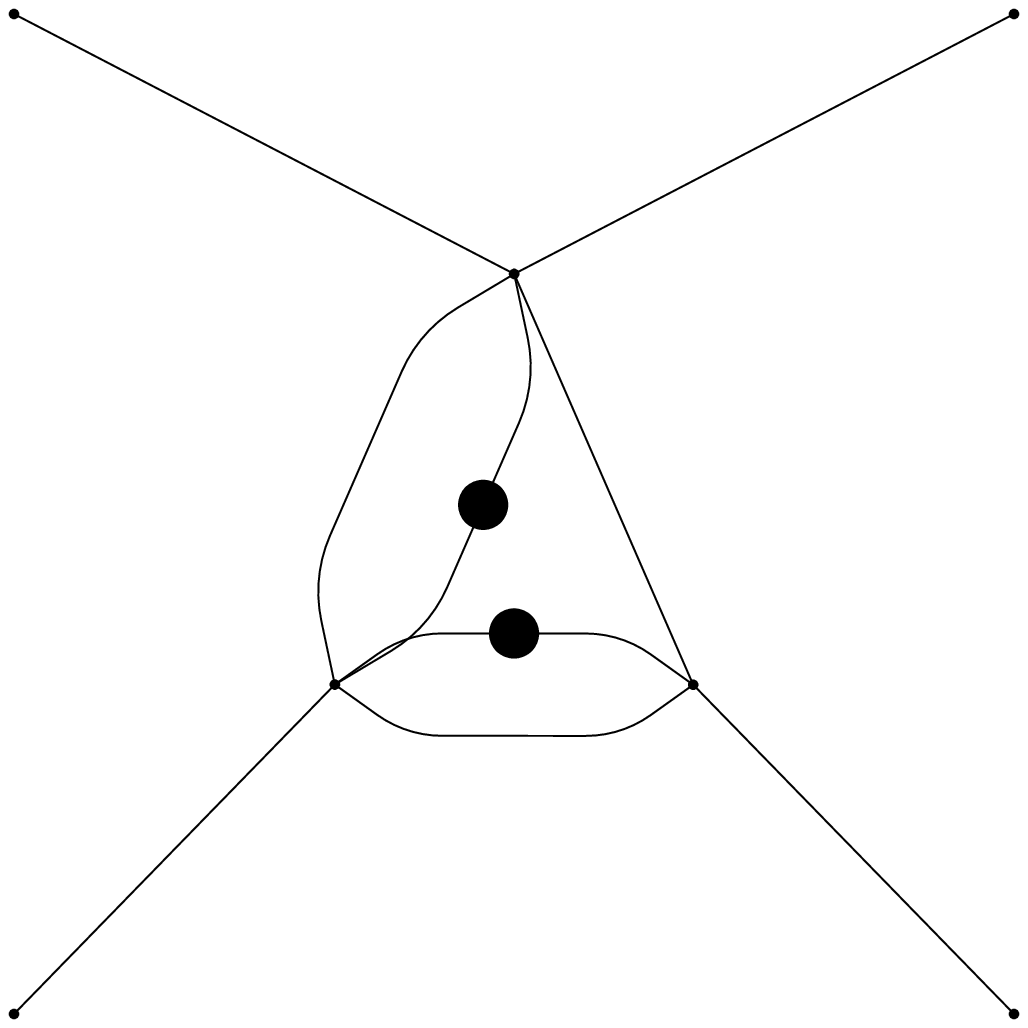}}
\subfloat[(5)]{\includegraphics[width=0.17\textwidth]{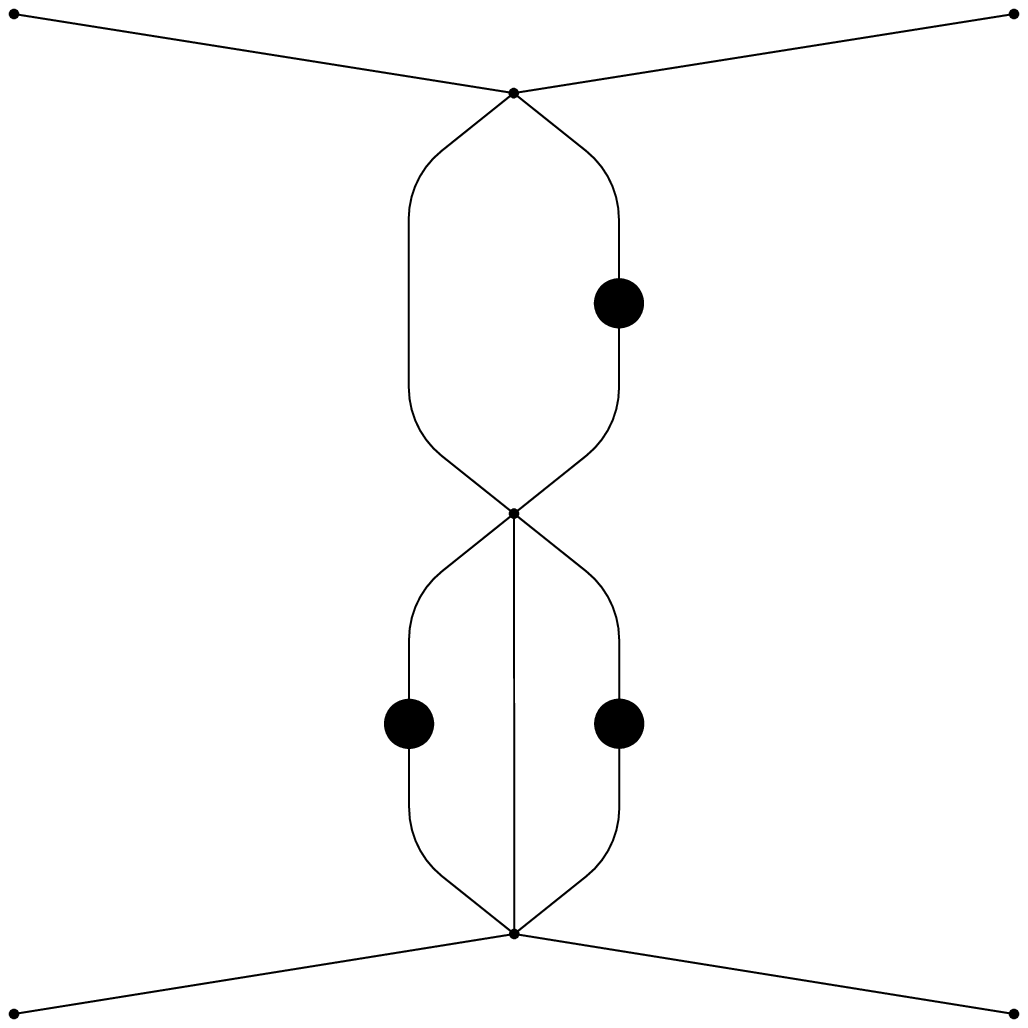}}
\subfloat[(6)*]{\includegraphics[width=0.17\textwidth]{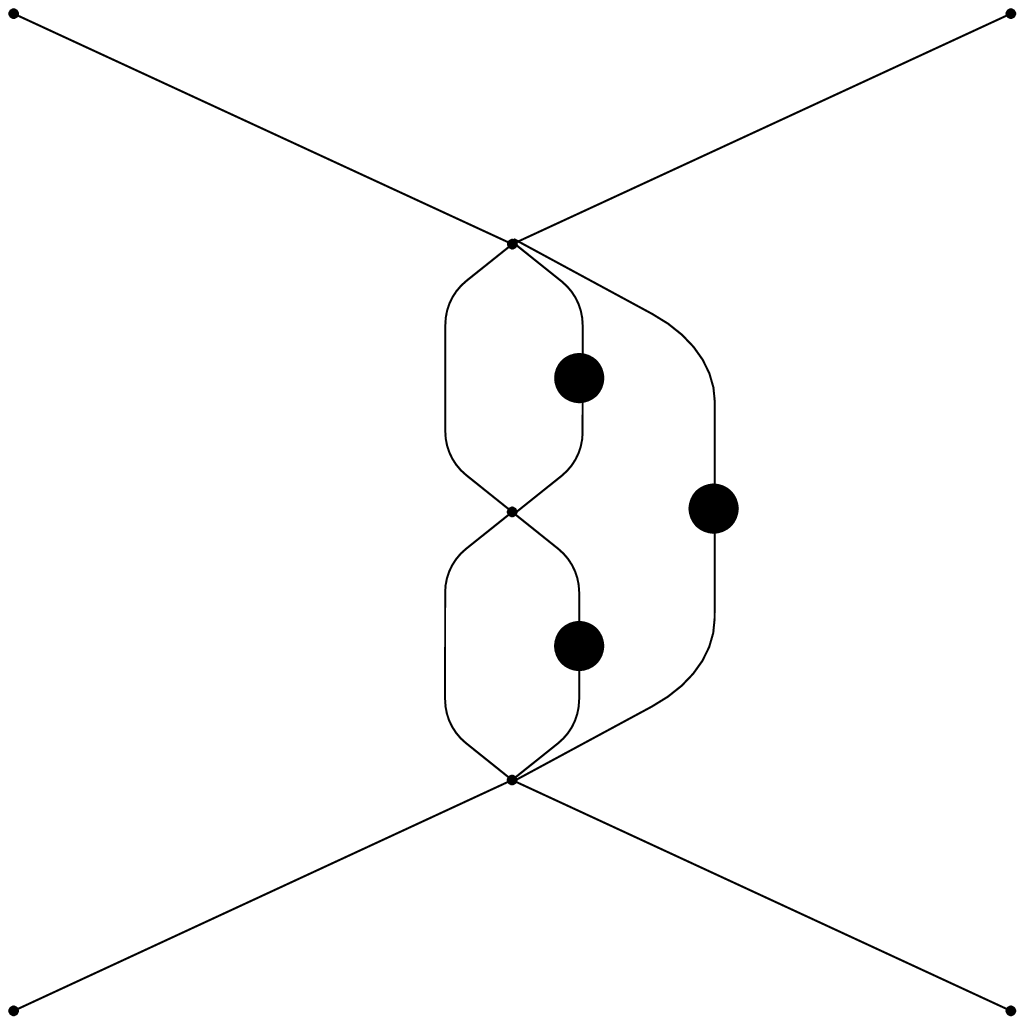}}
\newline
\subfloat[(7)]{\includegraphics[width=0.17\textwidth]{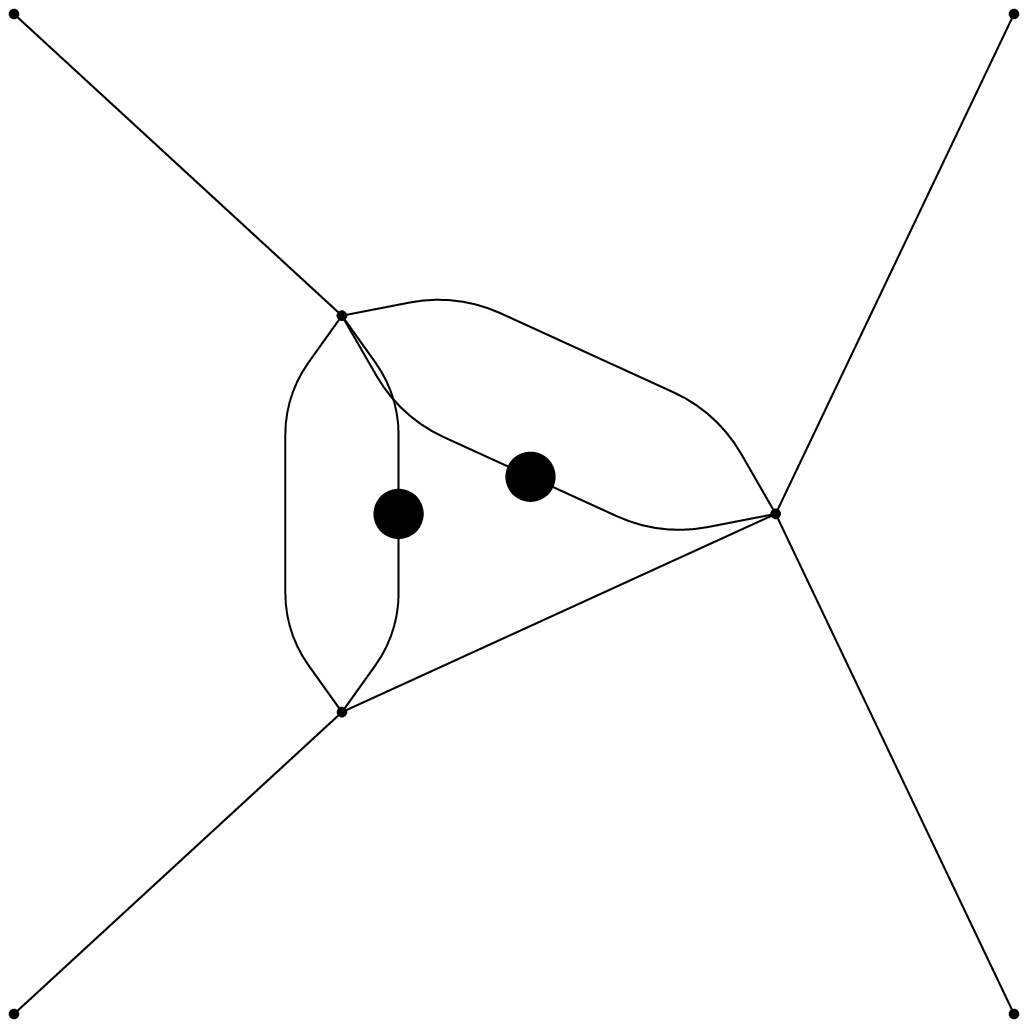}}
\subfloat[(8)*]{\includegraphics[width=0.17\textwidth]{tennis8n.eps}}
\subfloat[(9)]{\includegraphics[width=0.17\textwidth]{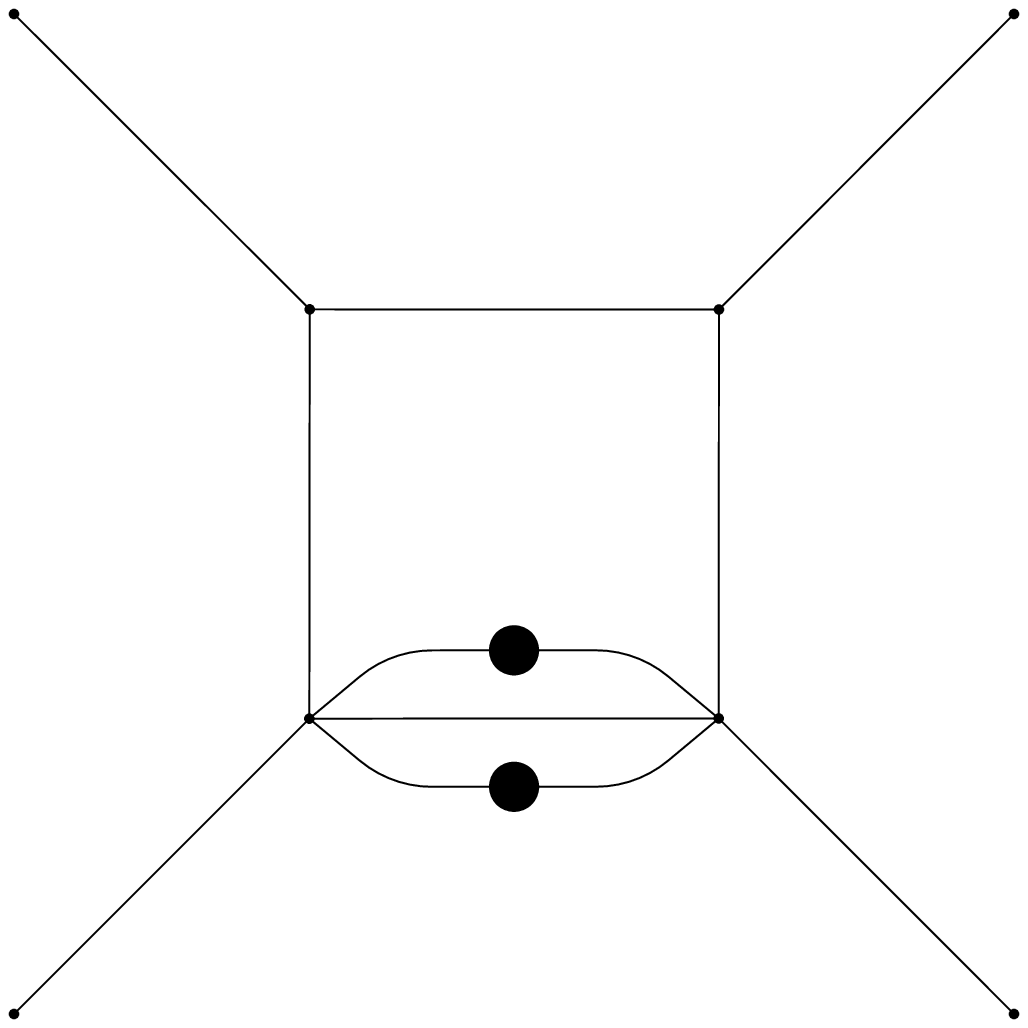}}
\subfloat[(10)]{\includegraphics[width=0.17\textwidth]{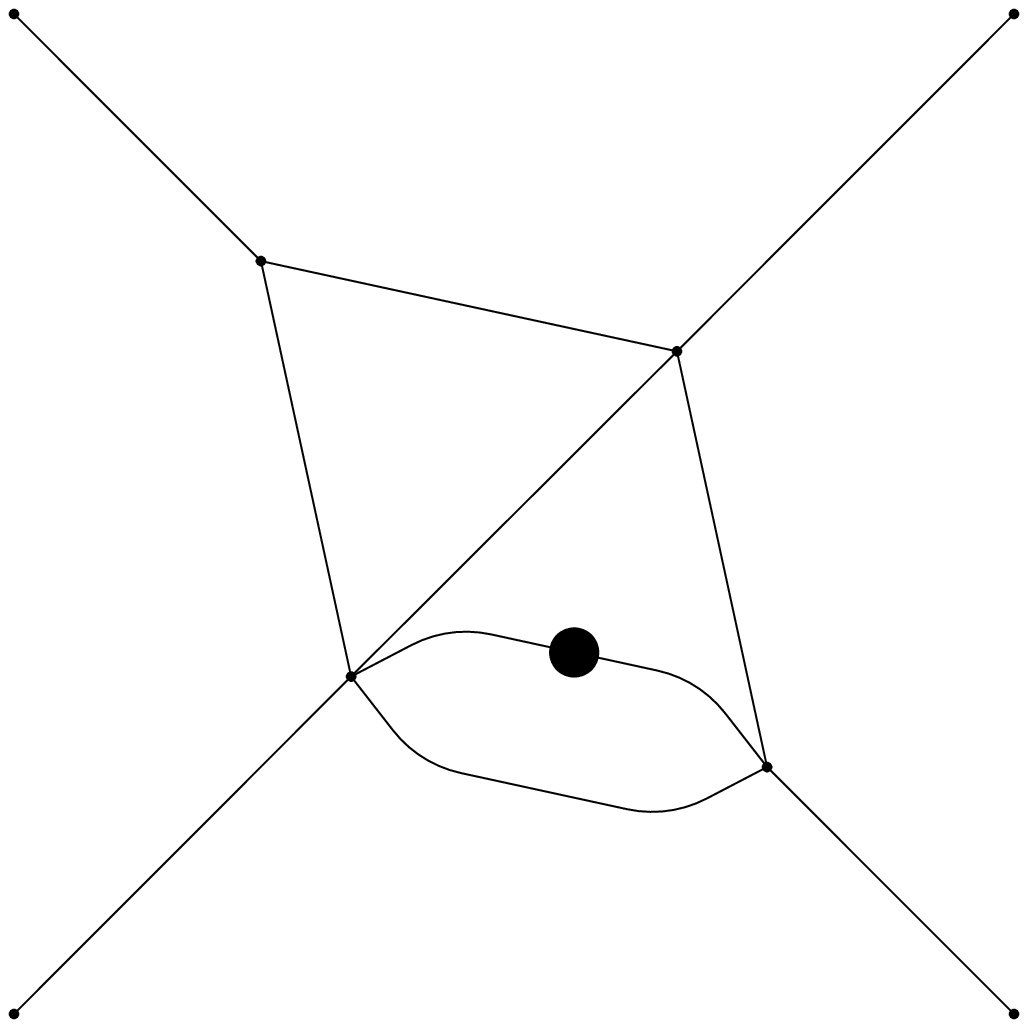}}
\subfloat[(11)]{\includegraphics[width=0.17\textwidth]{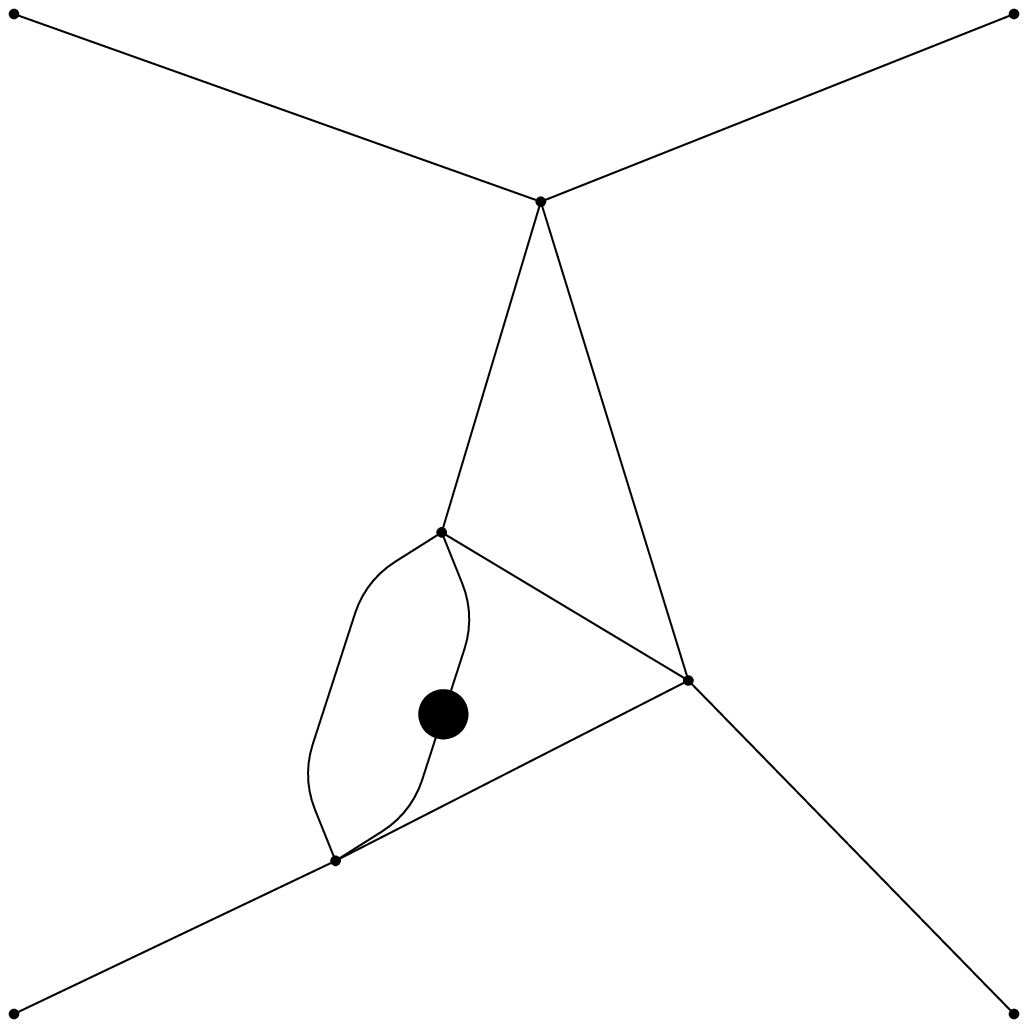}}
\subfloat[(12)]{\includegraphics[width=0.17\textwidth]{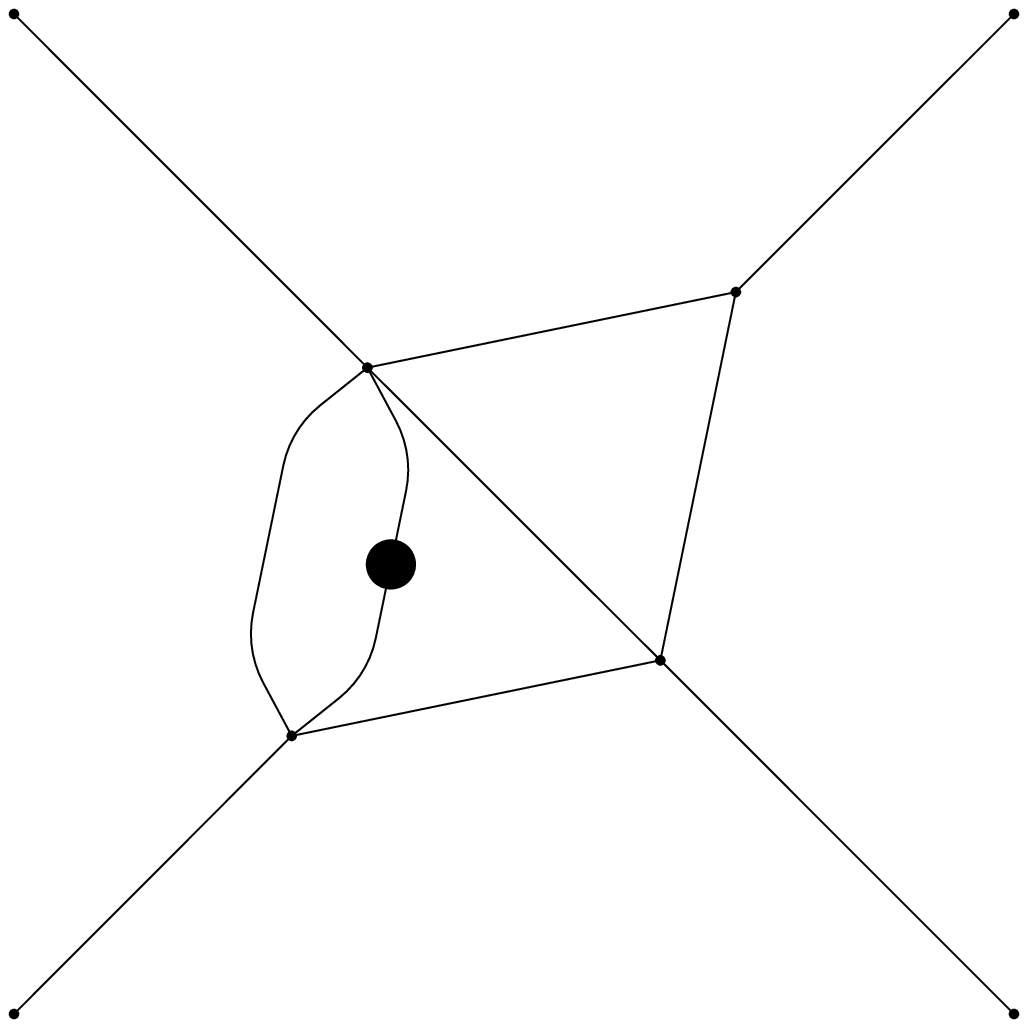}}
\newline
\subfloat[(13)]{\includegraphics[width=0.17\textwidth]{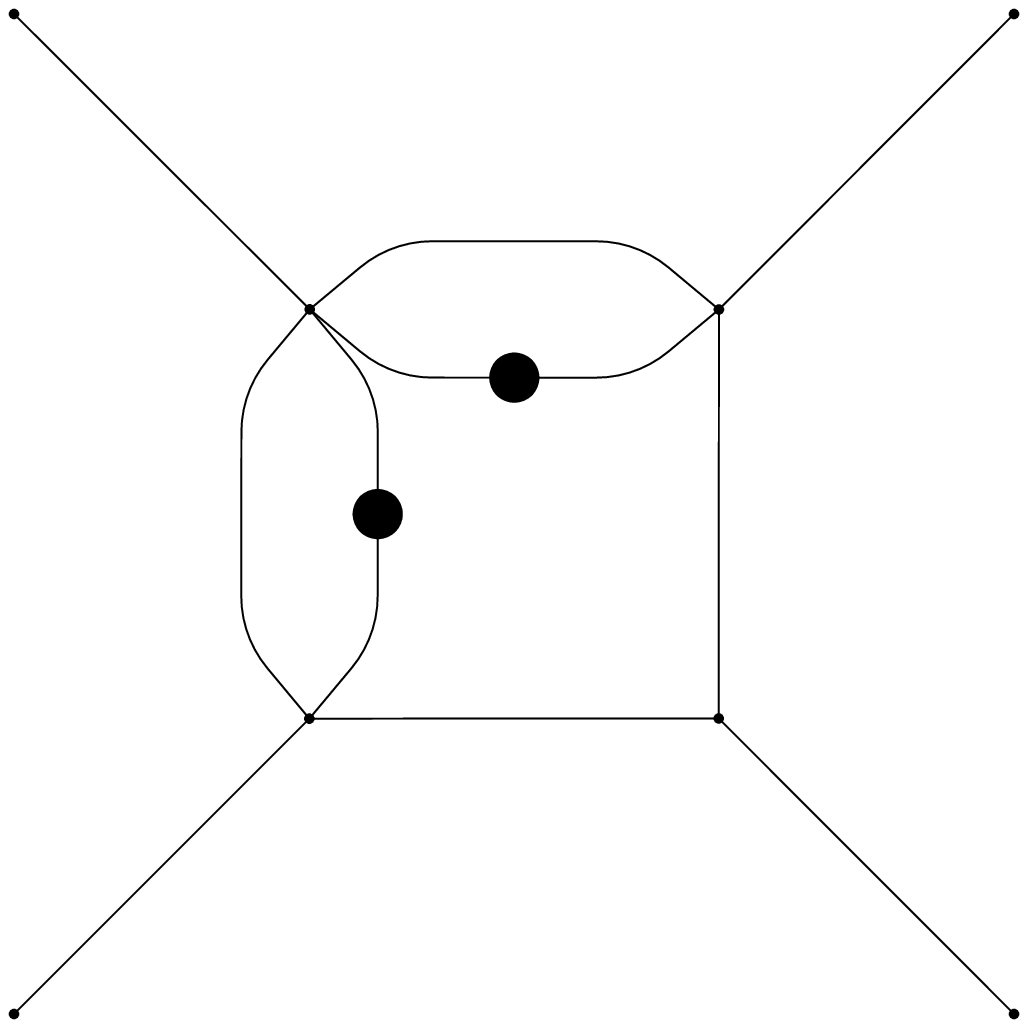}}
\subfloat[(14)]{\includegraphics[width=0.17\textwidth]{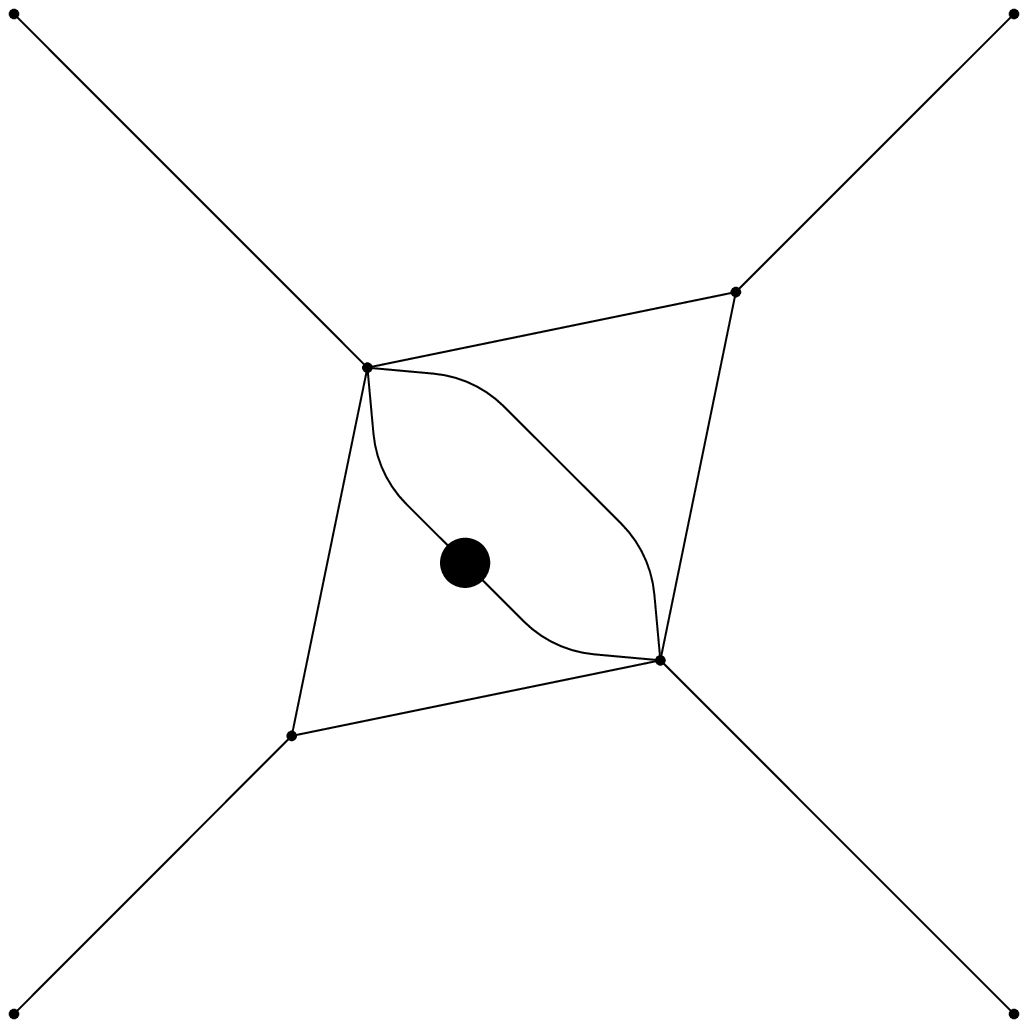}}
\subfloat[(17)*]{\includegraphics[width=0.17\textwidth]{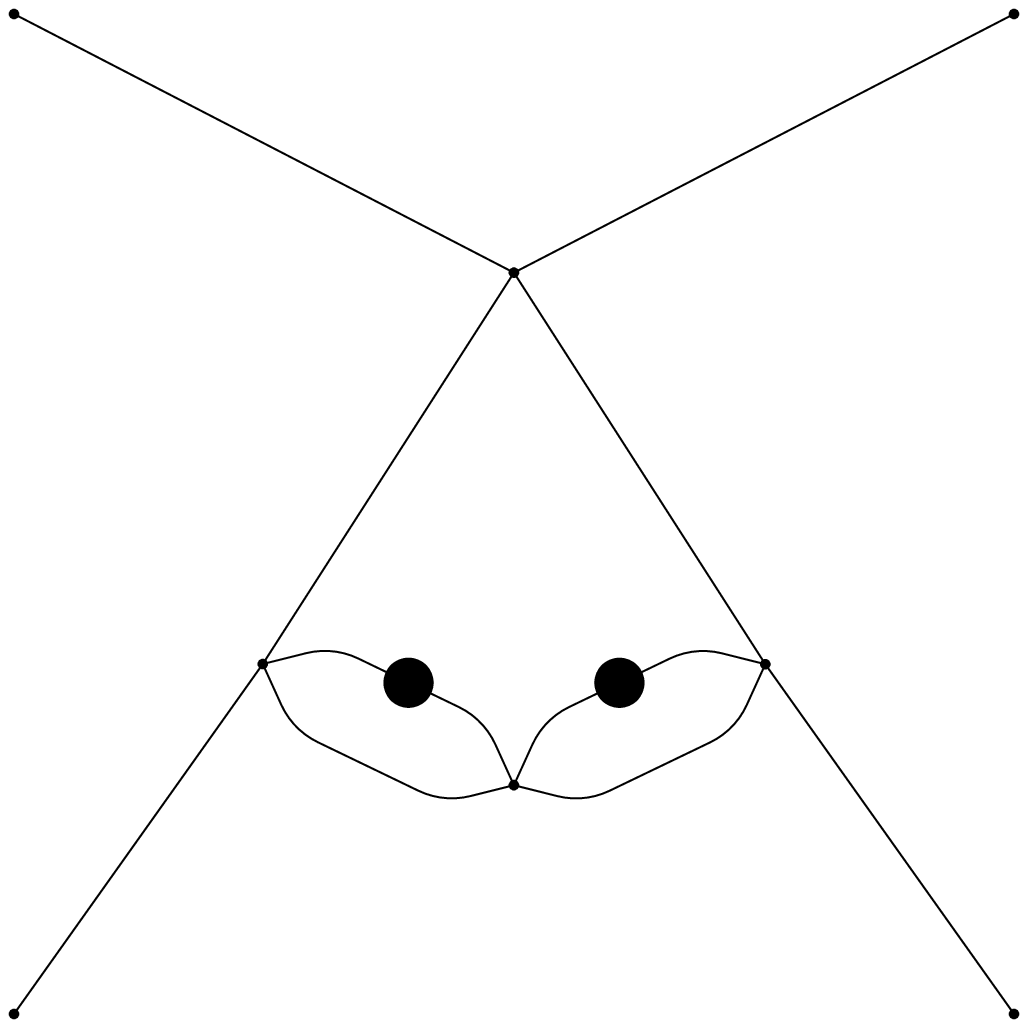}}
\subfloat[(18)]{\includegraphics[width=0.17\textwidth]{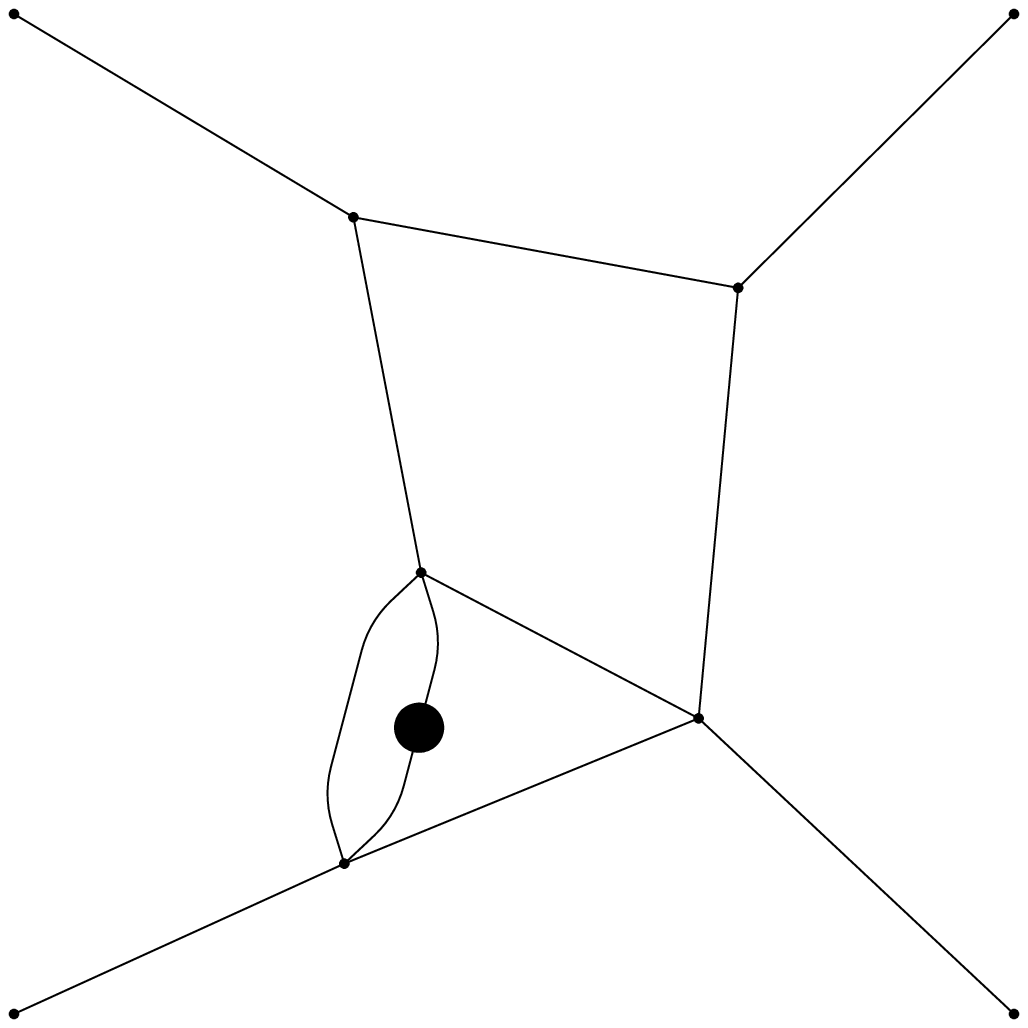}}
\subfloat[(19)]{\includegraphics[width=0.17\textwidth]{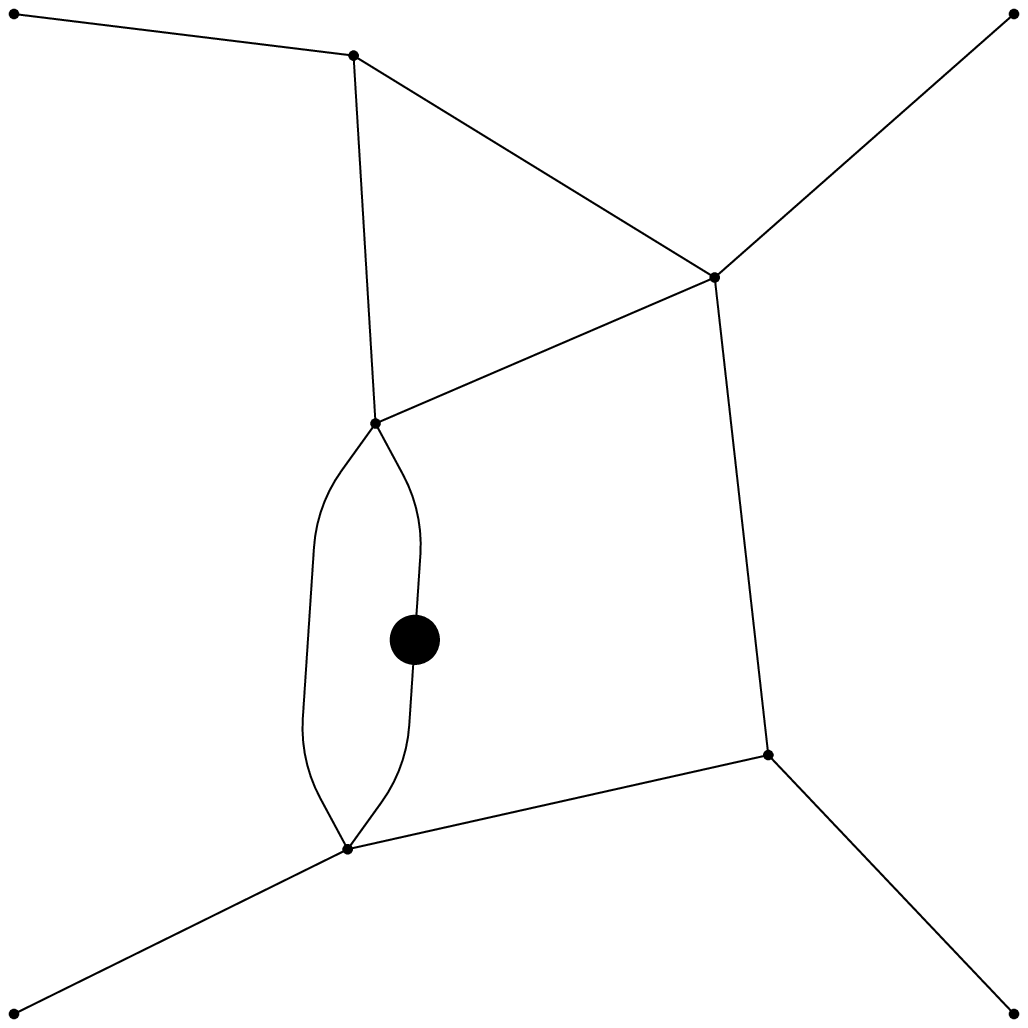}}
\subfloat[(25)*]{\includegraphics[width=0.17\textwidth]{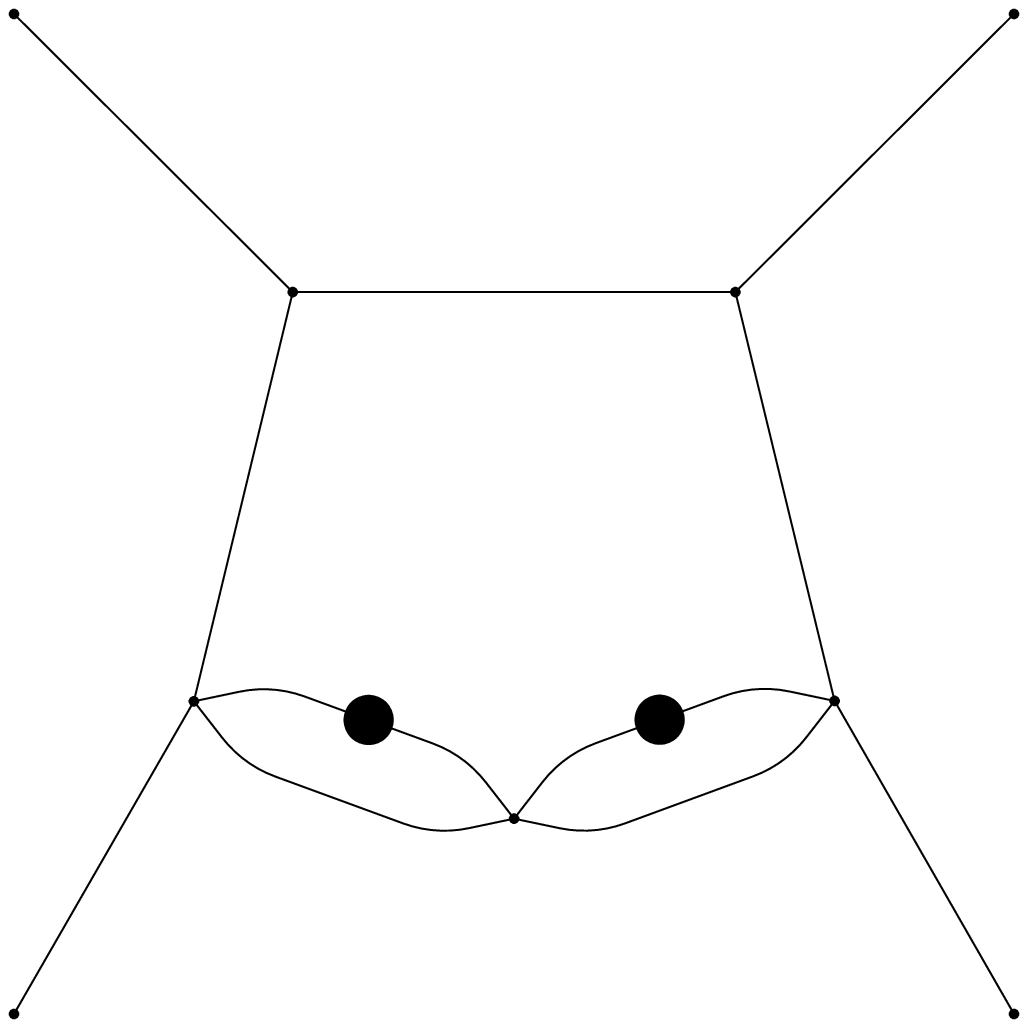}}
\newline
\subfloat[(26)]{\includegraphics[width=0.17\textwidth]{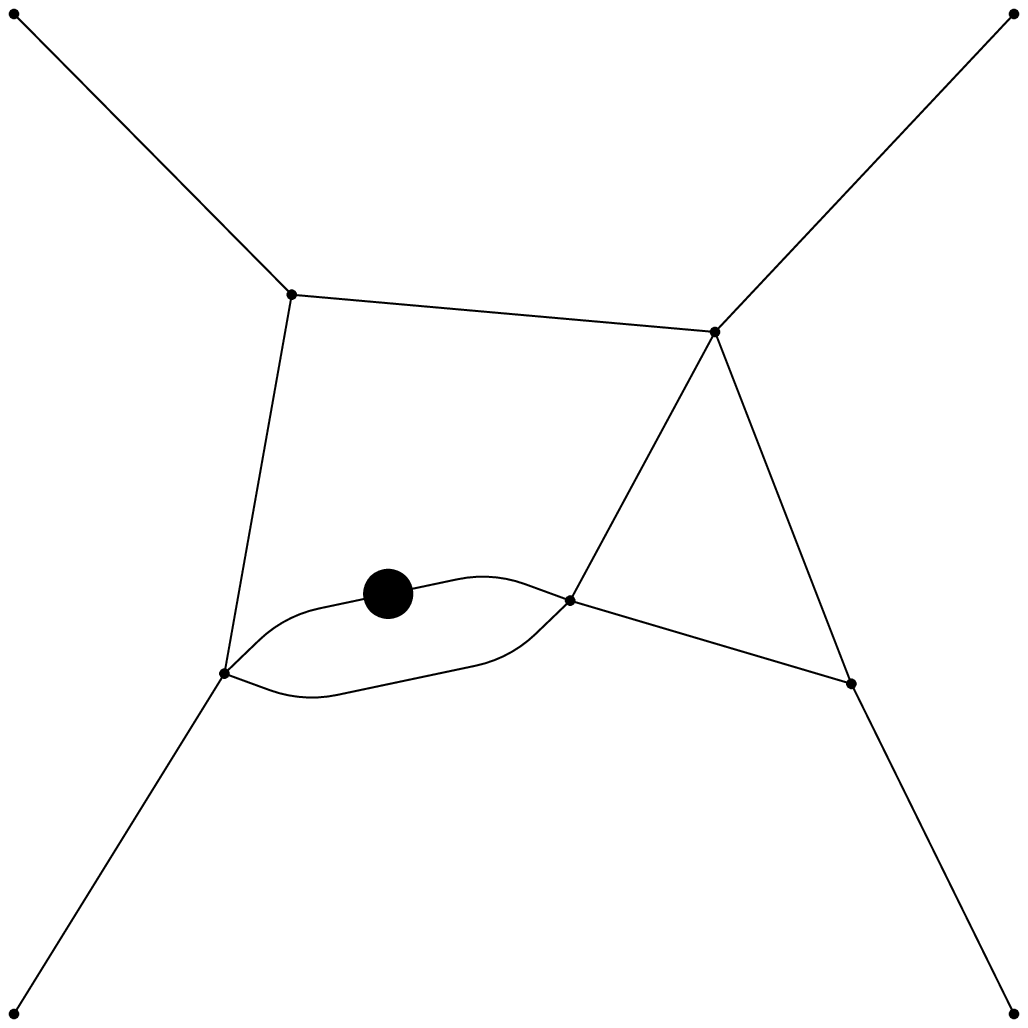}}
\subfloat[(29), (30)*]{\includegraphics[width=0.17\textwidth]{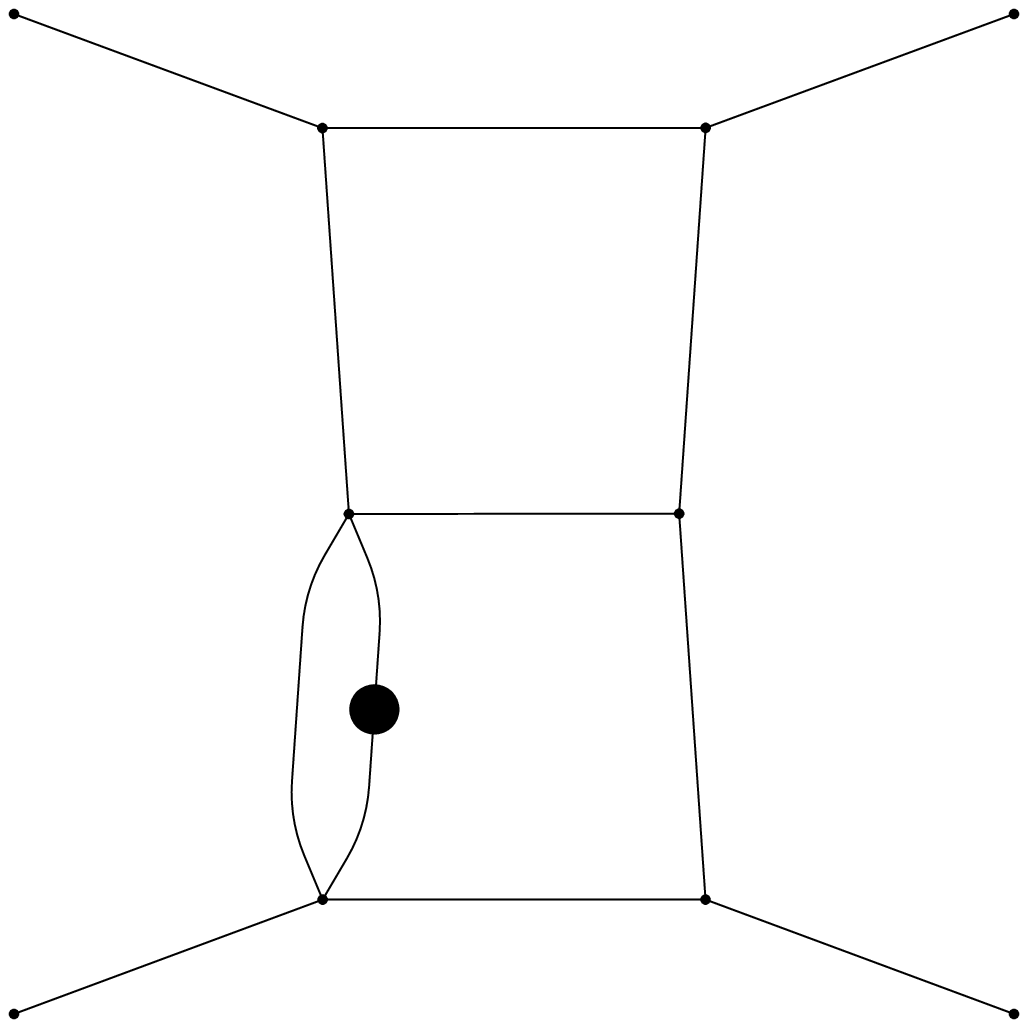}}
\caption{Master integrals for integral family E that have bubble subintegrals. 
Dots denote doubled propagators. 
An asterisk indicates that there are numerator factors not shown in the figure.
}
\label{fig:basistennis1}
\end{center}
\end{figure}

Many of the three-loop integrals we are interested in have bubble subintegrals (we will also sometimes refer to these as propagator-type subintegrals), i.e. 
they are lower-loop integrals with certain bubble insertions. In fact, the integrals of Fig.~\ref{fig:basisladders1} and Fig.~\ref{fig:basistennis1} are all of this type.
For definiteness, let us consider the specific case of integral $f^{A}_{19}$ of Fig.~\ref{fig:basisladders1}.

\begin{figure}[t] 
\captionsetup[subfigure]{labelformat=empty}
\begin{center}
\subfloat[(19)]{\includegraphics[width=0.3\textwidth]{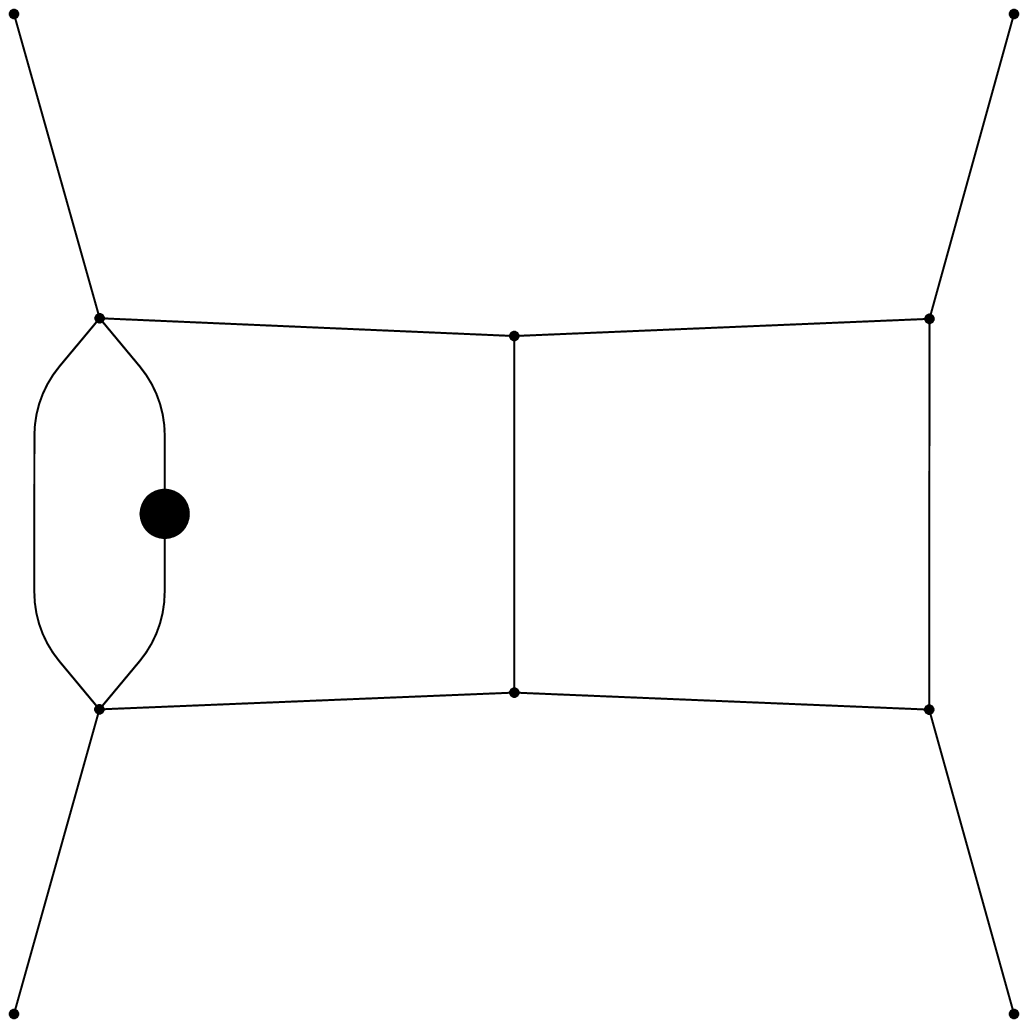}}
\hspace{1cm}
\hspace{1cm}
\subfloat[(19')]{\includegraphics[width=0.3\textwidth]{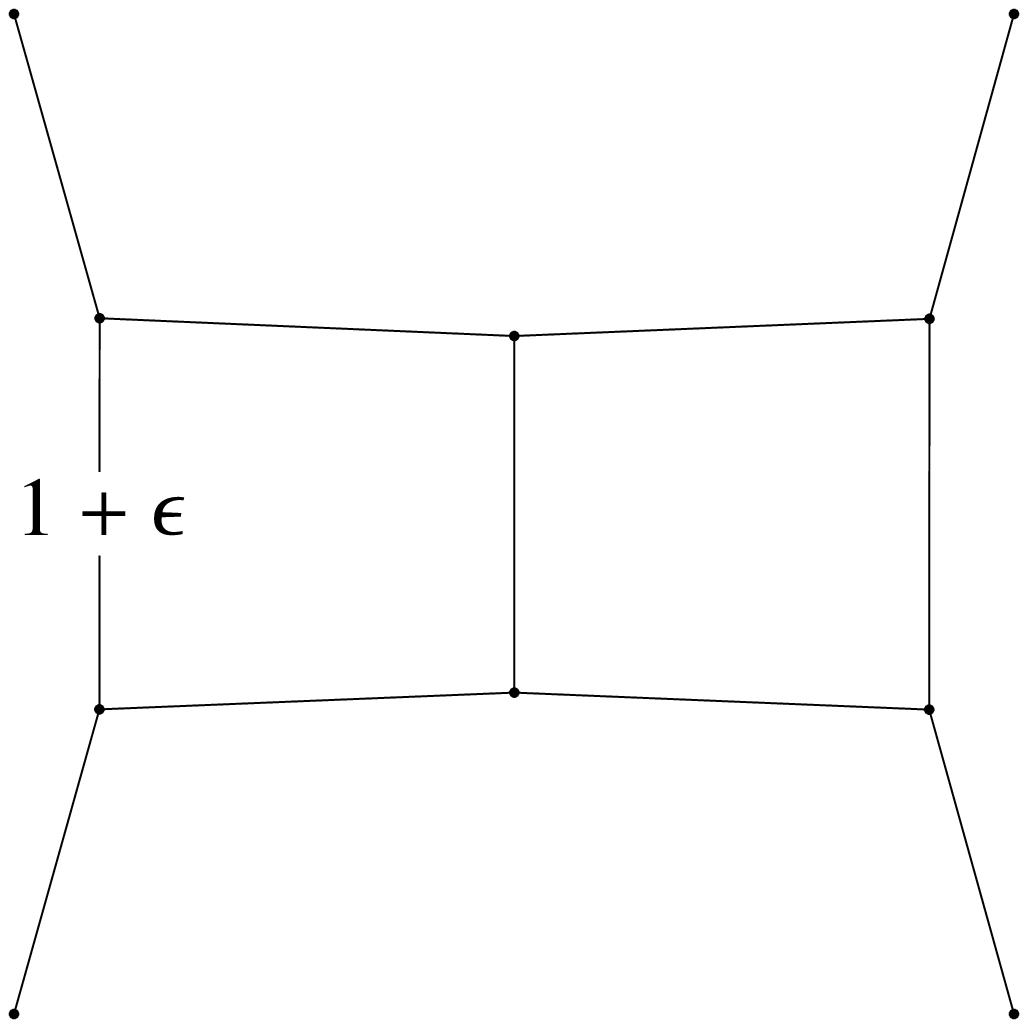}}
\caption{Integrating out propagator subintegrals related the basis choice at $(L+1)$ loops to the corresponding choice at $L$ loops, up to some trivial prefactors, and indices shifted by $\eps$.
}
\label{fig:bubblereduction}
\end{center}
\end{figure}

It is clear that we can always integrate out propagator subintegrals and obtain
a lower-loop integral, albeit with some power(s) shifted by $\eps$.
More concretely, we have 
\begin{align}
 \int  \frac{d^{D}k}{[-k^2]^{a_1} [-(k+p)^2]^{a_2}}  
= \frac{\Gamma(a -D/2) \Gamma(D/2-a_1) \Gamma(D/2-a_2 )}{\Gamma(a_1) \Gamma( a_2) \Gamma(D-a) } \frac{i \pi^{D/2}}{(-p^2)^{a-D/2}} \,,\end{align}
where $a=a_1 + a_2$.
In particular, if the indices $a_1$ and $a_2$ are equal to one and two, as in the present case, we see
that after integrating out the bubble subintegral, we obtain, up to some inessential prefactor, a double box
integral with one index shifted from $1$ to $1+\eps$, cf. Fig.~\ref{fig:bubblereduction}.

One might be worried about the effect of the shift of the power by $\eps$.
In fact, experience shows that in most cases the shifts in $\eps$ can be ignored 
for the purposes of uniform transcendentality. 
A qualitative explanation, which is applicable to many cases, is the following. 
Consider the integral
\begin{align}
I(x,\eps) := \int_0^1 \frac{1}{x+t} t^\eps \, dt \,.
\end{align}
For $\eps = 0$, this evaluates to a logarithm, and hence has degree one.
The full integral has a Taylor expansion in $\eps$. It is easy to see that the coefficient
of $\eps^{n}$ has weight $(n+1)$. Assigning weight $-1$ to $\eps$, 
we see that $I(x,\eps)$ is a function for which each term in the expansion in $\eps$ has uniform weight one.
We see that the presence of the factor $t^\eps$ had was inessential as far as the transcendental weight of the integral was concerned. 

We see that this reasoning motivates the choice for the master integrals shown in Figs.~\ref{fig:basisladders1},\ref{fig:basistennis1}.
Similarly, in the case of triangle subintegrals, explicit parametrizations can be useful.
In particular, whenever there is a triangle integral with an on-shell corner, a well-known trick is to use Feynman parameters to combine the two propagators adjacent to the on-shell leg. In this way, one obtains a one-fold integral over a configuration with a propagator subintegral, 
which was discussed above.

\begin{figure}[h] 
\captionsetup[subfigure]{labelformat=empty}
\begin{center}
\subfloat[(17)]{\includegraphics[width=0.17\textwidth]{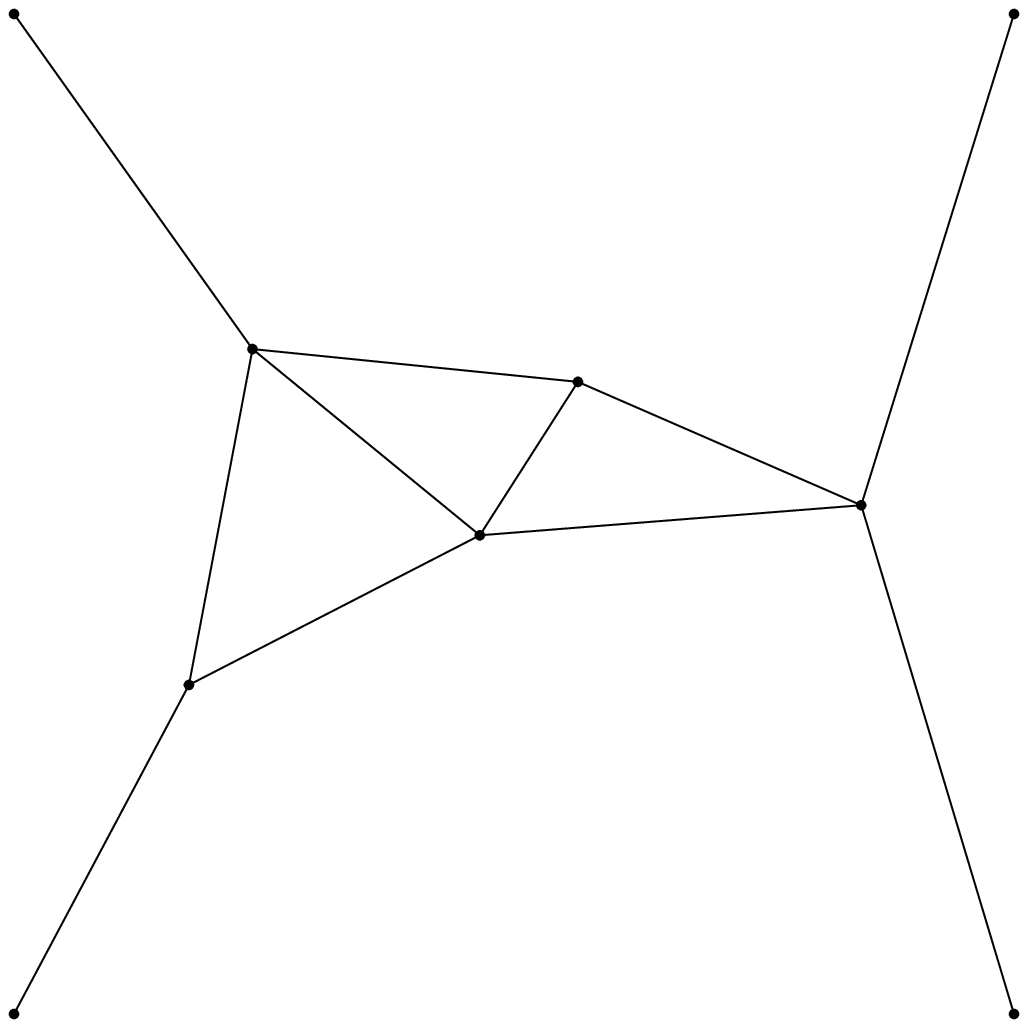}}
\subfloat[(20)]{\includegraphics[width=0.17\textwidth]{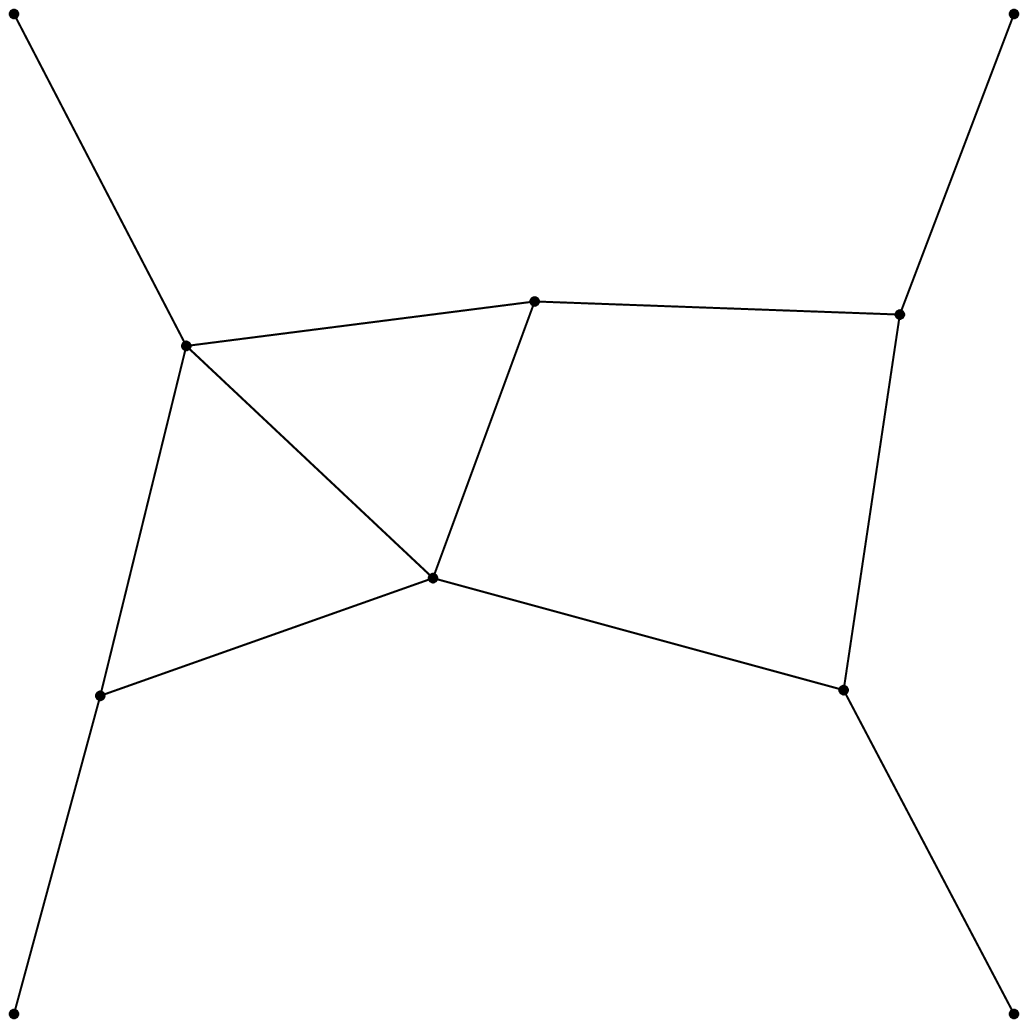}}
\subfloat[(21)]{\includegraphics[width=0.17\textwidth]{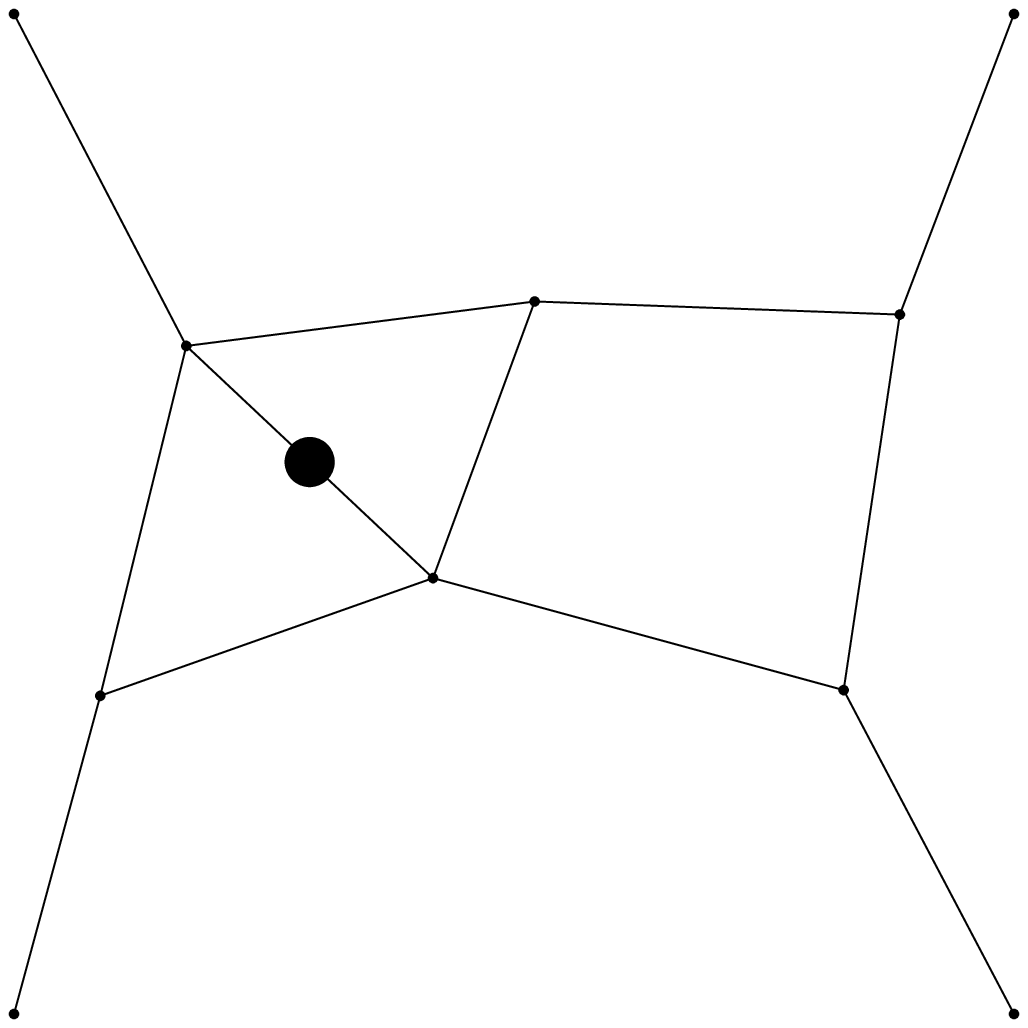}}
\subfloat[(15)]{\includegraphics[width=0.17\textwidth]{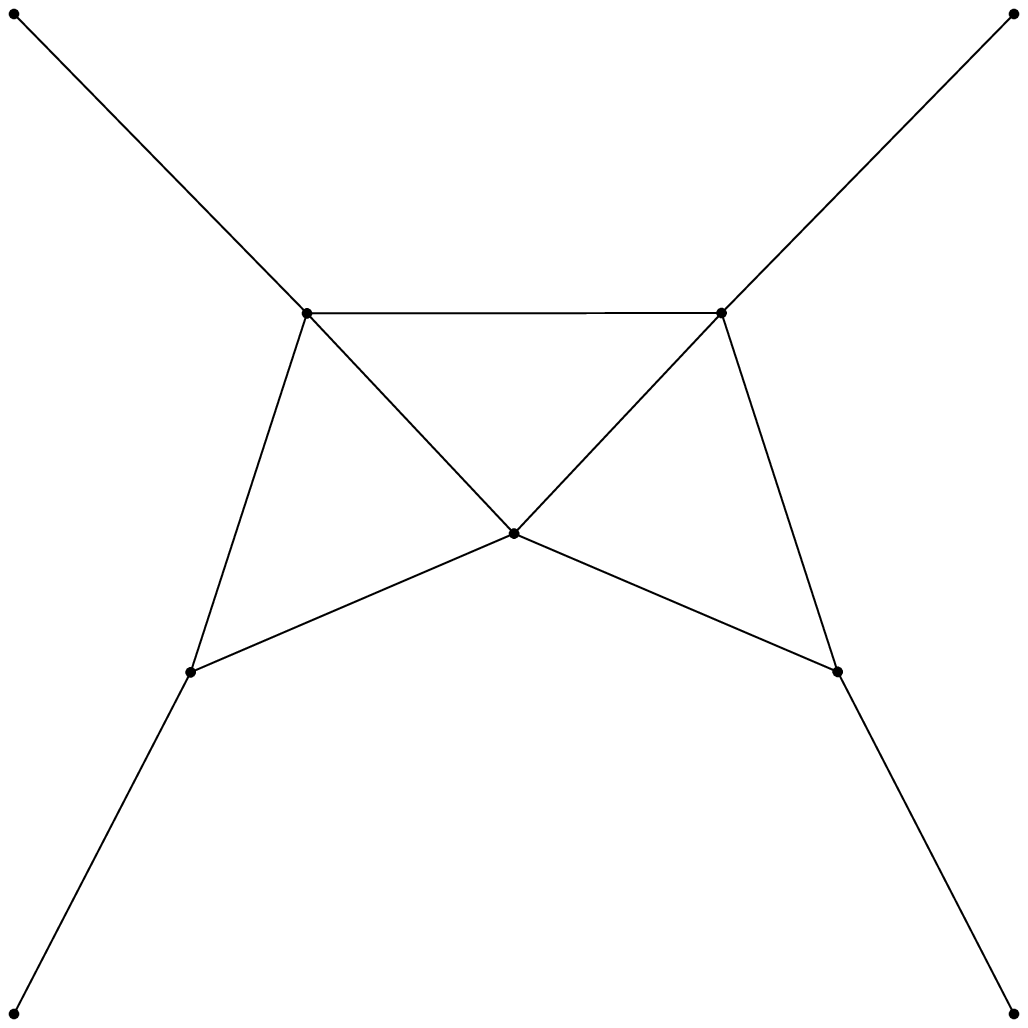}}
\subfloat[(16)]{\includegraphics[width=0.17\textwidth]{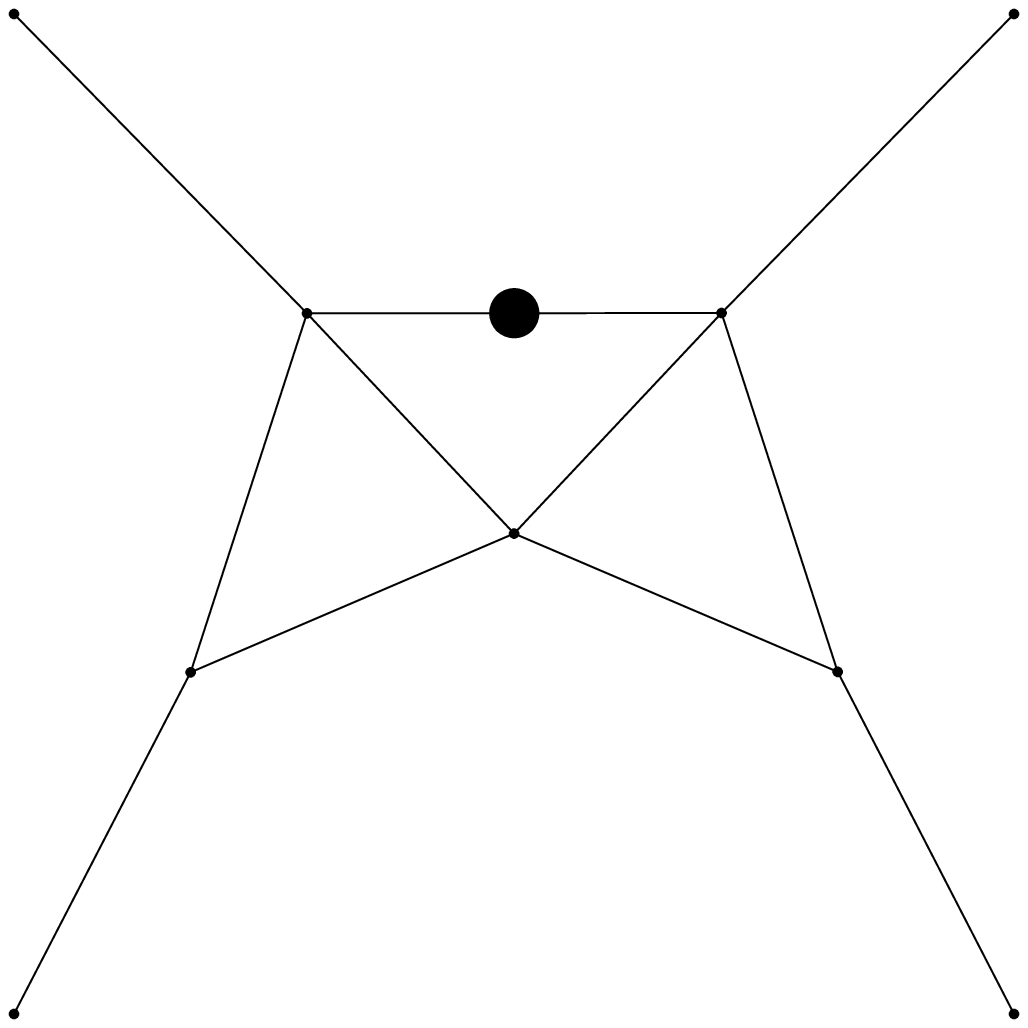}}
\subfloat[(24), (25)*, (26)*]{\includegraphics[width=0.17\textwidth]{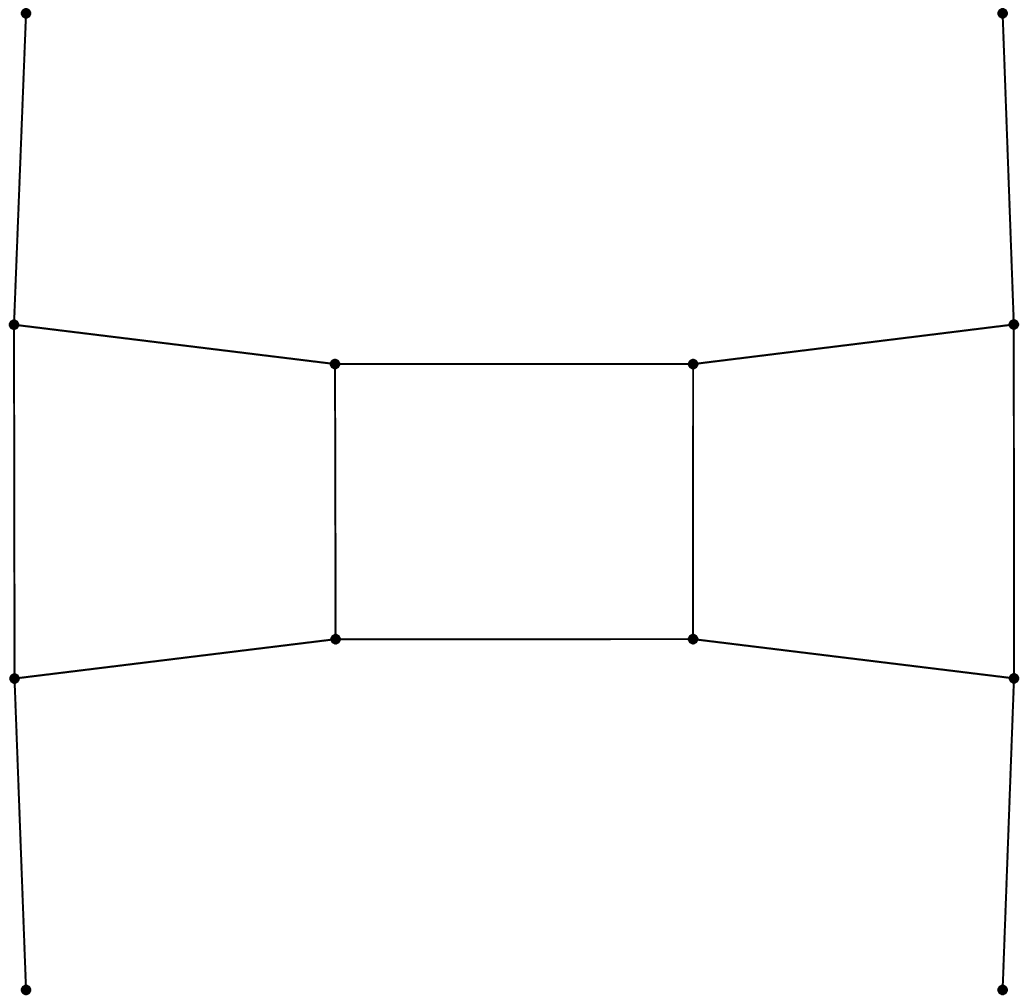}}
\caption{Master integrals for integral family A without bubble subintegrals. 
Dots denote doubled propagators. 
An asterisk indicates that there are numerator factors not shown in the figure.
}
\label{fig:basisladders2}
\end{center}
\end{figure}

\subsection{Example 2: leading singularities, (generalized) unitarity cuts}

A more general method is to study leading singularities or the closely related 
(generalized) unitarity cuts of loop integrals.
In particular, a very useful cut can be done whenever we have a box subintegral. 
In this case, we can consider the same integral with the four propagators of the box cut, 
i.e. replaced by delta functions. Alternatively, we may view this as replacing the integration
by contours in the complex plane around the poles of the propagators.
As a result, the subintegral is completely localized and can be easily evaluated.
In this way, we relate the $(L+1)$-loop integral to an $L$-loop integral.
The strategy is then to choose the integrals such that the resulting lower-loop integrals
that can be obtained by cutting lines have uniform transcendentality.

\subsection{General comments}

In summary, we can use these rules to generate candidate integrals that are expected to be pure functions
of uniform transcendentality. One can then use the IBP reduction to determine how many of the candidate integrals are linearly independent and can hence be used as master integrals. One then proceeds by writing out the system of differential equations in the new basis. As we discuss in the following section, this provides an immediate test of the basis choice -- 
when successful, the transcendentality properties of the basis functions are made manifest by the differential equations. 
Before presenting our choice of integral basis, we make a number of general comments
on the strategy of finding such a basis.

The discussion of unitarity cuts in the examples was four-dimensional.
Of course, in principle one can also analyze these cuts in $4-2\eps$ dimensions 
This is closely related to massive integrals. In practice, we have found that in most cases, 
the na{\"i}ve four-dimensional integrand analysis is sufficient.
See also the related discussion for example 1.

We would also like to mention another fact that makes this approach extremely efficient in practice:
for a given family of integrals, one can start working in sectors with fewer propagators (i.e. number of positive indices), which restricts the size of the basis.
This allows to verify the properties of the basis choice by looking at a small number
of integrals at a time, by inspecting the resulting differential equations. 

In some cases, it can happen that the candidate integrals one selects using the above criteria
do not yet have the desired properties, e.g. if not enough cuts were considered, or if there
are subtle effects that invalidate a na{\"i}ve four-dimensional analysis.
Often, a small modification to the candidate integral(s) is then sufficient to obtain the desired
properties. Exactly how to modify the integrals can be deduced by inspecting the differential
equations. An example of such a case are integrals $f^E_{34}$ and $f^E_{36}$ given below.

Finally, it should be obvious from the above discussion that the ideas for finding convenient basis elements do not rely on planarity, massless particles, four-point kinematics, etc., although all of those features lead to technical simplifications. More generally, we would also expect that generalizations of the 'd-log' representations 
of ref. \cite{ArkaniHamed:2012nw} can give insight into transcendentality properties of loop integrals. For example, in the slightly simpler setting of heavy quark effective theory (i.e. Wilson line) integrals, such representations were used successfully, see  \cite{Henn:2013wfa}.

\subsection{Integral basis for integral classes A and E}

\begin{figure}[t] 
\captionsetup[subfigure]{labelformat=empty}
\begin{center}
\subfloat[(15)]{\includegraphics[width=0.17\textwidth]{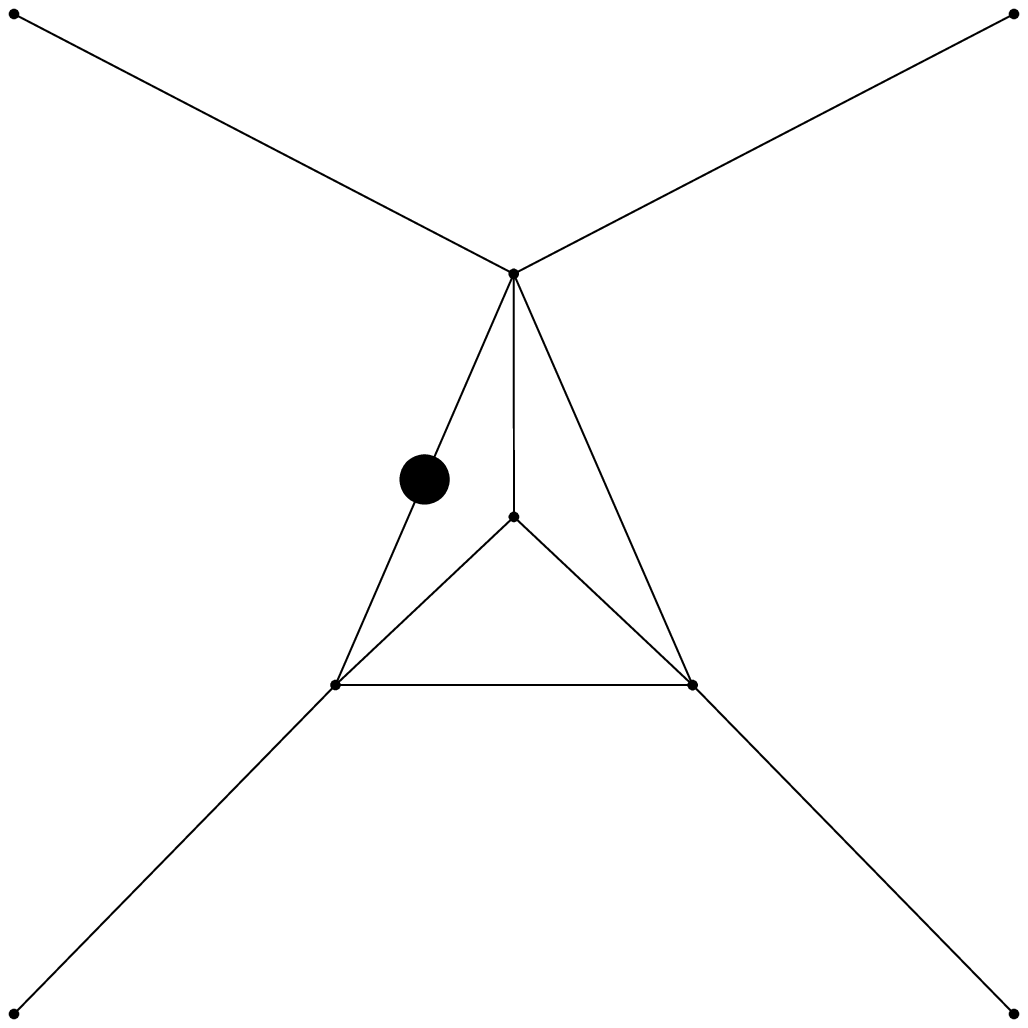}}
\subfloat[(16)]{\includegraphics[width=0.17\textwidth]{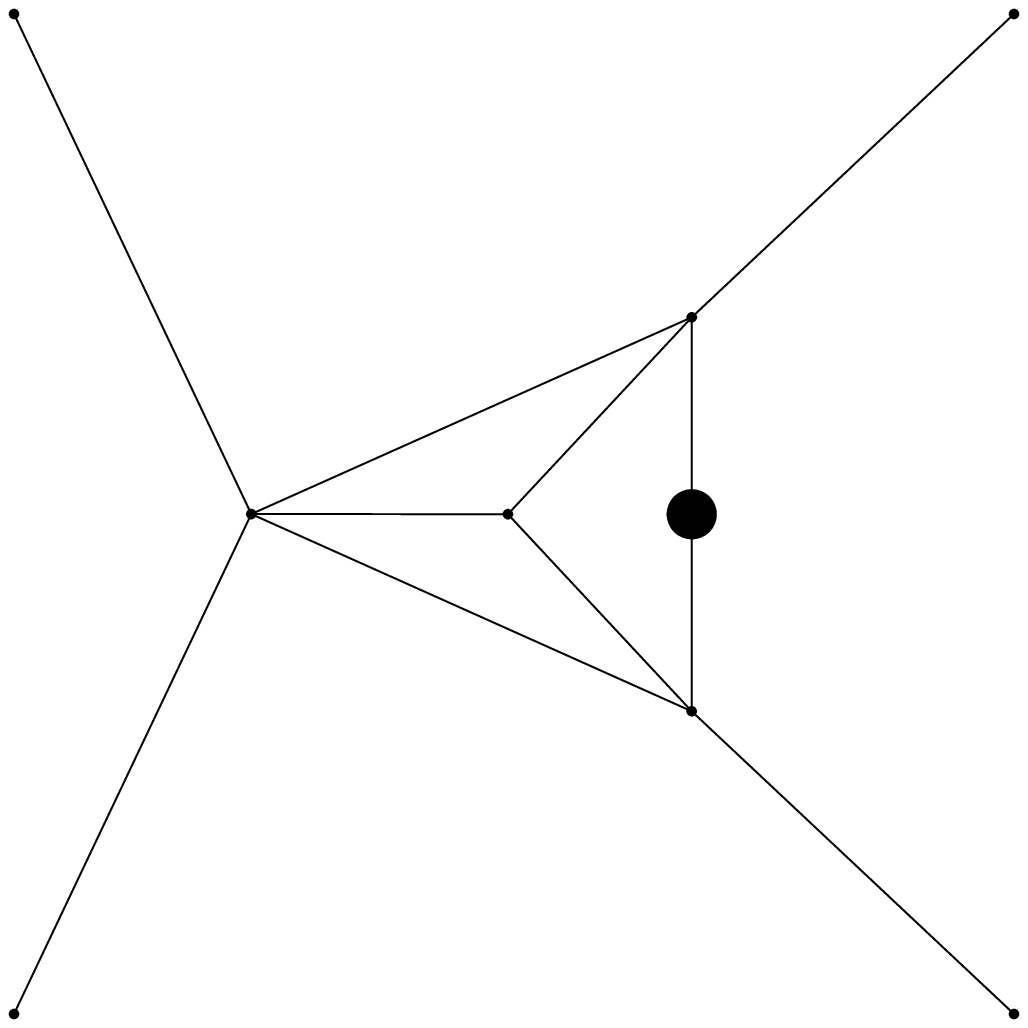}}
\subfloat[(20)]{\includegraphics[width=0.17\textwidth]{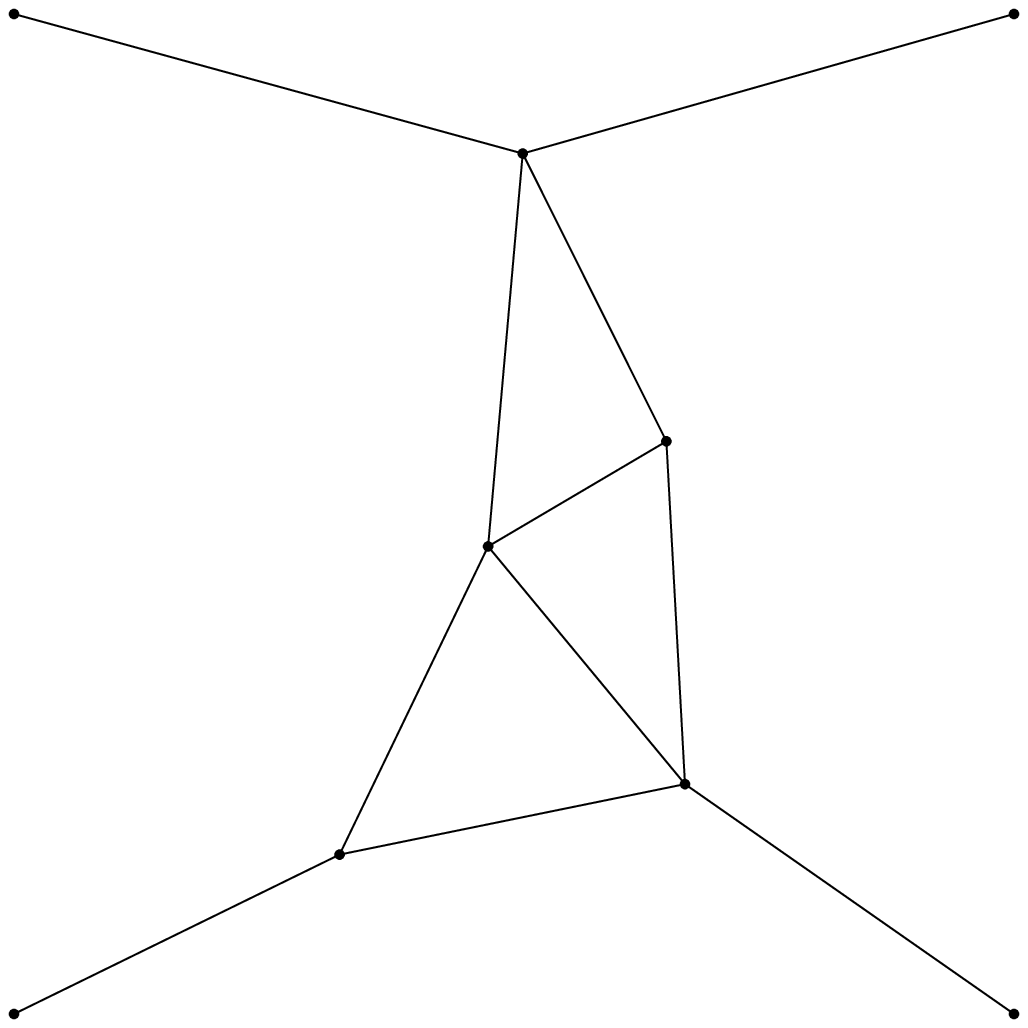}}
\subfloat[(21)]{\includegraphics[width=0.17\textwidth]{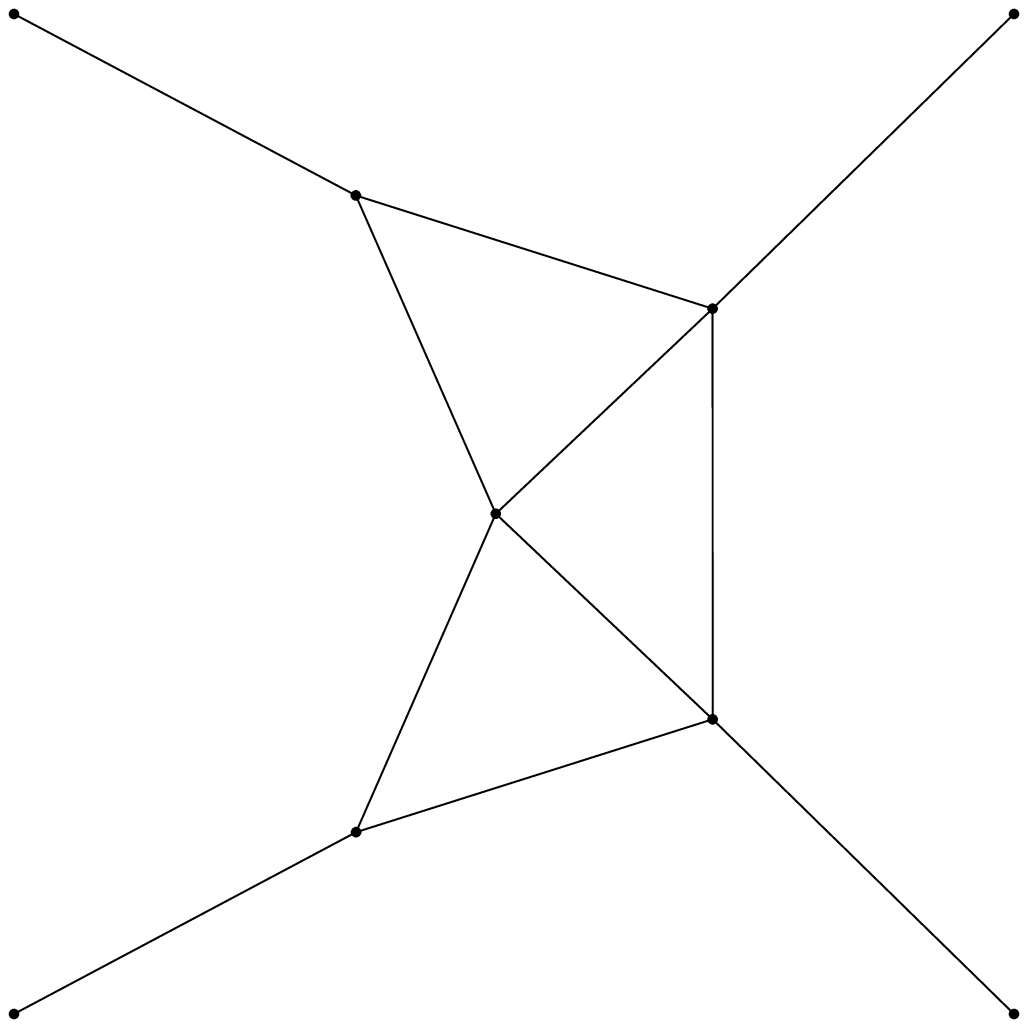}}
\subfloat[(22)]{\includegraphics[width=0.17\textwidth]{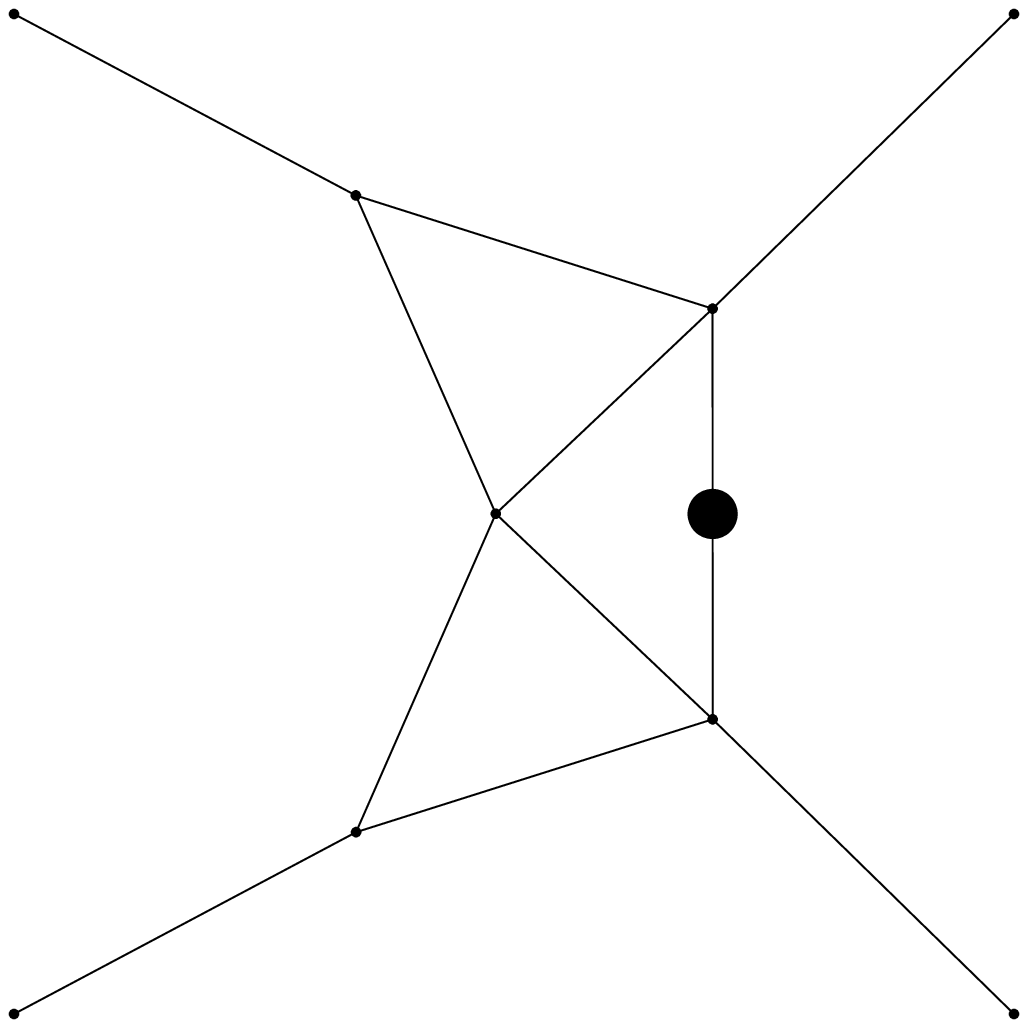}}
\subfloat[(23)]{\includegraphics[width=0.17\textwidth]{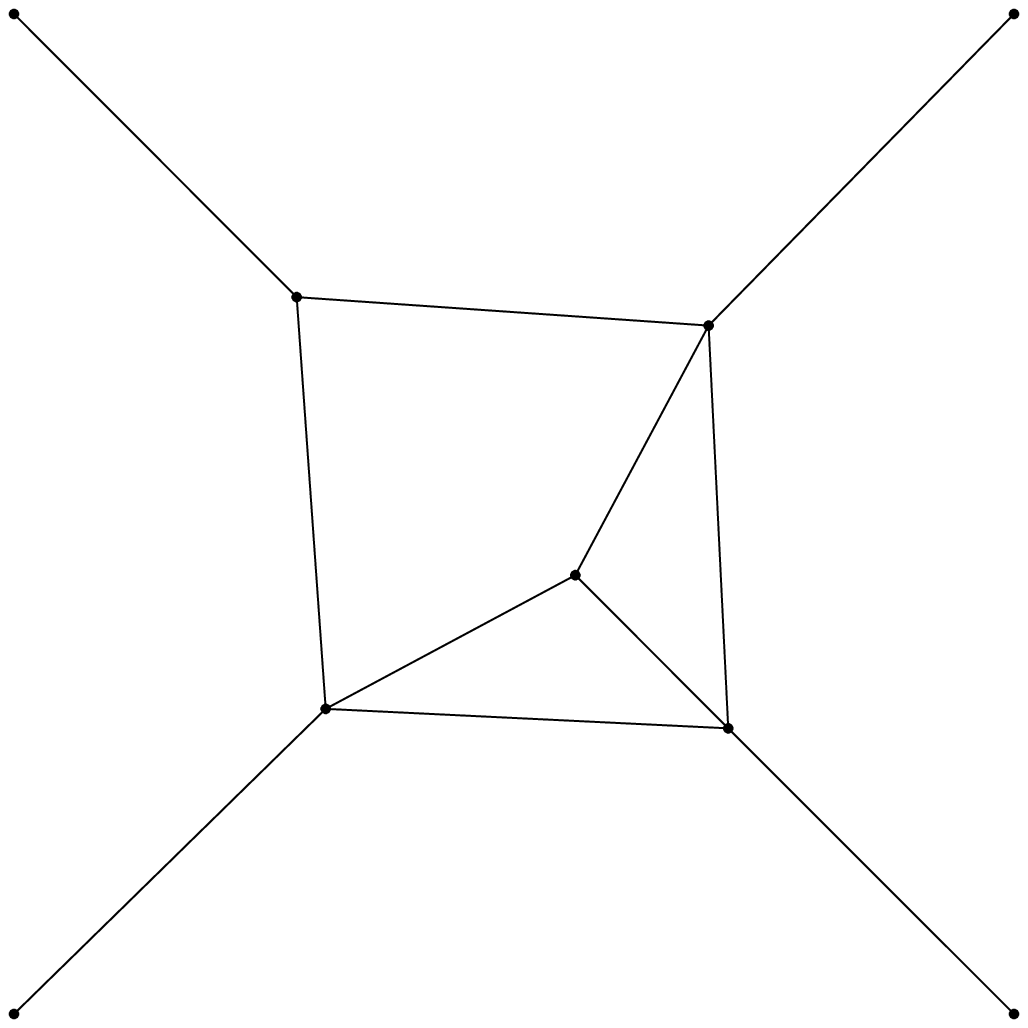}}
\newline
\subfloat[(24)]{\includegraphics[width=0.17\textwidth]{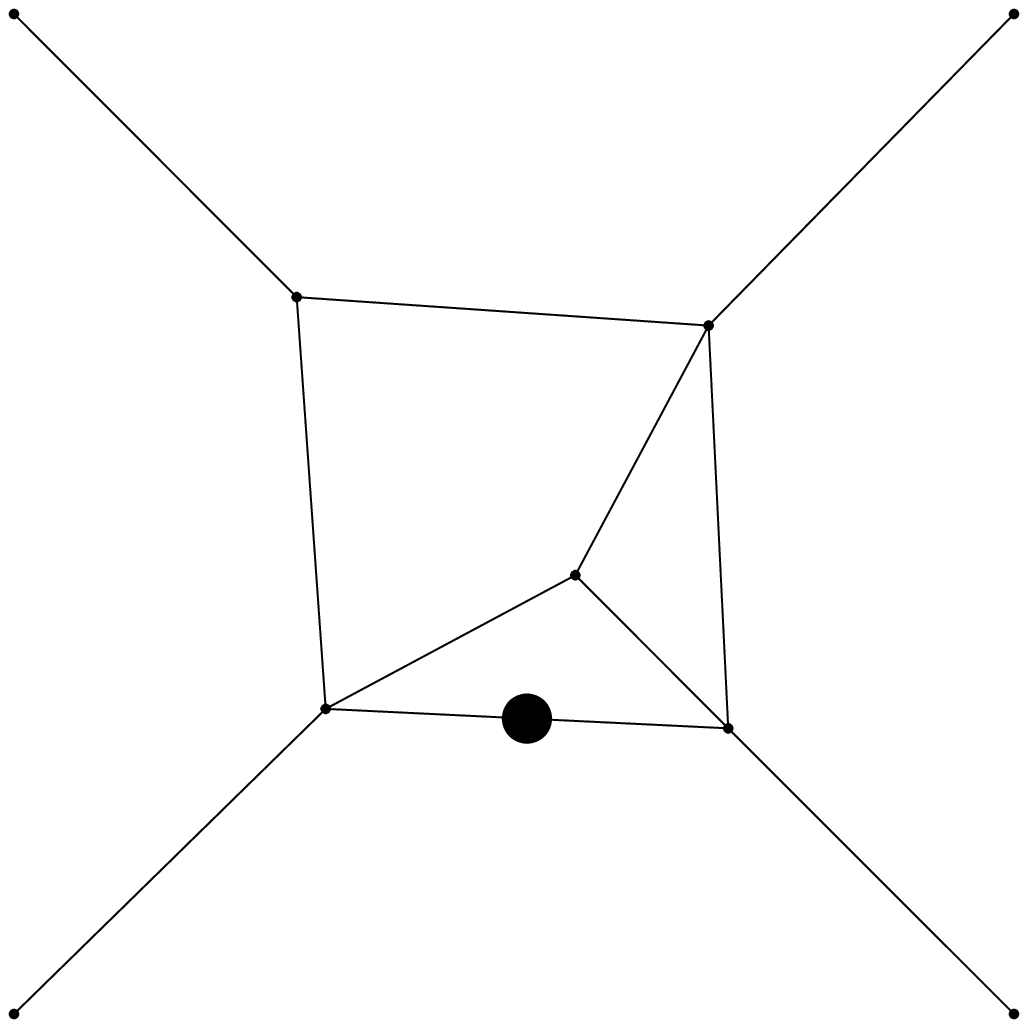}}
\subfloat[(27)]{\includegraphics[width=0.17\textwidth]{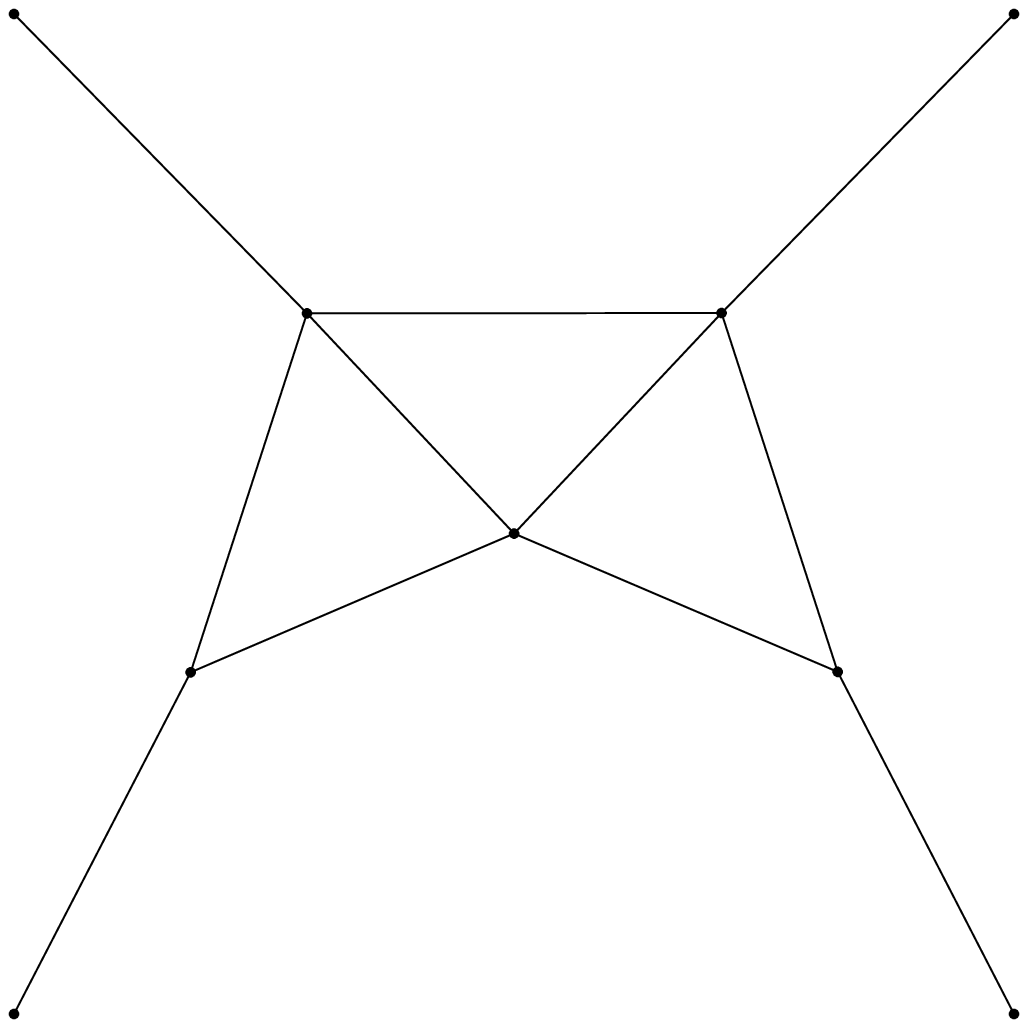}}
\subfloat[(28)]{\includegraphics[width=0.17\textwidth]{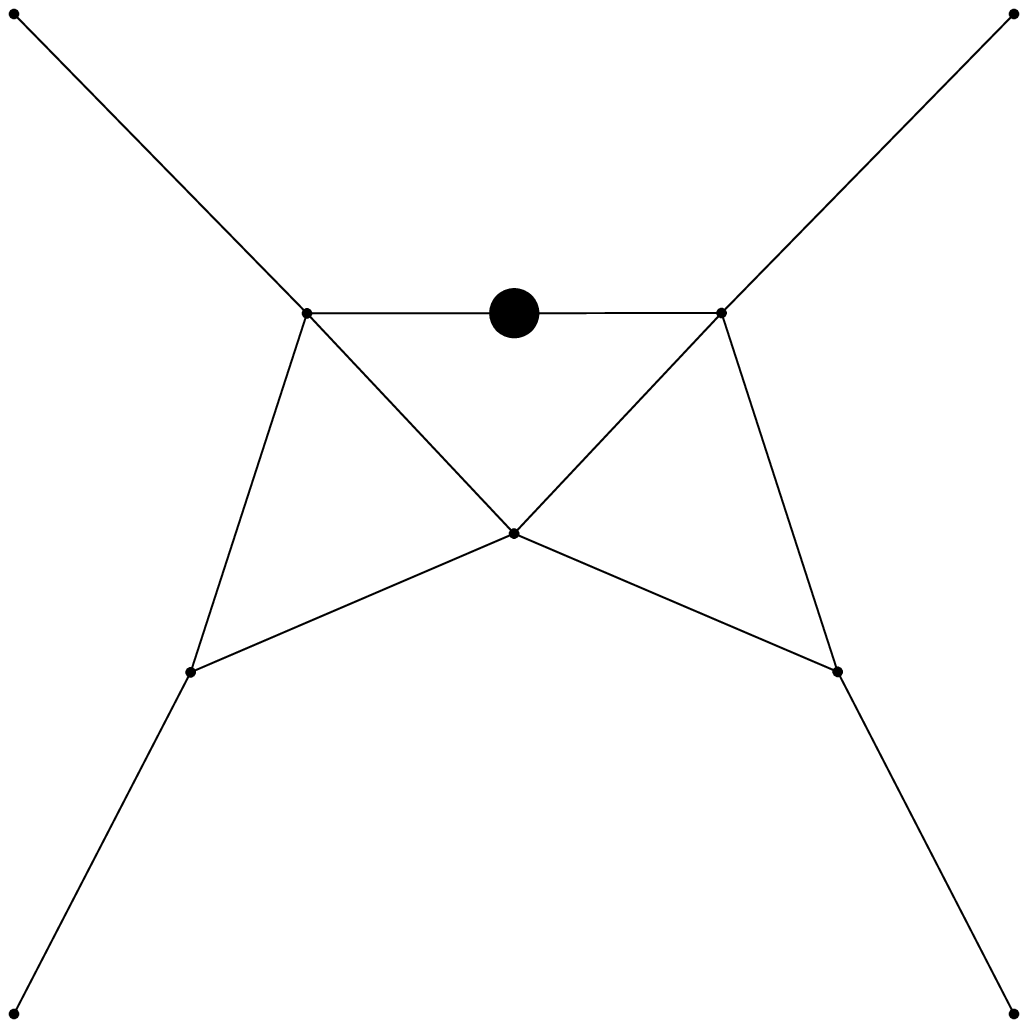}}
\subfloat[(31)]{\includegraphics[width=0.17\textwidth]{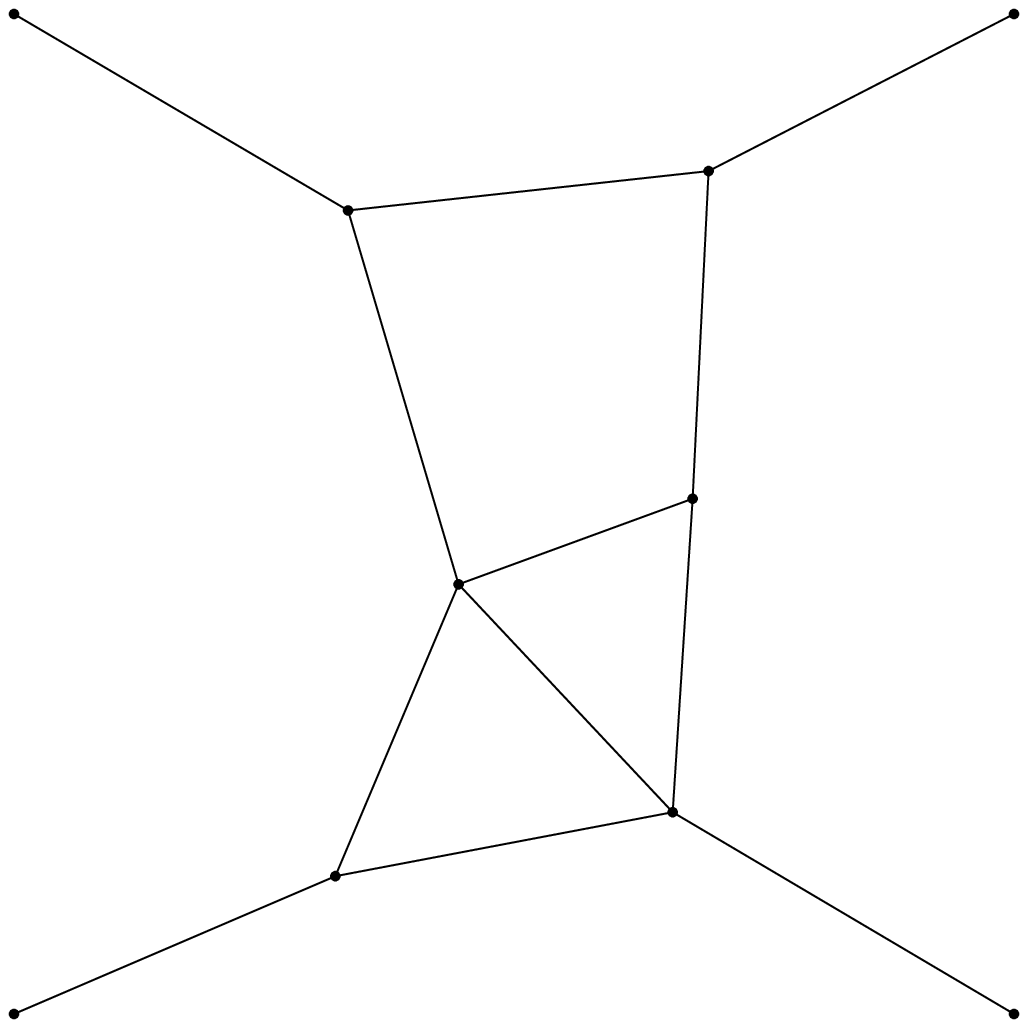}}
\subfloat[(32)]{\includegraphics[width=0.17\textwidth]{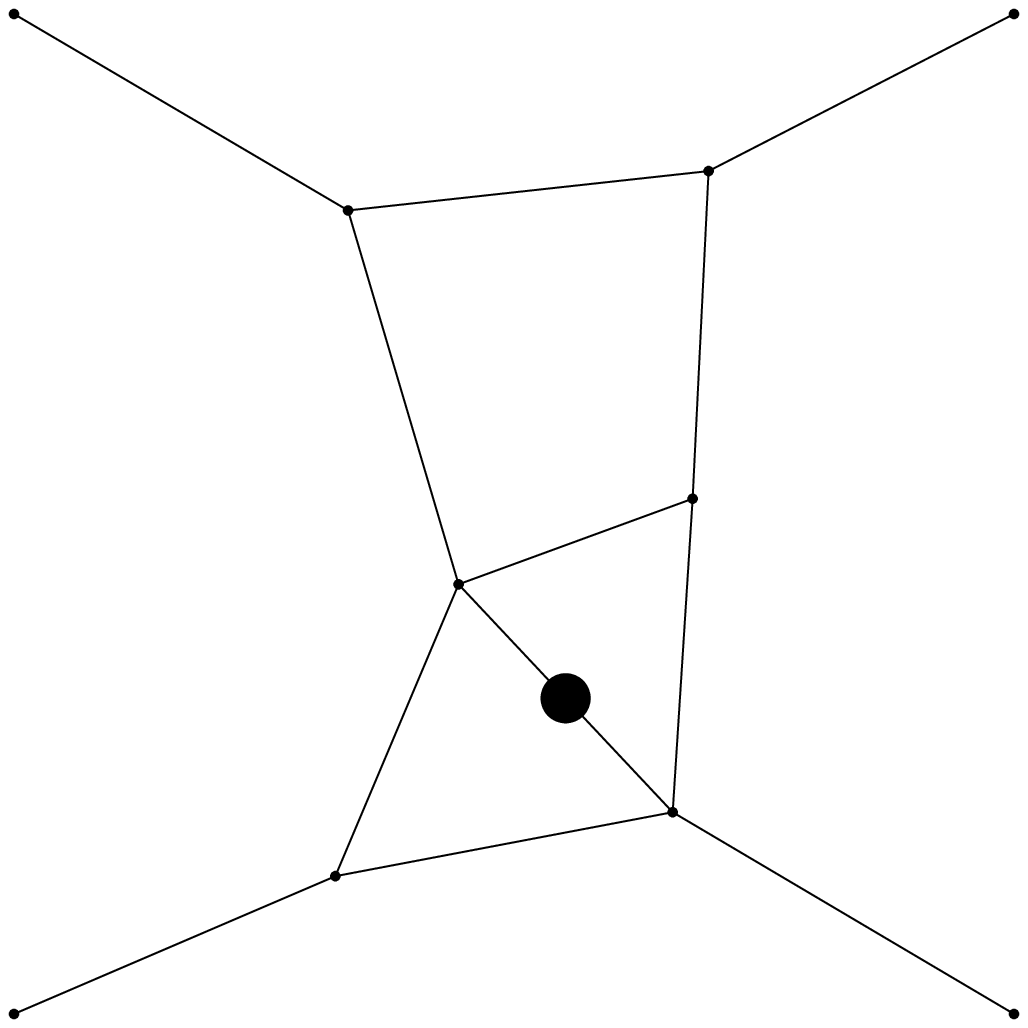}}
\subfloat[(33)*]{\includegraphics[width=0.17\textwidth]{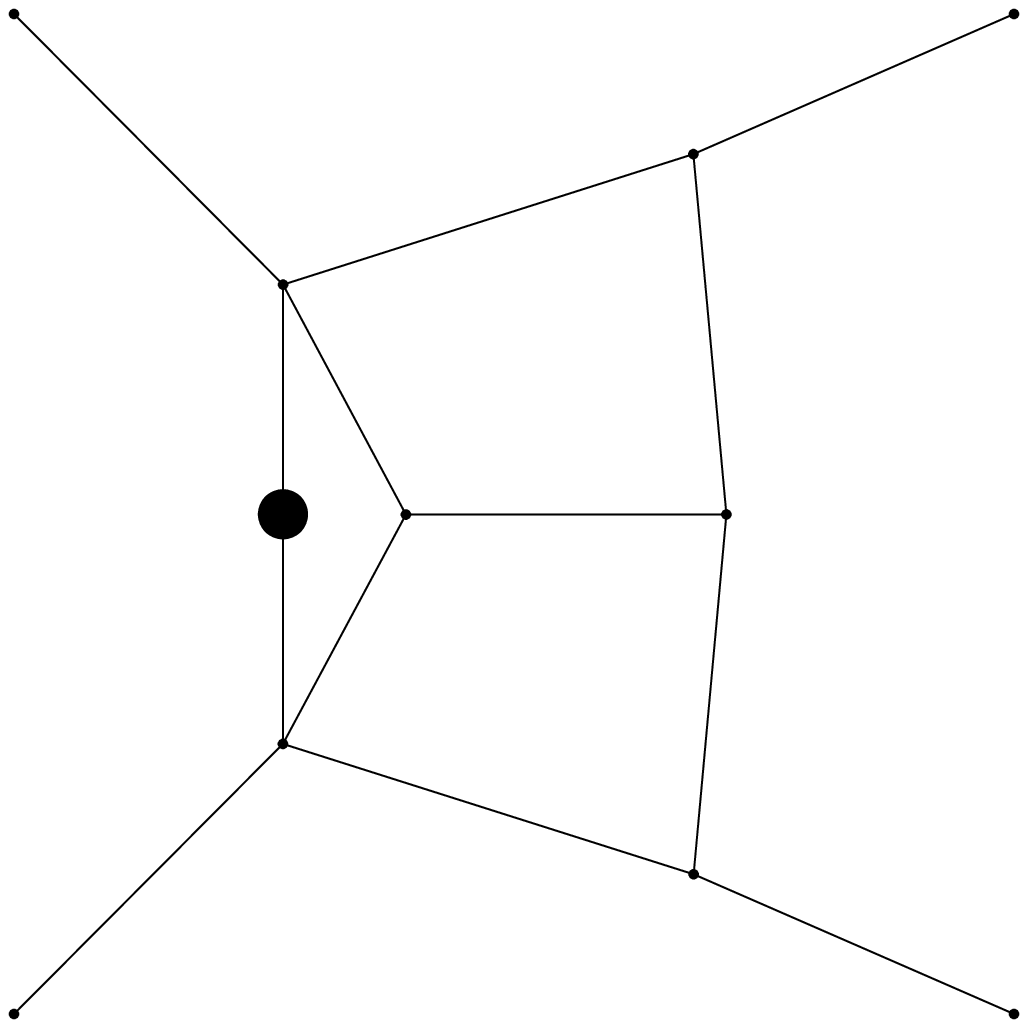}}
\newline
\subfloat[(34)*]{\includegraphics[width=0.17\textwidth]{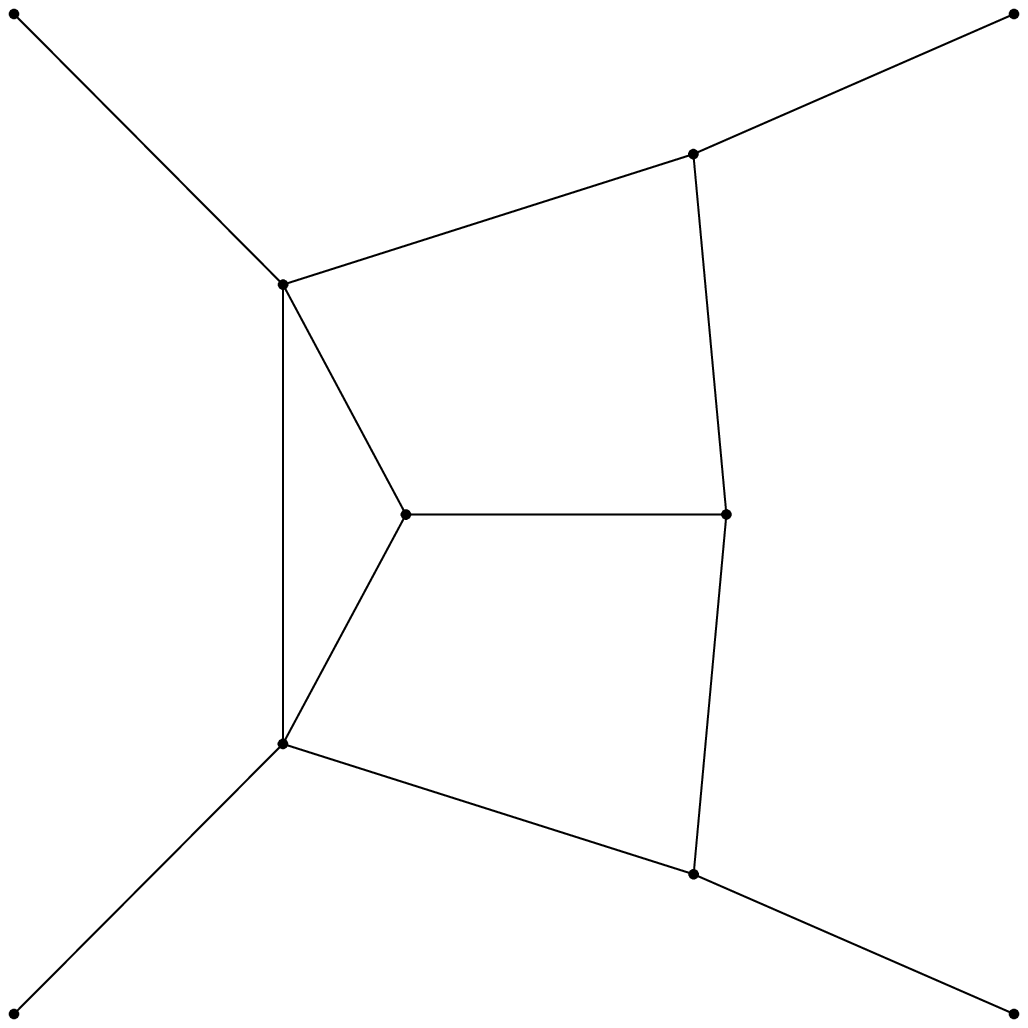}}
\subfloat[(35)*]{\includegraphics[width=0.17\textwidth]{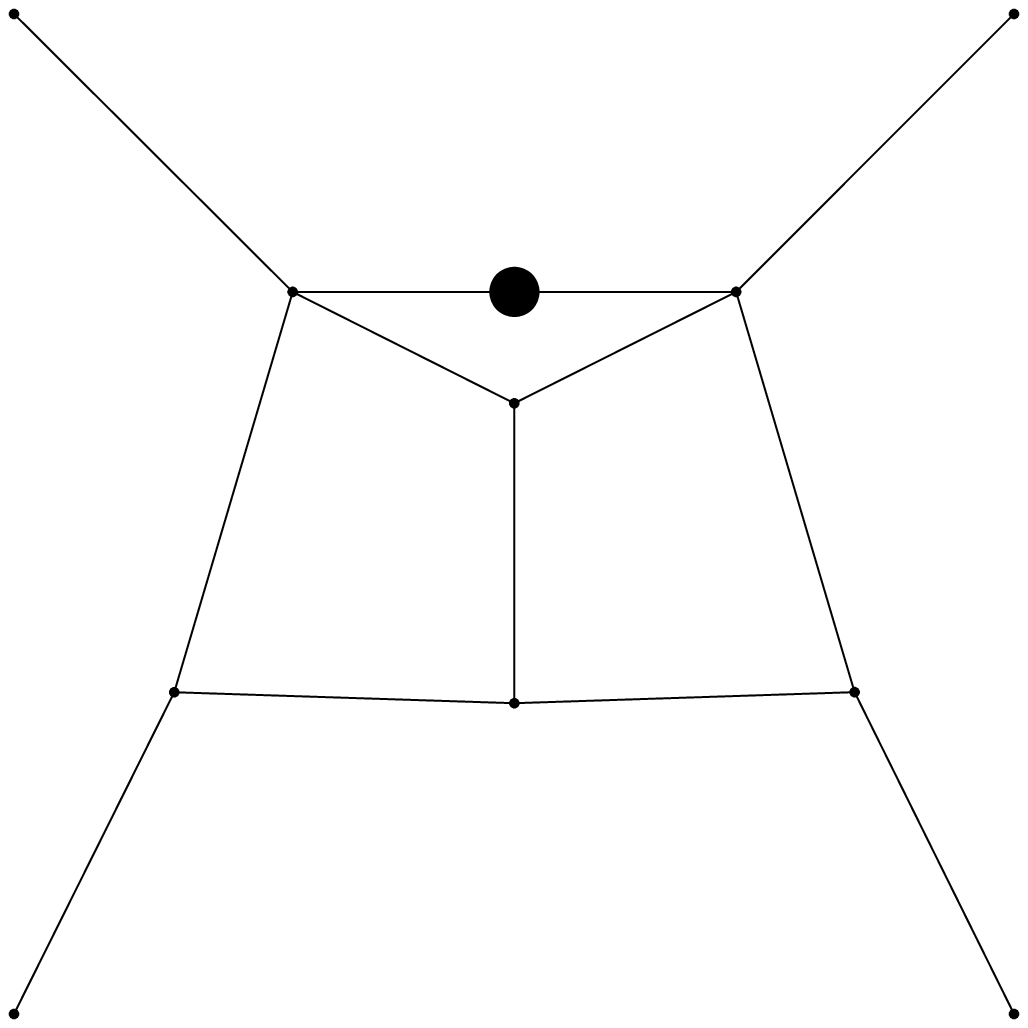}}
\subfloat[(36)*]{\includegraphics[width=0.17\textwidth]{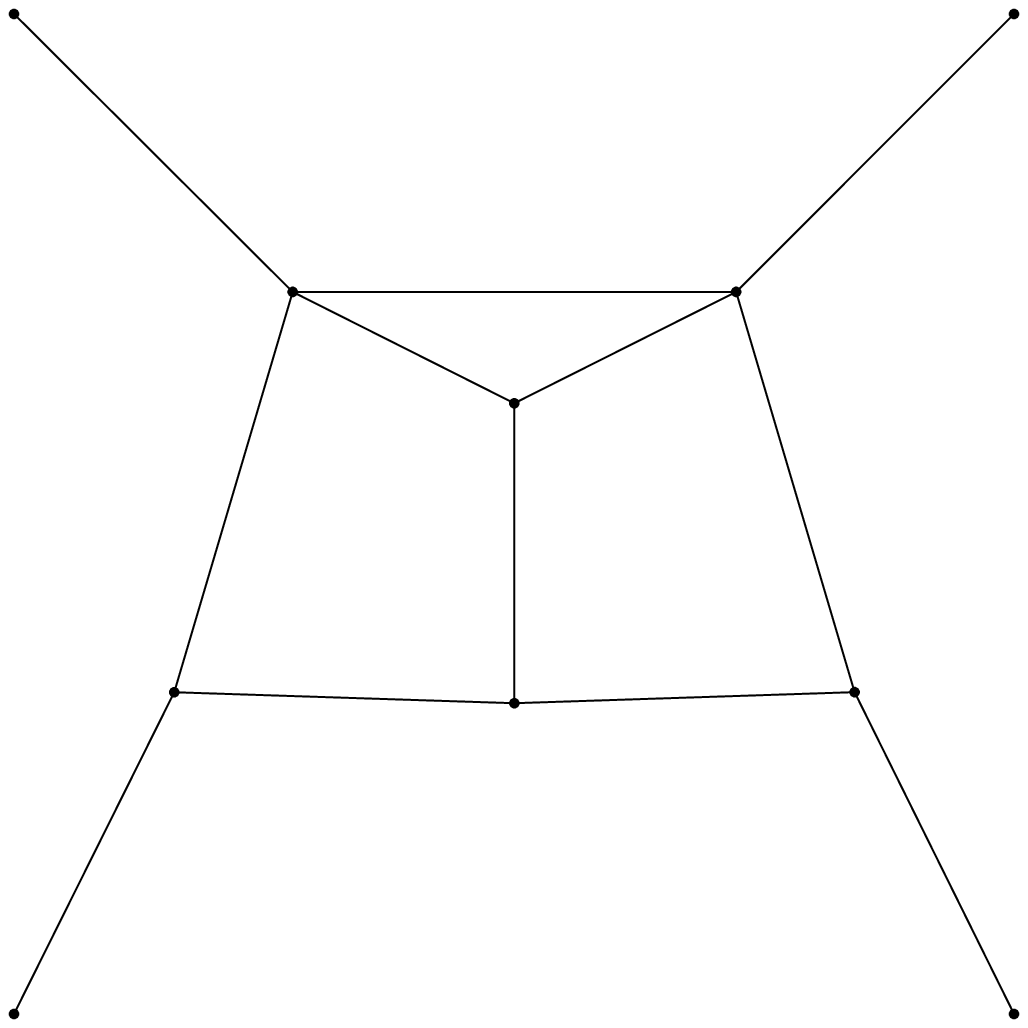}}
\subfloat[(37)*]{\includegraphics[width=0.17\textwidth]{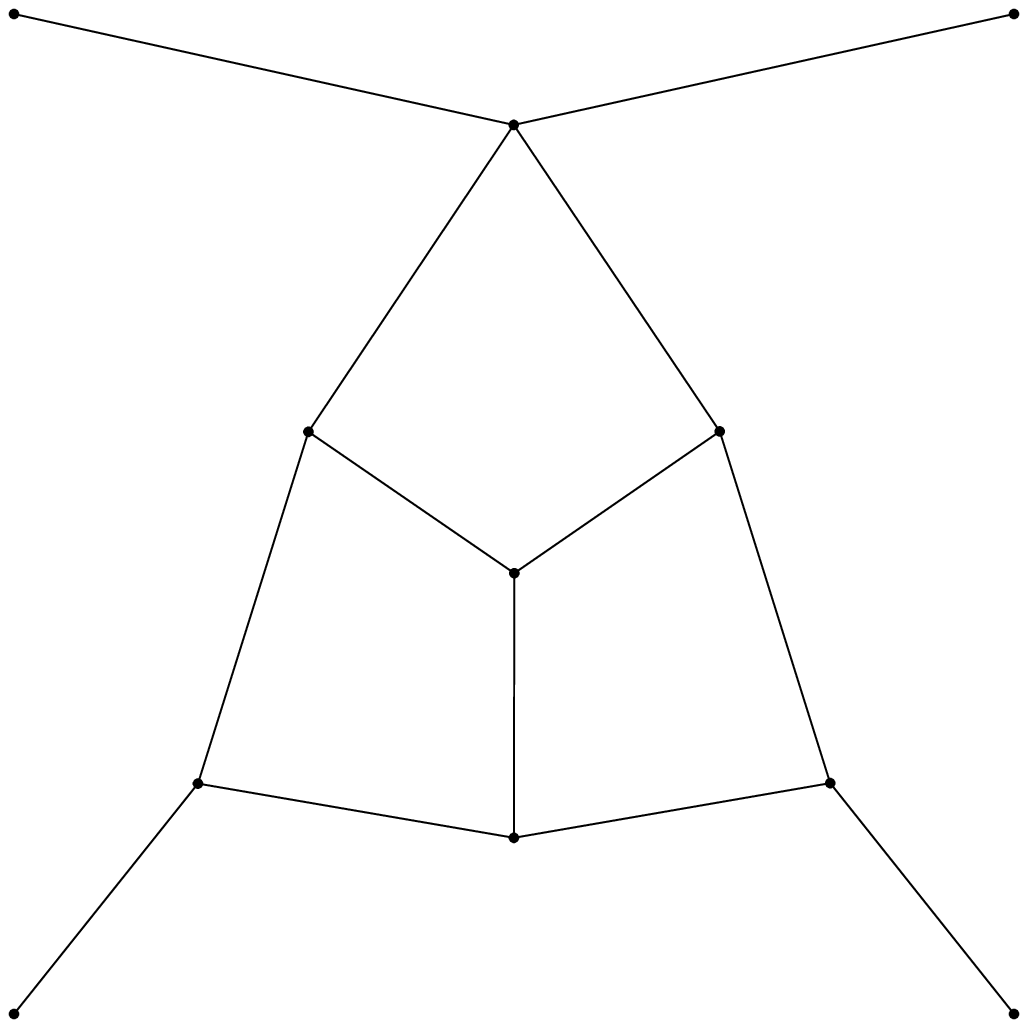}}
\subfloat[(38)*]{\includegraphics[width=0.17\textwidth]{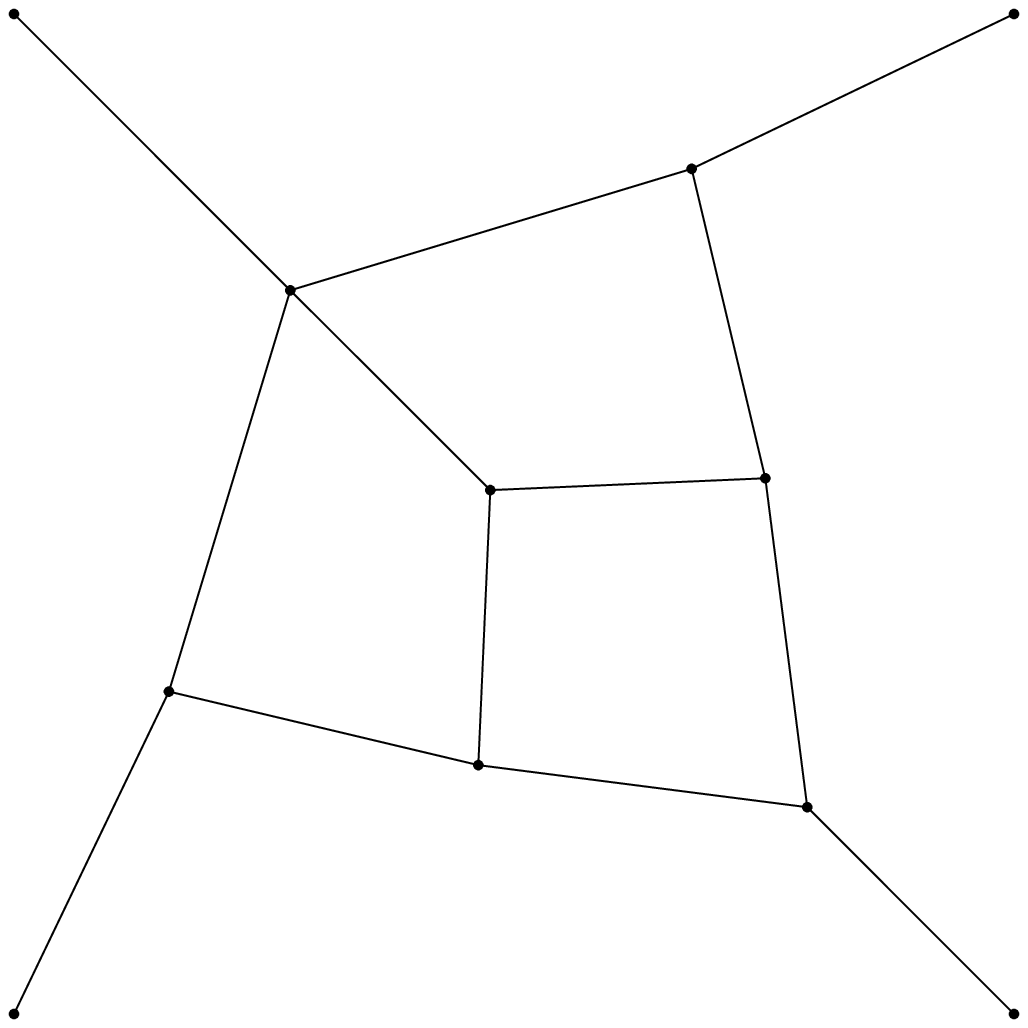}}
\subfloat[(39)*, (40)*, (41)*]{\includegraphics[width=0.17\textwidth]{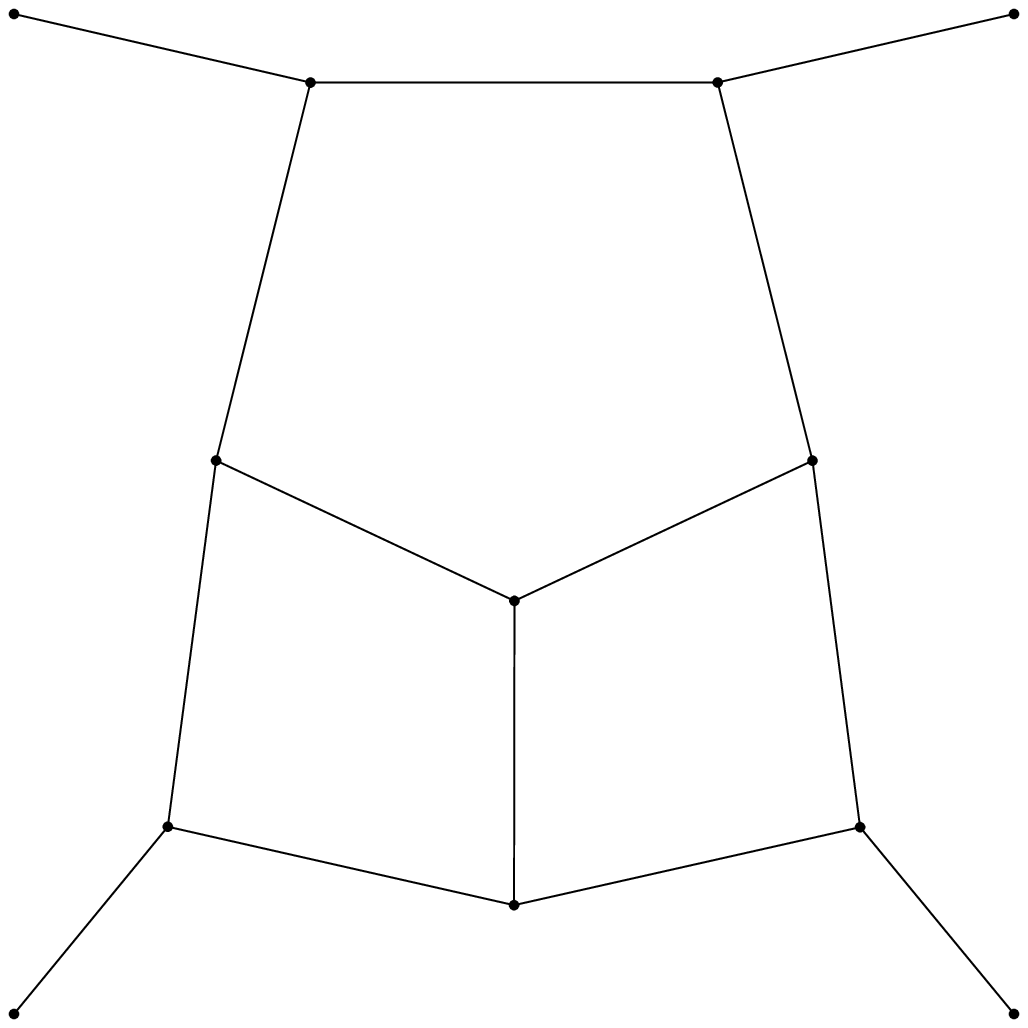}}
\caption{Master integrals for integral family E  without bubble subintegrals. 
Dots denote doubled propagators. 
An asterisk indicates that there are numerator factors not shown in the figure.
Integrals $34$ and $36$ involve an admixture of integral $23$, see eqs. in the main text.
}
\label{fig:basistennis2}
\end{center}
\end{figure}

In the way explained above we straightforwardly arrived at the basis choice
depicted in the Figs.~\ref{fig:basisladders1},\ref{fig:basisladders2},\ref{fig:basistennis1} and \ref{fig:basistennis2}.
There are $26$ master integrals in family A, and  $41$ in family E.
$7$ integrals are shared between the two families, so that we have a total of $60$ inequivalent integrals.
(Some further integrals can be obtained from interchanging $s$ and $t$.)

In formulas, we define
\begin{align}\label{defg}
f^{A}_i = \eps^3 \, (-s)^{3 \eps} \, \frac{ e^{3 \eps \gamma_{\rm E}}}{ (i \pi^{D/2})^3}  \,  \, g^{A}_i  \,.
\end{align}
This formula has three prefactors that we explain presently.
The factor $(-s)^{3 \eps}$ is there to make the basis functions $f^{A}_i$ dimensionless.
The factor $\eps^3$ ensures that all basis functions admit a Taylor expansion around $\eps=0$.
Finally, we have pulled out a standard conventional normalization factor for three-loop integrals.
The functions $g^A_i$ are defined as
\begin{align}
g^A_1=&t F^A_{0,0,0,0,0,0,2,2,2,1,0,0,0,0,0}, \qquad
g^A_2=s F^A_{0,2,0,0,1,0,0,2,2,0,0,0,0,0,0}, \\
g^A_3=&\eps s F^A_{0,0,0,0,1,1,2,2,1,0,0,0,0,0,0},\qquad
g^A_4=\eps s F^A_{0,0,0,1,2,0,2,1,1,0,0,0,0,0,0},\\
   g^A_5=&s F^A_{0,1,2,-1,0,1,0,2,2,0,0,0,0,0,0},\qquad
   g^A_6=s^2 F^A_{0,2,2,0,2,1,0,1,0,0,0,0,0,0,0},\\
   g^A_7=&\eps s t  F^A_{0,0,0,0,1,1,2,2,1,1,0,0,0,0,0},\qquad
   g^A_8=\eps^2 (s+t) F^A_{0,0,0,1,1,0,2,1,1,1,0,0,0,0,0},\\
   g^A_9=&\eps s t   F^A_{0,0,1,1,0,0,2,1,1,2,0,0,0,0,0},\qquad
   g^A_{10}=\eps s^2 F^A_{0,0,1,1,2,1,2,1,0,0,0,0,0,0,0},\\
   g^A_{11}=&\eps^2 (s+t) F^A_{0,1,0,0,1,0,1,1,2,1,0,0,0,0,0},\qquad
   g^A_{12}=-\eps (2 \eps-1) s F^A_{1,1,0,0,1,1,0,2,1,0,0,0,0,0,0},\\
   g^A_{13}=&s^3 F^A_{2,1,2,1,2,1,0,0,0,0,0,0,0,0,0},\qquad
   g^A_{14}=\eps s F^A_{0,0,1,1,0,0,2,1,1,2,0,0,-1,0,0},\\
   g^A_{15}=&\eps^3 t  F^A_{0,1,1,0,0,1,1,1,1,1,0,0,0,0,0},\qquad
   g^A_{16}=\eps^2 s^2 F^A_{0,1,2,0,0,1,1,1,1,1,0,0,0,0,0},\\
   g^A_{17}=&\eps^3 s F^A_{0,1,1,0,1,1,1,1,1,0,0,0,0,0,0},\qquad
   g^A_{18}=\eps^2 s^2 F^A_{0,0,1,1,1,1,2,1,1,1,0,0,-1,0,0},\\
   g^A_{19}=&\eps^2 s^2 t  F^A_{0,0,1,1,1,1,2,1,1,1,0,0,0,0,0},\qquad
   g^A_{20}=\eps^3 s (s+t) F^A_{0,1,1,0,1,1,1,1,1,1,0,0,0,0,0},\\
   g^A_{21}=&\eps^2 s^2 t F^A_{0,1,1,0,1,1,1,2,1,1,0,0,0,0,0},\qquad
   g^A_{22}=\eps^2 s^2 t F^A_{1,1,0,0,1,1,1,2,1,1,0,0,0,0,0},\\
   g^A_{23}=&\eps^2 s^2
   F^A_{1,1,0,0,1,1,1,2,1,1,-1,0,0,0,0},\qquad
   g^A_{24}=\eps^3 s^3 t F^A_{1,1,1,1,1,1,1,1,1,1,0,0,0,0,0},\\
   g^A_{25}=&\eps^3 s^3
   F^A_{1,1,1,1,1,1,1,1,1,1,-1,0,0,0,0},\qquad
   g^A_{26}=\eps^3 s^3 F^A_{1,1,1,1,1,1,1,1,1,1,0,0,-1,0,0}
\end{align}
For integral family E, we have (\ref{defg}) with 'A' replaced by 'E', and
\begin{align}\label{masters_ladder}
g^{E}_1 =& s \, F^E_{0,0,1,0,0,2,2,2,0,0,0,0,0,0,0}, \qquad
g^{E}_2 = t F^E_{0,1,0,0,0,2,2,0,0,2,0,0,0,0,0},\\
g^{E}_3  =&  -2 {\eps} t   F^E_{0,0,1,0,0,2,2,0,1,1,0,0,0,0,0},\qquad
g^{E}_4  =  -2 \eps t F^E_{0,0,1,0,1,2,2,0,1,0,0,0,0,0,0},\\
g^{E}_5  =&  t^2 F^E_{0,2,0,0,2,0,1,0,2,1,0,0,0,0,0}   ,\qquad
   g^{E}_6   =  t F^E_{0,1,0,0,2,2,2,0,1,0,0,-1,0,0,0},\\
   g^{E}_7  =&  -2 \eps s F^E_{0,1,0,1,0,2,2,1,0,0,0,0,0,0,0},\qquad
   g^{E}_8  =  +s F^E_{1,0,2,2,0,1,0,2,0,0,-1,0,0,0,0},\\
   g^{E}_9  =&  -2 \eps s t F^E_{0,0,1,0,0,2,2,1,1,1,0,0,0,0,0},\qquad
    g^{E}_{10} =  4 \eps^2 (s+t)   F^E_{0,0,2,0,1,1,1,1,1,0,0,0,0,0,0} ,\\
   g^{E}_{11}  =&  4 \eps^2 t F^E_{0,2,0,1,0,1,1,0,1,1,0,0,0,0,0} , \qquad
  g^{E}_{12}  =  4 \eps^2 (s+t)    F^E_{0,2,0,1,0,1,1,1,0,1,0,0,0,0,0} ,\\
   g^{E}_{13}  =&  -2 \eps s t F^E_{0,2,0,1,1,2,1,1,0,0,0,0,0,0,0},\qquad
   g^{E}_{14}  = 4  \eps^2 (s+t)    F^E_{0,1,1,0,0,2,1,1,0,1,0,0,0,0,0} ,\\
 g^{E}_{15}  =&   4 \eps^2 t F^E_{1,0,1,0,1,1,1,0,2,0,0,0,0,0,0} ,\qquad
   g^{E}_{16}  =    4 \eps^2 s F^E_{1,0,1,0,2,1,1,1,0,0,0,0,0,0,0} ,\\
   g^{E}_{17}  =&  -2 \eps t F^E_{2,0,1,2,0,1,0,0,1,1,-1,0,0,0,0},\qquad
   g^{E}_{18}  =   4  \eps^2 s t  F^E_{0,2,0,1,0,1,1,1,1,1,0,0,0,0,0} ,\\
  g^{E}_{19}  =&  4 \eps^2 s t F^E_{0,2,0,1,1,1,1,1,1,0,0,0,0,0,0} ,\qquad
   g^{E}_{20}  =  -8   \eps^3 t F^E_{0,1,1,0,1,1,1,0,1,1,0,0,0,0,0},\\
   g^{E}_{21}  =&  -8 \eps^3 s F^E_{0,1,1,0,1,1,1,1,1,0,0,0,0,0,0}, \qquad
  g^{E}_{22} = 4 \eps^2  t^2  F^E_{0,1,1,0,2,1,1,1,1,0,0,0,0,0,0} ,\\
   g^{E}_{23}  =&  -8 \eps^3 (s+t) F^E_{1,0,1,0,1,1,1,1,1,0,0,0,0,0,0},\qquad
  g^{E}_{24}  =   4 \eps^2 s t   F^E_{1,0,2,0,1,1,1,1,1,0,0,0,0,0,0} ,\\
   g^{E}_{25}  =&  -2 \eps s t F^E_{2,0,1,2,0,1,0,1,1,1,-1,0,0,0,0},\qquad
   g^{E}_{26}  =  4\eps^2 s t   F^E_{1,0,2,1,1,1,0,1,1,0,0,0,0,0,0} ,\\
   g^{E}_{27}  =&  -8 \eps^3 t F^E_{1,1,1,1,1,1,0,1,0,0,0,0,0,0,0}, \qquad
  g^{E}_{28}  =  4 \eps^2  s^2 F^E_{1,1,1,1,1,1,0,2,0,0,0,0,0,0,0} ,\\
  g^{E}_{29}  =&  4 \eps^2 s t^2 F^E_{0,2,0,1,1,1,1,1,1,1,0,0,0,0,0} ,\qquad
 g^{E}_{30}  =   4 \eps^2  t^2 F^E_{0,2,0,1,1,1,1,1,1,1,-1,0,0,0,0} ,\\
   g^{E}_{31}  =&  -8 \eps^3 t (s+t) F^E_{0,1,1,0,1,1,1,1,1,1,0,0,0,0,0},\qquad
  g^{E}_{32}  =  4 \eps^2 s  t^2 F^E_{0,1,1,0,1,1,2,1,1,1,0,0,0,0,0} ,\\
 g^{E}_{33}  =&   4 s t \eps^2 F^E_{1,2,0,1,1,1,1,1,0,1,0,0,-1,0,0} ,\qquad
   g^{E}_{35}  = 4 \eps^2 s t
   F^E_{1,1,1,1,1,1,1,2,0,0,-1,0,0,0,0} ,\\
  g^{E}_{34}  =&   {8 \eps^3 t \left(    F^E_{1, 0, 1, 0, 1, 1, 1, 1, 1, 0, 0, 0, 0, 0, 0}- F^E_{1, 1, 0, 1, 1, 1, 1, 1, 0, 1, 0, 0, 0, 0, -1}    \right) },\\
   g^{E}_{36}  =&  -{8 \eps^3  \left( t
   F^E_{1, 0, 1, 0, 1, 1, 1, 1, 1, 0, 0, 0, 0, 0, 0}+  s F^E_{1, 1, 1, 1, 1, 1, 1, 1, 0, 0, 0, -1, 0, 0, 0}\right)},\\
   %
   g^{E}_{37}  =&  -8 \eps^3 t^2
   F^E_{1,1,1,1,1,1,1,0,1,1,-1,0,0,0,0},\qquad
   g^{E}_{38} = -8 \eps^3 s t F^E_{1,1,1,1,1,1,1,1,0,1,-1,0,0,0,0},\\
   g^{E}_{39} =& -8 \eps^3 s t^2
   F^E_{1,1,1,1,1,1,1,1,1,1,-1,0,0,0,0}, \qquad
   g^{E}_{40} = -8 \eps^3 t^2 F^E_{1,1,1,1,1,1,1,1,1,1,-2,0,0,0,0},\\
   g^{E}_{41} =& -8 \eps^3 s t
   F^E_{1,1,1,1,1,1,1,1,1,1,-1,-1,0,0,0} \,.
\end{align}

Having found  a convenient set of master integrals, let us now study the system of differential 
equations they satisfy. We will find that the ladder indeed make all the properties that we were
looking for manifest.

\section{Knizhnik-Zamolodchikov equation for four-point integrals}
\label{sec:diffeq}

Here we study the differential equations satisfied by the master integrals.  
We find that with the above choice of basis, the differential equations take the form predicted
in ref. \cite{Henn:2013pwa}.
The basis integrals discussed in the previous section were normalized to be dimensionless, and hence
only depend on the ratio $x=t/s$.
In this variable, the differential equations take the following form,
\begin{align}\label{KZequation}
\partial_x \, f(x,\eps) = \eps \, \left[ \frac{a}{x} + \frac{b}{1+x} \right] \, f(x,\eps) \,.
\end{align}
This is a specialization of eq. (\ref{diffeq_special}) to one variable, with a specific form of the matrix $A(x)$.
Here $a$ and $b$ are $N\times N$ matrices with constant indices, with $N=26$ and $N=41$, respectively for cases A and E.
Explicit exressions for these matrices are presented in Appendix~A.
We obtain this system of equations for both the triple ladder and the tennis court family of integrals.

We wish to emphasize that the size of the system does not pose any problems when solving the equations,
since the solution is obtained in a completely algebraic way.

We see that equation (\ref{KZequation}) has three regular singularities, at $x=0$, $x=-1$, and $x=\infty$.
These three points correspond to the limits $s=0$, $u=0$, and $t=0$, respectively.
The absence of singularities of planar integrals as $u\to 0$ will provide an important boundary condition, as discussed in the next section.
We remark that equation (\ref{KZequation})  is a particular case of the Knizhnik-Zamolodchikov equations \cite{Knizhnik:1984nr}. It can also be described as a Fuchsian system of differential equations with three regular singular points.

Let us now discuss the solution of those equations.
The normalization of the master integrals in eq. (\ref{masters_ladder}) was 
chosen such that functions $f_{i}$ are finite as $\eps \to 0$.
We are interested in a solution near $D\approx 4$ dimensions, so we parametrize, e.g. for family $A$,
\begin{align}
f^{A}_{i}(x,\eps) = \sum_{j=0}^{6} \eps^j f_{i}^{A,j}(x) + \cO(\eps^7) \,.
\end{align}
{}From eq. (\ref{KZequation}) it is clear that the iterative solution in $\eps$ for all functions $f_{i}$ can be
expressed in terms of harmonic polylogarithms \cite{Remiddi:1999ew} of argument $x$ and with indices drawn from $0,-1$.
Equation (\ref{KZequation}) determines the solution up to boundary constants.
We will determine the latter in the next section. Here we would already like to mention that the boundary 
constants have the property of uniform weight, and this, together with the structure of eq. (\ref{KZequation}), implies that all basis functions are pure functions of uniform weight, as anticipated.

\subsection{Boundary conditions}

For planar graphs we expect the limit $u\to 0$, i.e. $x\to -1$ to be finite.
Another condition that we can impose it that the solution be real for $x>0$, i.e. when $s$ and $t$ have the same sign. For planar graphs, this is obvious from the Feynman parametrization.
As we will see, these assumptions fix almost all of the boundary constants in this problem, except for some 
elementary propagator-type integrals.

As can be seen from  (\ref{KZequation}), the entries $1/(1+x)$ can lead to terms singular as $x\to -1$, and
the regularity at $x\to -1$ therefore imposes constraints on the integration constants.
For example, at order $\eps$, this condition means that $H_{1}(x) = \log(1+x)$ must come with zero coefficient,
and this imposes constraints on the integration constants at order $\eps^0$.
The absence of the function ${\rm Li}_{2}$ at order $\eps^2$ in our results can be understand in this way.
Given these constraints, one might wonder how one can obtain functions different from logarithms. The answer is the following.
At higher orders, there can be an interplay between boundary constants at different orders, as the following example
shows,
\begin{align}
\pi^2 \int_{-1+\delta}^x \; d \log(1+y) - \int_{-1+\delta}^x \log^2 y \; d\log(1+y) \,,
\end{align}
which is finite as $\delta \to 0$, and hence there can be finite combinations of HPLs with indices $-1$.

In practice, we found that when computing up to order $\eps^n$, considering the consistency condition with $x\to -1$ at order $\eps^{(n+1)}$ and $\eps^{(n+2)}$ gives all constraints. These constraints are very powerful. We found that, together with condition that the solution be real for $x>0$, they determine most boundary conditions.

The only additional information needed can easily be obtained from the propagator-type 
integral $f_1$, which can be expressed in terms of $\Gamma$ functions,
\begin{align}\label{resultful}
f^{A}_1  =&  \; e^{3 \eps \gamma_{\rm E}}  { \Gamma^4(1-\eps) \Gamma(1+3 \eps) }/{\Gamma(1-4 \eps)}  \nonumber \\
=& \;1 - \eps^2 \, \frac{\pi^2}{4}  - 29 \eps^3 \, \zeta_3 - \eps^4\, \frac{71}{160}\pi^4 
+ \eps^5 \, \left( \frac{29}{4}\pi^2 \zeta_3-\frac{1263}{5} \zeta_5 \right) 
\nonumber \\
& \;+ \eps^6  \, \left(-\frac{11539}{24192}\pi^6+\frac{841}{2} \zeta_3^2\right) + \cO(\eps^7) \,.
\end{align}

\subsection{Summary and explicit results}

In summary, the equations (\ref{KZequation}), together with finiteness at $x\to -1$, reality of the solution in the region $x>0$, and the exact result for the trivial integral (\ref{resultful}) determines all basis functions to all orders in $\eps$.
The solution can be obtained in an algebraic way. 
At each order $\eps^n$, it is given by a linear combination of HPLs. 
The transcendental weight of each term is $n$.
In Appendix~B, we present explicit results for the ten-propagator integrals, 
up to order $\eps^6$, i.e. transcendental weight $6$.
Explicit results for all integrals, and up to weight $6$, can be found in the ancillary files {\tt resultA.m} and {\tt resultE.m}.

We performed a series of analytical and numerical checks of our results.
The highest poles in $\epsilon$ were evaluated using the general Mellin-Barnes representations
derived in refs.~\cite{Smirnov:2003vi,Bern:2005iz}. 
The two known analytical results  for the triple box without numerator~\cite{Smirnov:2003vi},
i.e. $f^A_{24}$
and for the tennis court diagram with a special numerator~\cite{Bern:2005iz}
i.e. $f^E_{39}$ also served as important checks.
All the master integrals (except for the ten-propagator
integrals of family E) were also numerically checked with {\tt FIESTA}
\cite{Smirnov:2008py,Smirnov:2009pb} with sufficient accuracy.

Finally, we wish to mention that the {\em symbol} \cite{arXiv:math/0606419,2009arXiv0908.2238G} 
of the terms in the solution can be obtained in an
even more straightforward way, and in that case the only information required in addition to eq.  (\ref{KZequation})
is the value of the first term in the $\eps$ expansion.
The latter follows from the boundary conditions, as explained above, but we give it here for convenience.
We have
\begin{align}
f^{A,0} =& \{1,1,-\frac{1}{9},-\frac{1}{6},1,-1,\frac{16}{9},0,1,\frac{1}{4},0,0,1,-\frac{1}{4},0,-\frac{1}{4},0,-\frac{4}{9},\frac{49}{36},0,\frac{7}{3},\frac{25}{9}, \nonumber \\ & -\frac{
   16}{9},\frac{16}{9},-\frac{49}{36},-\frac{4}{9}\}\,,
   \end{align}
and
\begin{align}
f^{E,0} =& \{1,1,\frac{2}{9},\frac{1}{3},-1,1,\frac{1}{3},1,-\frac{32}{9},0,0,0,-\frac{8}{3},0,0, 
\frac{1}{9},\frac{2}{9},-\frac{8}{9},-\frac{8}{9},0,0,-1,0, -\frac{4}{9},-\frac{32}{9},\nonumber \\
&-\frac{8}{9},0,-1,\frac{77}{9},-\frac{16}{3},0,\frac{28}{3},-\frac{49}{9},0,-\frac{49}{9},0,-\frac{2}{9},-\frac{14}{9},\frac{128}{9},-\frac{98}{9},-\frac{56
   }{9}\}\,.
\end{align}
This, together with the differential equations (\ref{KZequation}) and the explicit form of the matrices $a$ and $b$ given in eqs. (\ref{matrixAa}) - (\ref{matrixEb})  completely specifies the symbol of the answer, to any order in $\eps$.

\section{Discussion and outlook}
\label{sec:discussion}

In this work, we computed the master integrals for planar massless four-point integrals. Via IBP, they are sufficient to compute all integrals relevant for virtual corrections to $2 \rightarrow 2$ scattering at that order. We wrote out results in the small $\eps$ expansion up to weight six, and using the information provided here, higher-order results can be obtained at will.

It is interesting to note that as a by-product of our analysis, we also obtained result for
three-loop single scale integrals that na{\"i}rely cannot be obtained from differential equations.
We found that they were entirely determined from consistency of the system of differential
equations with the physical boundary conditions. 

We focused on the phenomenologically relevant expansion of the master integrals for $\eps \to 0$,
and solved this problem in principle to all orders in $\eps$.
It is interesting to ask if one can write down a solution for the master integrals valid for finite $\eps$.
The Knizhnik-Zamolodchikov equations should be a good starting point for such an analysis.
See for example ref. \cite{Kalmykov:2012rr,Kalmykov:2010xv} and references therein for cases where the solution can
be expressed in terms of (generalized) hypergeometric functions.

An obvious future direction is to apply this method to previously unknown non-planar integrals at three loops.
The latter are required in order to evaluate the three-loop non-planar contributions to supersymmetric Yang-Mills
and supergravity theories, where explicit representations in terms of loop integrals are available, see \cite{Bern:2007hh}
and references therein.

The knowledge that certain integrals are pure functions
of uniform transcendentality, can also be of practical advantage independently of the
differential equations methods. 
Apart from serving as an important check of calculations,
this property simplifies very much the application
of the so-called {\tt PSLQ} algorithm \cite{PSLQ} because one then
needs to consider only transcendental numbers of a given weight,
and not numbers of lower weights. Another characteristic example of
uniform transcendentality is within the method suggested in ref.~\cite{Fleischer:1998nb}, 
where the dependence of the coefficient at the $n$-th term of a Taylor
series is revealed from the information about finite number of terms and
the uniform transcendentality essentially restricts the number of terms
in the corresponding Ansatz.

It would be interesting to understand further criteria for integrals to be pure functions of uniform transcendentally. It is possible that this might also be of interest for mathematicians, who have been
investigating transcendental properties of Feynman integrals, see e.g. \cite{Brown:2009rc,Marcolli} and references therein, albeit usually for particular classes of single-scale off-shell integrals in strictly four dimensions.

\vspace{0.2 cm}
{\em Acknowledgments.}
We thank T.~Gehrmann for useful discussions, and P.~Marquard for help in producing the figures.
J.M.H. thanks the organizers of LoopFest XII, where this work was presented, for their invitation.
J.M.H. was supported in part by the Department of Energy grant DE-SC0009988,
and by the IAS AMIAS fund. 
 The work by A.S. and V.S. was supported by the Russian Foundation for Basic Research through grant
11-02-01196. The work of A.S. was also supported by the DFG through SFB/TR~9 ``Computational Particle Physics''.

\appendix

\section{Matrices in Knizhnik-Zamolodchikov equation}

The non-zero matrix elements of $a$ and $b$ in (\ref{KZequation})
for both cases are given by the following relations:
\bea\label{matrixAa}
a^A_{1,1}  &=& -3, a^A_{7,1} = 4/3, a^A_{7,7} = -3, a^A_{8,1} = -1/6, a^A_{8,4} = 
-1, a^A_{8,8} = -3, a^A_{9,1} = 1, 
\nonumber \\
a^A_{9,9} &=& -3, a^A_{11,1} = -1/3, 
a^A_{11,2} = 1/3, a^A_{11,11} = -3, a^A_{14,1} = -1/4, a^A_{14,9} = 1/2, 
\nonumber \\
a^A_{15,15}  &=& 3, a^A_{16,1} = 1/3, a^A_{16,4} = -8, a^A_{16,5} = -8/3, 
a^A_{16,15} = 12, a^A_{16,16} = -3, 
\nonumber \\
a^A_{18,1}  &=& 2, a^A_{18,3} = -2/3, 
a^A_{18,4} = -40/9, a^A_{18,7} = -1, a^A_{18,8} = -24, a^A_{18,9} = -2, 
\nonumber \\
a^A_{18,10}  &=& 4/3, a^A_{18,14} = -8/3, a^A_{18,18} = 1, a^A_{18,19} = 2/3, 
a^A_{19,1} = -2, a^A_{19,4} = 8, 
\nonumber \\
a^A_{19,7}  &=& 3/2, a^A_{19,8} = 24, a^A_{19,9} 
= 2, a^A_{19,19} = -3, a^A_{20,1} = 23/27, a^A_{20,2} = 17/54, 
\nonumber \\
a^A_{20,3}  &=& 
-1/6, a^A_{20,4} = -56/9, a^A_{20,5} = -14/9, a^A_{20,6} = 1/6, a^A_{20,7} = 
-1, a^A_{20,8} = -20/3, 
\nonumber \\
a^A_{20,11}  &=& -2, a^A_{20,15} = 8/3, a^A_{20,16} = 
-2, a^A_{20,17} = -2/3, a^A_{20,20} = 1, a^A_{20,21} = 1/3, 
\nonumber \\
a^A_{21,1}  &=& -4/3, a^A_{21,2} = -4/3, a^A_{21,7} = 3, a^A_{21,11} = 12, a^A_{21,21} = -3, 
a^A_{22,1} = -4/3, 
\nonumber \\
a^A_{22,2}  &=& -4/3, a^A_{22,7} = 3, a^A_{22,11} = 12, 
a^A_{22,22} = -3, a^A_{23,1} = 20/9, a^A_{23,2} = 19/9, 
\nonumber \\
a^A_{23,3}  &=& -2, 
a^A_{23,7} = -3, a^A_{23,11} = -20, a^A_{23,12} = 1, a^A_{23,22} = 2, 
a^A_{23,23} = 1, 
\nonumber \\
a^A_{24,19}  &=& 4, a^A_{24,21} = -4, a^A_{24,22} = 2, 
a^A_{24,24} = -3, a^A_{25,1} = -8/3, a^A_{25,2} = 41/18, 
\nonumber \\
a^A_{25,3}  &=& -7/2, 
a^A_{25,4} = 68/3, a^A_{25,5} = 14/9, a^A_{25,6} = 7/2, a^A_{25,8} = 48, 
a^A_{25,9} = 4, 
\nonumber \\
a^A_{25,10}  &=& 3, a^A_{25,11} = -12, a^A_{25,12} = 3, 
a^A_{25,13} = 1, a^A_{25,17} = -2, a^A_{25,19} = -6, 
\nonumber \\
a^A_{25,21}  &=& 6, 
a^A_{25,22} = -2, a^A_{25,24} = 2, a^A_{25,25} = 1, a^A_{26,1} = -28/9, 
a^A_{26,2} = -7/6, 
\nonumber \\
a^A_{26,3}  &=& 9/2, a^A_{26,4} = 20/3, a^A_{26,5} = 22/9, 
a^A_{26,6} = 3/2, a^A_{26,7} = 3, a^A_{26,8} = 16, 
\nonumber \\
a^A_{26,10}  &=& 3, a^A_{26,11} = 12, a^A_{26,13} = 1, a^A_{26,15} = -16, a^A_{26,16} = 4, 
\nonumber \\
a^A_{26,17}  &=& 6, 
a^A_{26,19} = -2, a^A_{26,20} = -12, a^A_{26,21} = 2, a^A_{26,22} = -3, 
\nonumber \\
a^A_{26,23}  &=& -3, a^A_{26,24} = 1, a^A_{26,25} = 1 
\,,
\eea

\bea\label{matrixAb}
b^A_{7,1} &=& -4/3, b^A_{7,3} = 4, b^A_{7,7} = 1, b^A_{8,8} = 2, b^A_{9,1} = -1, 
b^A_{9,9} = 2, b^A_{9,14} = 4, 
\nonumber\\
b^A_{11,11} &=& 3, b^A_{14,1} = 1/4, b^A_{14,9} = 
-1/2, b^A_{14,14} = -1, b^A_{15,1} = -1/12, b^A_{15,4} = 2, 
\nonumber\\
b^A_{15,5} &=& 2/3, 
b^A_{15,8} = 2, b^A_{15,15} = -3, b^A_{15,16} = 1, b^A_{16,1} = -1/3, 
b^A_{16,4} = 8, 
\nonumber\\
b^A_{16,5} &=& 8/3, b^A_{16,15} = -12, b^A_{16,16} = 4, 
b^A_{18,1} = -2, b^A_{18,3} = 2/3, b^A_{18,4} = 40/9, 
\nonumber\\
b^A_{18,7} &=& 1, 
b^A_{18,8} = 24, b^A_{18,9} = 2, b^A_{18,10} = 2/3, b^A_{18,14} = 8/3, 
b^A_{18,18} = -1, 
\nonumber\\
b^A_{18,19} &=& -2/3, b^A_{19,1} = 2, b^A_{19,3} = 4, 
b^A_{19,4} = -40/3, b^A_{19,7} = -3/2, b^A_{19,8} = -24, 
\nonumber\\
b^A_{19,9} &=& -2, 
b^A_{19,10} = -2, b^A_{19,18} = 3, b^A_{19,19} = 2, b^A_{20,20} = 1, b^A_{21,1} 
= -16/9, 
\nonumber\\
b^A_{21,2} &=& 13/9, b^A_{21,3} = 7, b^A_{21,4} = 40/3, b^A_{21,5} = 
4, b^A_{21,6} = 2, b^A_{21,8} = 16, 
\nonumber\\
b^A_{21,15} &=& -16, b^A_{21,16} = 4, 
b^A_{21,17} = 4, b^A_{21,20} = -12, b^A_{21,21} = 1, b^A_{22,1} = 4/3, 
\nonumber\\
b^A_{22,2} &=& 5/3, b^A_{22,3} = 6, b^A_{22,7} = -3, b^A_{22,11} = -12, 
b^A_{22,12} = -3, b^A_{22,22} = 3, 
\nonumber\\
b^A_{22,23} &=& 3, b^A_{23,1} = -20/9, 
b^A_{23,2} = -10/9, b^A_{23,7} = 3, b^A_{23,11} = 20, b^A_{23,12} = 2, 
\nonumber\\
b^A_{23,22} &=& -2, b^A_{23,23} = -2, b^A_{24,2} = -17/9, b^A_{24,3} = 7, 
b^A_{24,4} = -40/3, b^A_{24,5} = -28/9, 
\nonumber\\
b^A_{24,6} &=& -7, b^A_{24,10} = -6, 
b^A_{24,12} = -6, b^A_{24,13} = -2, b^A_{24,17} = 4, b^A_{24,19} = -4, 
\nonumber\\
b^A_{24,21} &=& 4, b^A_{24,22} = -2, b^A_{24,24} = 3, b^A_{24,25} = 2, 
b^A_{24,26} = 2, b^A_{25,1} = 52/9, 
\nonumber\\
b^A_{25,2} &=& -1/2, b^A_{25,3} = -5/2, 
b^A_{25,4} = -100/3, b^A_{25,5} = -22/9, b^A_{25,6} = 3/2, b^A_{25,7} = -3, 
\nonumber\\
b^A_{25,8} &=& -64, b^A_{25,9} = -4, b^A_{25,10} = -1, b^A_{25,12} = 3, 
b^A_{25,13} = 1, b^A_{25,15} = 16, 
\nonumber\\
b^A_{25,16} &=& -4, b^A_{25,17} = -6, 
b^A_{25,18} = 6, b^A_{25,19} = 6, b^A_{25,20} = 12, b^A_{25,21} = -4, 
\nonumber\\
b^A_{25,22} &=& 2, b^A_{25,24} = -2, b^A_{25,25} = -1, b^A_{25,26} = -2, 
b^A_{26,1} = 28/9, b^A_{26,2} = 7/6, 
\nonumber\\
b^A_{26,3} &=& -9/2, b^A_{26,4} = -20/3, 
b^A_{26,5} = -22/9, b^A_{26,6} = 3/2, 
\nonumber\\
b^A_{26,7} &=& -3, b^A_{26,8} = -16, 
b^A_{26,10} = 3, b^A_{26,11} = -12, b^A_{26,13} = 1, b^A_{26,15} = 16, 
\nonumber\\
b^A_{26,16} &=& -4, b^A_{26,17} = -6, b^A_{26,19} = 2, b^A_{26,20} = 12, 
b^A_{26,21} = -2, b^A_{26,22} = 3, 
\nonumber\\
b^A_{26,23} &=& 3, b^A_{26,24} = -1, 
b^A_{26,25} = -1
\,,
\eea

\bea\label{matrixEa}
a^E_{2,2} &=& -3, a^E_{3,3} = -3, a^E_{4,4} = -3, a^E_{5,5} = -3, a^E_{6,6} = -3, 
a^E_{9,3} = -4, a^E_{9,9} = -1, 
\nonumber\\
a^E_{10,1} &=& 2/3, a^E_{10,4} = -2, a^E_{10,10} 
= -2, a^E_{11,11} = -3, a^E_{12,2} = -2/3, a^E_{12,7} = 2, 
\nonumber\\
a^E_{12,12} &=& -3, 
a^E_{13,4} = -4, a^E_{13,13} = -2, a^E_{14,1} = 4/3, a^E_{14,2} = -4/3, 
a^E_{14,14} = -3, 
\nonumber\\
a^E_{15,15} &=& -3, a^E_{17,17} = -3, a^E_{18,7} = -8/3, 
a^E_{18,11} = -4, a^E_{18,12} = 4, a^E_{18,18} = -1, 
\nonumber\\
a^E_{19,4} &=& -8, 
a^E_{19,6} = 16/3, a^E_{19,13} = 2, a^E_{19,19} = -3, a^E_{20,20} = -3, 
a^E_{21,1} = -2/3, 
\nonumber\\
a^E_{21,4} &=& -8, a^E_{21,6} = 16/3, a^E_{21,10} = 4, 
a^E_{21,21} = -3, a^E_{21,22} = 2, a^E_{22,22} = -4, 
\nonumber\\
a^E_{23,1} &=& 1/3, 
a^E_{23,4} = 8/3, a^E_{23,7} = 4, a^E_{23,10} = -4, a^E_{23,13} = 1, 
a^E_{23,15} = -10/3, 
\nonumber\\
a^E_{23,16} &=& -3, a^E_{23,23} = -3, a^E_{23,24} = -1, 
a^E_{24,24} = -3, a^E_{25,17} = -4, a^E_{25,25} = -1, 
\nonumber\\
a^E_{26,26}&=& -3, 
a^E_{27,27} = 3, a^E_{28,2} = 4/3, a^E_{28,7} = 16, a^E_{28,8} = -32/3, 
a^E_{28,27} = -6, 
\nonumber\\
a^E_{28,28} &=& -3, a^E_{29,2} = -4, a^E_{29,4} = 16, 
a^E_{29,5} = -6, a^E_{29,6} = -16/3, a^E_{29,29} = -3, 
\nonumber\\
a^E_{29,30}&=&-2, 
a^E_{30,5} = 6, a^E_{30,30} = -3, a^E_{31,1} = 184/27, a^E_{31,2} = 68/27, 
a^E_{31,3} = 2/3, 
\nonumber\\
a^E_{31,4} &=& 224/9, a^E_{31,5} = 4/3, a^E_{31,6} = -112/9, 
a^E_{31,9} = 4, a^E_{31,10} = -40/3, a^E_{31,14} = -4, 
\nonumber\\
a^E_{31,20} &=& 2/3, 
a^E_{31,21} = -8/3, a^E_{31,22} = -4, a^E_{31,31} = -5, a^E_{31,32} = 2/3, 
a^E_{32,1} = 112/9, 
\nonumber\\
a^E_{32,2}&=& -4/9, a^E_{32,3} = 14, a^E_{32,4} = 80/3, 
a^E_{32,5} = -8, a^E_{32,6} = -16, a^E_{32,9} = 6, 
\nonumber\\
a^E_{32,10} &=&-16, 
a^E_{32,14} = -12, a^E_{32,20} = 2, a^E_{32,21} = -8, a^E_{32,22} = -4, 
a^E_{32,31} = -6, 
\nonumber\\
a^E_{32,32} &=& -1, a^E_{33,2} = 38/3, a^E_{33,4} = 22/3, 
a^E_{33,6} = -24, a^E_{33,15} = 25/3, a^E_{33,22} = -4, 
\nonumber\\
a^E_{33,24}&=& 11/2, 
a^E_{33,33} = -3, a^E_{33,34} = 5, a^E_{34,34} = -3, a^E_{35,2} = -8/3, 
a^E_{35,4} = 8/3, 
\nonumber\\
a^E_{35,15} &=& 20/3, a^E_{35,24} = 4, a^E_{35,35} = -2, 
a^E_{36,1} = 29/6, a^E_{36,2} = 12/5, a^E_{36,4} = 40/3, 
\nonumber\\
a^E_{36,7} &=& 28, 
a^E_{36,8} = -40/3, a^E_{36,10} = -18, a^E_{36,12} = -8, a^E_{36,13} = 27/5, 
a^E_{36,15} = -8/3, 
\nonumber\\
a^E_{36,16} &=& -3/2, a^E_{36,24} = -8/5, a^E_{36,27} = 
-24/5, a^E_{36,28} = -4, a^E_{36,35} = -2/5, a^E_{36,36} = 1, 
\nonumber\\
a^E_{37,37} &=& 
-3, a^E_{38,2} = -4, a^E_{38,4} = -12, a^E_{38,6} = 16, a^E_{38,15} = -10, 
a^E_{38,22} = 4, a^E_{38,24} = -5, 
\nonumber\\
a^E_{38,34} &=& -10, a^E_{38,35} = 2, 
a^E_{38,38} = -3, a^E_{39,1} = -224/9, a^E_{39,2} = -64/9, a^E_{39,3} = -28, 
\nonumber\\
a^E_{39,4}&=& 104/3, a^E_{39,5} = -8, a^E_{39,6} = -64/3, a^E_{39,9} = -12, 
a^E_{39,10} = 32, a^E_{39,14} = 24, 
\nonumber\\
a^E_{39,15} &=&20, a^E_{39,20} = -4, 
a^E_{39,21} = 16, a^E_{39,24} = 10, a^E_{39,30} = -8, a^E_{39,31} = 12, 
\nonumber\\
a^E_{39,32} &=& -4, a^E_{39,34} = 20, a^E_{39,39} = -3, a^E_{39,40} = -2, 
a^E_{40,1} = 224/9, a^E_{40,2} = 136/9, 
\nonumber\\
a^E_{40,3} &=& 28, a^E_{40,4} = -32/3, 
a^E_{40,5} = 8, a^E_{40,6} = -32/3, a^E_{40,9} = 12, a^E_{40,10} = -32, 
\nonumber\\
a^E_{40,14} &=& -24, a^E_{40,20} = 4, a^E_{40,21} = -16, a^E_{40,22} = -8, 
a^E_{40,30} = 8, a^E_{40,31} = -12, 
\nonumber\\
a^E_{40,32}&=& 4, a^E_{40,40} = -3, 
a^E_{41,1} = -26/9, a^E_{41,2} = 124/3, a^E_{41,3} = -28, a^E_{41,4} = 208/3, 
\nonumber\\
a^E_{41,5} &=& 4, a^E_{41,6} = -32, a^E_{41,7} = 208/3, a^E_{41,8} = 128/9, 
a^E_{41,10} = 40, a^E_{41,12} = -140, 
\nonumber\\
a^E_{41,13} &=& 46, a^E_{41,14} = -18, 
a^E_{41,15} = -4/3, a^E_{41,16} = 2, a^E_{41,17} = 16, a^E_{41,18} = -6, 
\nonumber\\
a^E_{41,19} &=& -16, a^E_{41,20} = -4, a^E_{41,21} = -8, a^E_{41,22} = -12, 
a^E_{41,23} = 2, a^E_{41,24} = -3, 
\nonumber\\
a^E_{41,25} &=& -4, a^E_{41,26} = -2, 
a^E_{41,27} = 4, a^E_{41,28} = 4, a^E_{41,29} = -4, 
\nonumber\\
a^E_{41,30} &=& -4, 
a^E_{41,31} = 6, a^E_{41,32} = 2, a^E_{41,33} = -4, a^E_{41,35} = 2, 
\nonumber\\
a^E_{41,37} &=& 1, a^E_{41,38} = -1, a^E_{41,39} = 1, a^E_{41,40} = 1
\,,
\eea

\bea\label{matrixEb}
b^E_{9,1} &=& 8/3, b^E_{9,3} = 4, b^E_{9,9} = 1, b^E_{10,10} = 2, b^E_{12,12} = 
2, b^E_{13,4} = 4, 
\nonumber\\
b^E_{13,7} &=& 4, b^E_{13,13} = 1, b^E_{14,14} = 3, b^E_{18,7} 
= 8/3, b^E_{18,11} = 4, b^E_{18,12} = -4, 
\nonumber\\
b^E_{18,18} &=& 1, b^E_{19,4} = 8, 
b^E_{19,6} = -16/3, b^E_{19,13} = -2, b^E_{19,19} = 3, b^E_{21,1} = 2/3, 
\nonumber\\
b^E_{21,4} &=& 8, b^E_{21,6} = -16/3, b^E_{21,10} = -4, b^E_{21,21} = -3, 
b^E_{21,22} = -2, b^E_{22,1} = -4/3, 
\nonumber\\
b^E_{22,4} &=& -16, b^E_{22,6} = 32/3, 
b^E_{22,21} = 6, b^E_{22,22} = 4, b^E_{23,23} = 5, b^E_{24,1} = -2, 
\nonumber\\
b^E_{24,2} 
&=& -4/3, b^E_{24,4} = 16/3, b^E_{24,7} = 8, b^E_{24,10} = -8, b^E_{24,12} = 4, 
b^E_{24,15} = -8/3, 
\nonumber\\
b^E_{24,16} &=& -6, b^E_{24,23} = 4, b^E_{24,24} = 1, 
b^E_{25,8} = 8/3, b^E_{25,17} = 4, b^E_{25,25} = 1, 
\nonumber\\
b^E_{26,7} &=& 8, b^E_{26,8} 
= -16/3, b^E_{26,13} = -2, b^E_{26,26} = 3, b^E_{27,2} = 2/3, b^E_{27,7} = 8, 
\nonumber\\
b^E_{27,8} &=&-16/3, b^E_{27,12} = -4, b^E_{27,27} = -3, b^E_{27,28} = -2, 
b^E_{28,2} = -4/3, b^E_{28,7} = -16, 
\nonumber\\
b^E_{28,8} &=& 32/3, b^E_{28,27} = 6, 
b^E_{28,28} = 4, b^E_{29,2} = 4, b^E_{29,4} = -16, b^E_{29,5} = 6, 
\nonumber\\
b^E_{29,6} &=& 
16/3, b^E_{29,18} = 3, b^E_{29,19} = 2, b^E_{29,29} = 2, b^E_{29,30} = 2, 
b^E_{30,2} = -2, 
\nonumber\\
b^E_{30,5} &=& -3, b^E_{30,6} = 8/3, b^E_{30,7} = -4, 
b^E_{30,11} = -6, b^E_{30,12} = 6, b^E_{30,13} = 2, 
\nonumber\\
b^E_{30,18} &=& -3, 
b^E_{30,19} = -4, b^E_{30,29} = -1, b^E_{30,30} = -1, b^E_{31,31} = 1, 
b^E_{32,1} = -64/9, 
\nonumber\\
b^E_{32,2} &=& 52/9, b^E_{32,3} = -14, b^E_{32,4} = -80/3, 
b^E_{32,5} = 8, b^E_{32,6} = 16, b^E_{32,10} = 16, 
\nonumber\\
b^E_{32,20} &=& -2, 
b^E_{32,21} = 8, b^E_{32,22} = 4, b^E_{32,31} = 6, b^E_{32,32} = 1, b^E_{33,1} 
= -2, 
\nonumber\\
b^E_{33,2} &=& -18, b^E_{33,4} = -18, b^E_{33,6} = 24, b^E_{33,7} = -16, 
b^E_{33,10} = 16, b^E_{33,12} = 16, 
\nonumber\\
b^E_{33,13} &=&-8, b^E_{33,15} = -3, 
b^E_{33,16} = 2, b^E_{33,22} = 4, b^E_{33,23} = -8, b^E_{33,24} = -3/2, 
\nonumber\\
b^E_{33,33} &=& 2, b^E_{33,34} = -5, b^E_{34,1} = -13/15, b^E_{34,2} = -16/5, 
b^E_{34,4} = -122/5, b^E_{34,6} = 40/3, 
\nonumber\\
b^E_{34,7} &=& -12, b^E_{34,10} = 28/5, 
b^E_{34,12} = 58/5, b^E_{34,13} = -16/5, b^E_{34,15} = 1/5, b^E_{34,16} = 
11/5, 
\nonumber\\
b^E_{34,21}&=& 24/5, b^E_{34,22} = 4, b^E_{34,23} = -14/5, b^E_{34,24} = 
-3/5, b^E_{34,33} = 2/5, b^E_{34,34} = -1, 
\nonumber\\
b^E_{35,1} &=& -39/2, b^E_{35,2} = 
-14/3, b^E_{35,4} = -52/3, b^E_{35,7} = -24, b^E_{35,8} = 24, b^E_{35,10} = 22, 
\nonumber\\
b^E_{35,12} &=& 22, b^E_{35,13} = -11, b^E_{35,15} = 2/3, b^E_{35,16} = 
3/2, b^E_{35,23} = -6, b^E_{35,24} = 3/2, 
\nonumber\\
b^E_{35,28} &=& 4, b^E_{35,35} = 2, 
b^E_{35,36} = -5, b^E_{36,1} = -9/2, b^E_{36,2} = -12/5, b^E_{36,4} = -32/3, 
\nonumber\\
b^E_{36,7} &=& -24, b^E_{36,8} = 40/3, b^E_{36,10} = 14, b^E_{36,12} = 8, 
b^E_{36,13} = -22/5, b^E_{36,15} = -2/3, 
\nonumber\\
b^E_{36,16} &=& -3/2, b^E_{36,23} = 2, 
b^E_{36,24} = 3/5, b^E_{36,27} = 24/5, b^E_{36,28} = 4, b^E_{36,35} = 2/5, 
\nonumber\\
b^E_{36,36} &=& -1, b^E_{38,2} = 4, b^E_{38,4} = 12, b^E_{38,6} = -16, 
b^E_{38,15} = 10, b^E_{38,22} = -4, 
\nonumber\\
b^E_{38,24} &=& 5, b^E_{38,34} = 10, 
b^E_{38,35} = -2, b^E_{38,38} = 3, b^E_{39,1} = 68/9, b^E_{39,2} = -8/9, 
\nonumber\\
b^E_{39,3} &=& 28, b^E_{39,4} = -56, b^E_{39,5} = 8, b^E_{39,6} = 64/3, 
b^E_{39,12} = 8, b^E_{39,13} = -12, 
\nonumber\\
b^E_{39,14} &=& -12, b^E_{39,15} = -52/3, 
b^E_{39,16} = -4, b^E_{39,18} = 12, b^E_{39,20} = 4, b^E_{39,21} = -16, 
\nonumber\\
b^E_{39,23} &=& -4, b^E_{39,24} = -8, b^E_{39,25} = 8, b^E_{39,26} = 4, 
b^E_{39,30} = 8, b^E_{39,31} = -12, 
\nonumber\\
b^E_{39,32} &=& 4, b^E_{39,34} = -20, 
b^E_{39,38} = 2, b^E_{39,39} = 3, b^E_{39,40} = 2, b^E_{39,41} = 2, 
\nonumber\\
b^E_{40,1} &=& -68/9, 
b^E_{40,2} = -64/9, b^E_{40,3} = -28, b^E_{40,4} = 32, b^E_{40,5} = 
-8, b^E_{40,6} = 32/3, 
\nonumber\\
b^E_{40,12} &=& -8, b^E_{40,13} = 12, b^E_{40,14} = 12, 
b^E_{40,15} = -8/3, b^E_{40,16} = 4, b^E_{40,18} = -12, 
\nonumber\\
b^E_{40,20} &=& -4, 
b^E_{40,21} = 16, b^E_{40,22} = 8, b^E_{40,23} = 4, b^E_{40,24} = -2, 
b^E_{40,25} = -8, 
\nonumber\\
b^E_{40,26} &=& -4, b^E_{40,30} = -8, b^E_{40,31} = 12, 
b^E_{40,32} = -4, b^E_{40,35} = 4, b^E_{40,37} = 1, 
\nonumber\\
b^E_{40,38} &=& -6, 
b^E_{40,39} = -2, b^E_{40,40} = -1, b^E_{40,41} = -2, b^E_{41,1} = 26/9, 
b^E_{41,2} = -332/9, 
\nonumber\\
b^E_{41,3} &=& 28, b^E_{41,4} = -208/3, b^E_{41,5} = -4, 
b^E_{41,6} = 32, b^E_{41,7} = -80, b^E_{41,10} = -40, 
\nonumber\\
b^E_{41,12} &=& 124, 
b^E_{41,13} = -30, b^E_{41,14} = 18, b^E_{41,15} = 4/3, b^E_{41,16} = -2, 
b^E_{41,17} = -16, 
\nonumber\\
b^E_{41,18} &=& 6, b^E_{41,19} = 16, b^E_{41,20} = 4, 
b^E_{41,21} = 8, b^E_{41,22} = 12, b^E_{41,23} = -2, 
\nonumber\\
b^E_{41,24} &=& 3, 
b^E_{41,25} = -4, b^E_{41,26} = -6, b^E_{41,29} = 4, b^E_{41,30} = 4, 
b^E_{41,31} = -6, 
\nonumber\\
b^E_{41,32} &=& -2, b^E_{41,33} = 4, b^E_{41,35} = -2, 
b^E_{41,37} = -1, b^E_{41,38} = 1, 
\nonumber\\
b^E_{41,39}&=& -1, b^E_{41,40} = -1
\,.
\eea

\section{Explicit results up to weight six}

Here are results for master integrals with ten propagators. 
We denote harmonic polylogarithms \cite{Remiddi:1999ew} by $H_{\vec{w}} = H_{\vec{w}}(x)$.
All the other results can be found in the ancillary files {\tt resultA.m} and {\tt resultE.m}.

\subsection{Triple ladder master integrals}

\bea
f^{A}_{24}(x,\eps)=
\frac{16}{9}-\frac{11}{3} \epsilon  H_{0}
   +\epsilon ^2 \biggl(-\frac{3 \pi ^2}{2}+6 H_{0,0}\biggr)
   +\epsilon ^3 \biggl(-\frac{3}{2} \pi ^2 H_{-1}+\frac{65}{12} \pi ^2 H_{0}-3 H_{-1,0,0}
&& \nn \\ &&  \hspace*{-140mm}   
   -3 H_{0,0,0}-\frac{131 \zeta _3}{9}\biggr)
   +\epsilon ^4 \biggl(-\frac{1411 \pi ^4}{1080}-\frac{3}{2} \pi ^2 H_{-1,-1}+\frac{7}{2} \pi ^2 H_{-1,0}
   +\frac{23}{2} \pi ^2 H_{0,-1}-19 \pi ^2 H_{0,0}
\nn \\ &&  \hspace*{-140mm}    
   -3 H_{-1,-1,0,0}+18 H_{-1,0,0,0}+23 H_{0,-1,0,0}-36 H_{0,0,0,0}-3 H_{-1} \zeta _3+\frac{82}{3} H_{0} \zeta _3\biggr)
\nn \\ &&  \hspace*{-140mm}    
   +\epsilon ^5 \biggl(-\frac{13}{8} \pi ^4 H_{-1}+\frac{683}{160} \pi ^4 H_{0}-\frac{3}{2} \pi ^2 H_{-1,-1,-1}
   +\frac{7}{2} \pi ^2 H_{-1,-1,0}+\frac{35}{2} \pi ^2 H_{-1,0,-1}
\nn \\ &&  \hspace*{-140mm}    
   -\frac{55}{4} \pi ^2 H_{-1,0,0}+\frac{47}{2} \pi ^2 H_{0,-1,-1}-\frac{185}{6} \pi ^2 H_{0,-1,0}
   -\frac{119}{2} \pi ^2 H_{0,0,-1}+\frac{261}{4} \pi ^2 H_{0,0,0}
\nn \\ &&  \hspace*{-140mm}    
   -3 H_{-1,-1,-1,0,0}+18 H_{-1,-1,0,0,0}+35 H_{-1,0,-1,0,0}-81 H_{-1,0,0,0,0}+47 H_{0,-1,-1,0,0}
\nn \\ &&  \hspace*{-140mm}    
   -138 H_{0,-1,0,0,0}-119 H_{0,0,-1,0,0}+243 H_{0,0,0,0,0}+\frac{73 \pi ^2 \zeta _3}{4}-3 H_{-1,-1} \zeta _3-49 H_{-1,0} \zeta _3
\nn \\ &&  \hspace*{-140mm}    
   +47 H_{0,-1} \zeta _3-33 H_{0,0} \zeta _3-\frac{301 \zeta _5}{15}\biggr)
   +\epsilon ^6 \biggl(-\frac{624607 \pi ^6}{544320}-\frac{13}{8} \pi ^4 H_{-1,-1}+\frac{323}{120} \pi ^4 H_{-1,0}
\nn \\ &&  \hspace*{-140mm}    
   +\frac{641}{72} \pi ^4 H_{0,-1}-\frac{665}{48} \pi ^4 H_{0,0}-\frac{3}{2} \pi ^2 H_{-1,-1,-1,-1}
   +\frac{7}{2} \pi ^2 H_{-1,-1,-1,0}+\frac{35}{2} \pi ^2 H_{-1,-1,0,-1}
\nn \\ &&  \hspace*{-140mm}    
   -\frac{55}{4} \pi ^2 H_{-1,-1,0,0}+\frac{107}{2} \pi ^2 H_{-1,0,-1,-1}
   -\frac{317}{6} \pi ^2 H_{-1,0,-1,0}-\frac{151}{2} \pi ^2 H_{-1,0,0,-1}
\nn \\ &&  \hspace*{-140mm}    
   +51 \pi ^2 H_{-1,0,0,0}+\frac{71}{2} \pi ^2 H_{0,-1,-1,-1}-\frac{353}{6} \pi ^2 H_{0,-1,-1,0}-\frac{247}{2} \pi ^2 H_{0,-1,0,-1}
\nn \\ &&  \hspace*{-140mm}    
   +\frac{427}{4} \pi ^2 H_{0,-1,0,0}-\frac{311}{2} \pi ^2 H_{0,0,-1,-1}
   +\frac{1025}{6} \pi ^2 H_{0,0,-1,0}+\frac{531}{2} \pi ^2 H_{0,0,0,-1}
\nn \\ &&  \hspace*{-140mm}    
   -\frac{441}{2} \pi ^2 H_{0,0,0,0}-3 H_{-1,-1,-1,-1,0,0}+18 H_{-1,-1,-1,0,0,0}+35 H_{-1,-1,0,-1,0,0}
\nn \\ &&  \hspace*{-140mm}    
   -81 H_{-1,-1,0,0,0,0}+107 H_{-1,0,-1,-1,0,0}-210 H_{-1,0,-1,0,0,0}-151 H_{-1,0,0,-1,0,0}
\nn \\ &&  \hspace*{-140mm}    
   +324 H_{-1,0,0,0,0,0}+71 H_{0,-1,-1,-1,0,0}-282 H_{0,-1,-1,0,0,0}-247 H_{0,-1,0,-1,0,0}
\nn \\ &&  \hspace*{-140mm}    
   +621 H_{0,-1,0,0,0,0}-311 H_{0,0,-1,-1,0,0}+714 H_{0,0,-1,0,0,0}+531 H_{0,0,0,-1,0,0}
\nn \\ &&  \hspace*{-140mm}    
   -1134 H_{0,0,0,0,0,0}-\frac{37}{12} \pi ^2 H_{-1} \zeta _3
   -\frac{220}{3} \pi ^2 H_{0} \zeta _3-3 H_{-1,-1,-1} \zeta _3-49 H_{-1,-1,0} \zeta _3
\nn \\ &&  \hspace*{-140mm}    
   +107 H_{-1,0,-1} \zeta _3+138 H_{-1,0,0} \zeta _3+71 H_{0,-1,-1} \zeta _3+141 H_{0,-1,0} \zeta _3-311 H_{0,0,-1} \zeta _3
\nn \\ &&  \hspace*{-140mm}    
   -48 H_{0,0,0} \zeta _3+\frac{167 \zeta _3^2}{9}+57 H_{-1} \zeta _5-\frac{444}{5} H_{0} \zeta _5\biggr)
+ \cO(\eps^7) \,.
\eea
\bea
f^{A}_{25}(x,\eps)=
-\frac{49}{36}+\frac{5}{2} \epsilon  H_{0}
   +\epsilon ^2 \biggl(\frac{241 \pi ^2}{144}-3 H_{0,0}\biggr)
   +\epsilon ^3 \biggl(\frac{11}{4} \pi ^2 H_{-1}-\frac{47}{8} \pi ^2 H_{0}+\frac{11}{2} H_{-1,0,0}
&& \nn \\ &&  \hspace*{-144mm}      
   -\frac{9}{2} H_{0,0,0}+\frac{641 \zeta _3}{36}\biggr)
   +\epsilon ^4 \biggl(\frac{847 \pi ^4}{640}  
   +\frac{23}{4} \pi ^2 H_{-1,-1}-\frac{89}{12} \pi ^2 H_{-1,0}-\frac{63}{4} \pi ^2 H_{0,-1}+\frac{39}{2} \pi ^2 H_{0,0}
\nn \\ &&  \hspace*{-144mm}    
   +\frac{23}{2} H_{-1,-1,0,0}-33 H_{-1,0,0,0}-\frac{63}{2} H_{0,-1,0,0}+54 H_{0,0,0,0}+\frac{23}{2} H_{-1} \zeta _3-39 H_{0} \zeta _3\biggr)
\nn \\ &&  \hspace*{-144mm}    
   +\epsilon ^5 \biggl(\frac{1609}{720} \pi ^4 H_{-1}-\frac{4141}{960} \pi ^4 H_{0}
   +\frac{35}{4} \pi ^2 H_{-1,-1,-1}-\frac{173}{12} \pi ^2 H_{-1,-1,0}-\frac{119}{4} \pi ^2 H_{-1,0,-1}
\nn \\ &&  \hspace*{-144mm}    
   +\frac{207}{8} \pi ^2 H_{-1,0,0}-\frac{171}{4} \pi ^2 H_{0,-1,-1}+\frac{183}{4} \pi ^2 H_{0,-1,0}
   +\frac{287}{4} \pi ^2 H_{0,0,-1}-\frac{513}{8} \pi ^2 H_{0,0,0}
\nn \\ &&  \hspace*{-144mm}    
   +\frac{35}{2} H_{-1,-1,-1,0,0}-69 H_{-1,-1,0,0,0}-\frac{119}{2} H_{-1,0,-1,0,0}+\frac{297}{2} H_{-1,0,0,0,0}
   -\frac{171}{2} H_{0,-1,-1,0,0}
\nn \\ &&  \hspace*{-144mm}    
   +189 H_{0,-1,0,0,0}+\frac{287}{2} H_{0,0,-1,0,0}-\frac{567}{2} H_{0,0,0,0,0}-\frac{3737 \pi ^2 \zeta _3}{144}
   +\frac{35}{2} H_{-1,-1} \zeta _3+\frac{65}{2} H_{-1,0} \zeta _3
\nn \\ &&  \hspace*{-144mm}    
   -\frac{171}{2} H_{0,-1} \zeta _3+\frac{117}{2} H_{0,0} \zeta _3+\frac{1143 \zeta _5}{20}\biggr)
   +\epsilon ^6 \biggl(\frac{3710783 \pi ^6}{4354560}+\frac{3181}{720} \pi ^4 H_{-1,-1}
\nn \\ &&  \hspace*{-144mm}    
   -\frac{4181}{720} \pi ^4 H_{-1,0}-\frac{185}{16} \pi ^4 H_{0,-1}+\frac{2111}{160} \pi ^4 H_{0,0}
   +\frac{47}{4} \pi ^2 H_{-1,-1,-1,-1}-\frac{257}{12} \pi ^2 H_{-1,-1,-1,0}
\nn \\ &&  \hspace*{-144mm}    
   -\frac{203}{4} \pi ^2 H_{-1,-1,0,-1}+\frac{395}{8} \pi ^2 H_{-1,-1,0,0}
   -\frac{371}{4} \pi ^2 H_{-1,0,-1,-1}+\frac{1085}{12} \pi ^2 H_{-1,0,-1,0}
\nn \\ &&  \hspace*{-144mm}    
   +\frac{531}{4} \pi ^2 H_{-1,0,0,-1}-\frac{177}{2} \pi ^2 H_{-1,0,0,0}-\frac{471}{4} \pi ^2 H_{0,-1,-1,-1}
   +\frac{499}{4} \pi ^2 H_{0,-1,-1,0}
\nn \\ &&  \hspace*{-144mm}    
   +\frac{755}{4} \pi ^2 H_{0,-1,0,-1}-\frac{1203}{8} \pi ^2 H_{0,-1,0,0}+\frac{923}{4} \pi ^2 H_{0,0,-1,-1}
   -\frac{2645}{12} \pi ^2 H_{0,0,-1,0}
\nn \\ &&  \hspace*{-144mm}    
   -\frac{1179}{4} \pi ^2 H_{0,0,0,-1}+\frac{837}{4} \pi ^2 H_{0,0,0,0}+\frac{47}{2} H_{-1,-1,-1,-1,0,0}-105 H_{-1,-1,-1,0,0,0}
\nn \\ &&  \hspace*{-144mm}    
   -\frac{203}{2} H_{-1,-1,0,-1,0,0}+\frac{621}{2} H_{-1,-1,0,0,0,0}-\frac{371}{2} H_{-1,0,-1,-1,0,0}+357 H_{-1,0,-1,0,0,0}
\nn \\ &&  \hspace*{-144mm}    
   +\frac{531}{2} H_{-1,0,0,-1,0,0}-594 H_{-1,0,0,0,0,0}-\frac{471}{2} H_{0,-1,-1,-1,0,0}+513 H_{0,-1,-1,0,0,0}
\nn \\ &&  \hspace*{-144mm}    
   +\frac{755}{2} H_{0,-1,0,-1,0,0}-\frac{1701}{2} H_{0,-1,0,0,0,0}+\frac{923}{2} H_{0,0,-1,-1,0,0}-861 H_{0,0,-1,0,0,0}
\nn \\ &&  \hspace*{-144mm}    
   -\frac{1179}{2} H_{0,0,0,-1,0,0}+1215 H_{0,0,0,0,0,0}-\frac{703}{24} \pi ^2 H_{-1} \zeta _3
   +93 \pi ^2 H_{0} \zeta _3+\frac{47}{2} H_{-1,-1,-1} \zeta _3
\nn \\ &&  \hspace*{-144mm}    
   +\frac{149}{2} H_{-1,-1,0} \zeta _3-\frac{371}{2} H_{-1,0,-1} \zeta _3-137 H_{-1,0,0} \zeta _3
   -\frac{471}{2} H_{0,-1,-1} \zeta _3-\frac{13}{2} H_{0,-1,0} \zeta _3
\nn \\ &&  \hspace*{-144mm}   
   +\frac{923}{2} H_{0,0,-1} \zeta _3-\frac{9901 \zeta _3^2}{72}+\frac{163}{2} H_{-1} \zeta _5-82 H_{0} \zeta _5\biggr)
+ \cO(\eps^7) \,.
\eea
\bea
f^{A}_{26}(x,\eps)=
-\frac{4}{9}+\frac{13 \pi ^2 \epsilon ^2}{36}+\frac{1}{2} \epsilon  H_{0}
   +\epsilon ^3 \biggl(\frac{9}{4} \pi ^2 H_{-1}-\frac{15}{8} \pi ^2 H_{0}+\frac{9}{2} H_{-1,0,0}-\frac{9}{2} H_{0,0,0}-\frac{71 \zeta _3}{18}\biggr)
&& \nn \\ &&  \hspace*{-150mm}
   +\epsilon ^4 \biggl(\frac{61 \pi ^4}{720}+\frac{21}{4} \pi ^2 H_{-1,-1}-\frac{25}{4} \pi ^2 H_{-1,0}
   -\frac{21}{4} \pi ^2 H_{0,-1}+\frac{25}{4} \pi ^2 H_{0,0}+\frac{21}{2} H_{-1,-1,0,0}
\nn \\ &&  \hspace*{-150mm}   
   -27 H_{-1,0,0,0}-\frac{21}{2} H_{0,-1,0,0}+27 H_{0,0,0,0}+\frac{21}{2} H_{-1} \zeta _3-2 H_{0} \zeta _3\biggr)
\nn \\ &&  \hspace*{-150mm} 
   +\epsilon ^5 \biggl(\frac{337}{240} \pi ^4 H_{-1}-\frac{1217}{960} \pi ^4 H_{0}+\frac{33}{4} \pi ^2 H_{-1,-1,-1}
   -\frac{53}{4} \pi ^2 H_{-1,-1,0}-\frac{93}{4} \pi ^2 H_{-1,0,-1}
\nn \\ &&  \hspace*{-150mm}    
   +\frac{165}{8} \pi ^2 H_{-1,0,0}-\frac{33}{4} \pi ^2 H_{0,-1,-1}+\frac{53}{4} \pi ^2 H_{0,-1,0}
   +\frac{93}{4} \pi ^2 H_{0,0,-1}-\frac{165}{8} \pi ^2 H_{0,0,0}
\nn \\ &&  \hspace*{-150mm}    
   +\frac{33}{2} H_{-1,-1,-1,0,0}-63 H_{-1,-1,0,0,0}-\frac{93}{2} H_{-1,0,-1,0,0}+\frac{243}{2} H_{-1,0,0,0,0}
   -\frac{33}{2} H_{0,-1,-1,0,0}
\nn \\ &&  \hspace*{-150mm}    
   +63 H_{0,-1,0,0,0}+\frac{93}{2} H_{0,0,-1,0,0}-\frac{243}{2} H_{0,0,0,0,0}-\frac{859 \pi ^2 \zeta _3}{72}
   +\frac{33}{2} H_{-1,-1} \zeta _3+\frac{27}{2} H_{-1,0} \zeta _3
\nn \\ &&  \hspace*{-150mm}    
   -\frac{33}{2} H_{0,-1} \zeta _3-\frac{27}{2} H_{0,0} \zeta _3-\frac{1457 \zeta _5}{30}\biggr)
   +\epsilon ^6 \biggl(\frac{2029 \pi ^6}{217728}+\frac{287}{80} \pi ^4 H_{-1,-1}-\frac{311}{80} \pi ^4 H_{-1,0}
\nn \\ &&  \hspace*{-150mm}    
   -\frac{287}{80} \pi ^4 H_{0,-1}+\frac{311}{80} \pi ^4 H_{0,0}+\frac{45}{4} \pi ^2 H_{-1,-1,-1,-1}
   -\frac{81}{4} \pi ^2 H_{-1,-1,-1,0}-\frac{177}{4} \pi ^2 H_{-1,-1,0,-1}
\nn \\ &&  \hspace*{-150mm}    
   +\frac{353}{8} \pi ^2 H_{-1,-1,0,0}-\frac{249}{4} \pi ^2 H_{-1,0,-1,-1}+\frac{269}{4} \pi ^2 H_{-1,0,-1,0}
   +\frac{377}{4} \pi ^2 H_{-1,0,0,-1}
\nn \\ &&  \hspace*{-150mm}    
   -\frac{135}{2} \pi ^2 H_{-1,0,0,0}-\frac{45}{4} \pi ^2 H_{0,-1,-1,-1}+\frac{81}{4} \pi ^2 H_{0,-1,-1,0}
   +\frac{177}{4} \pi ^2 H_{0,-1,0,-1}
\nn \\ &&  \hspace*{-150mm}    
   -\frac{353}{8} \pi ^2 H_{0,-1,0,0}+\frac{249}{4} \pi ^2 H_{0,0,-1,-1}-\frac{269}{4} \pi ^2 H_{0,0,-1,0}
   -\frac{377}{4} \pi ^2 H_{0,0,0,-1}+\frac{135}{2} \pi ^2 H_{0,0,0,0}
\nn \\ &&  \hspace*{-150mm}    
   +\frac{45}{2} H_{-1,-1,-1,-1,0,0}-99 H_{-1,-1,-1,0,0,0}-\frac{177}{2} H_{-1,-1,0,-1,0,0}+\frac{567}{2} H_{-1,-1,0,0,0,0}
\nn \\ &&  \hspace*{-150mm}    
   -\frac{249}{2} H_{-1,0,-1,-1,0,0}+279 H_{-1,0,-1,0,0,0}+\frac{377}{2} H_{-1,0,0,-1,0,0}-486 H_{-1,0,0,0,0,0}
\nn \\ &&  \hspace*{-150mm}    
   -\frac{45}{2} H_{0,-1,-1,-1,0,0}+99 H_{0,-1,-1,0,0,0}+\frac{177}{2} H_{0,-1,0,-1,0,0}-\frac{567}{2} H_{0,-1,0,0,0,0}
\nn \\ &&  \hspace*{-150mm}    
   +\frac{249}{2} H_{0,0,-1,-1,0,0}-279 H_{0,0,-1,0,0,0}-\frac{377}{2} H_{0,0,0,-1,0,0}+486 H_{0,0,0,0,0,0}
   -\frac{255}{8} \pi ^2 H_{-1} \zeta _3
\nn \\ &&  \hspace*{-150mm}    
   +\frac{97}{4} \pi ^2 H_{0} \zeta _3+\frac{45}{2} H_{-1,-1,-1} \zeta _3+\frac{111}{2} H_{-1,-1,0} \zeta _3
   -\frac{249}{2} H_{-1,0,-1} \zeta _3-99 H_{-1,0,0} \zeta _3
\nn \\ &&  \hspace*{-150mm}    
   -\frac{45}{2} H_{0,-1,-1} \zeta _3-\frac{111}{2} H_{0,-1,0} \zeta _3+\frac{249}{2} H_{0,0,-1} \zeta _3
   +99 H_{0,0,0} \zeta _3+\frac{275 \zeta _3^2}{18}
\nn \\ &&  \hspace*{-150mm}    
   -\frac{15}{2} H_{-1} \zeta _5+\frac{351}{5} H_{0} \zeta _5\biggr)
+ \cO(\eps^7) \,.
\eea

\subsection{Tennis court master integrals}

\bea
f^{E}_{39}(x,\eps)=
\frac{128}{9}-\frac{52}{3} \epsilon  H_{0}
   +\epsilon ^2 \biggl(-\frac{38 \pi ^2}{3}+8 H_{0,0}\biggr)
   +\epsilon ^3 \biggl(-10 \pi ^2 H_{-1}+\frac{157}{9} \pi ^2 H_{0}-20 H_{-1,0,0}
&& \nn \\ &&  \hspace*{-150mm}   
   +28 H_{0,0,0}-\frac{964 \zeta _3}{9}\biggr)
   +\epsilon ^4 \biggl(\frac{2429 \pi ^4}{810}-10 \pi ^2 H_{-1,-1}+\frac{50}{3} \pi ^2 H_{-1,0}+6 \pi ^2 H_{0,-1}
   -4 \pi ^2 H_{0,0}
\nn \\ &&  \hspace*{-150mm}   
   -20 H_{-1,-1,0,0}+80 H_{-1,0,0,0}+12 H_{0,-1,0,0}-64 H_{0,0,0,0} 
   -20 H_{-1} \zeta _3+\frac{328}{3} H_{0} \zeta _3\biggr)
\nn \\ &&  \hspace*{-150mm}
   +\epsilon ^5 \biggl(\frac{5}{18} \pi ^4 H_{-1}-\frac{10913 \pi ^4 H_{0}}{1080}-10 \pi ^2 H_{-1,-1,-1}
   +\frac{50}{3} \pi ^2 H_{-1,-1,0}+30 \pi ^2 H_{-1,0,-1}
\nn \\ &&  \hspace*{-150mm}   
   -\frac{71}{3} \pi ^2 H_{-1,0,0}-26 \pi ^2 H_{0,-1,-1}+\frac{82}{3} \pi ^2 H_{0,-1,0}+70 \pi ^2 H_{0,0,-1}-\frac{227}{3} \pi ^2 H_{0,0,0}
\nn \\ &&  \hspace*{-150mm}   
   -20 H_{-1,-1,-1,0,0}+80 H_{-1,-1,0,0,0}+60 H_{-1,0,-1,0,0}-172 H_{-1,0,0,0,0}-52 H_{0,-1,-1,0,0}
\nn \\ &&  \hspace*{-150mm}   
   +112 H_{0,-1,0,0,0}+140 H_{0,0,-1,0,0}-140 H_{0,0,0,0,0}+\frac{3257 \pi ^2 \zeta _3}{27}-20 H_{-1,-1} \zeta _3-20 H_{-1,0} \zeta _3
\nn \\ &&  \hspace*{-150mm}   
   -52 H_{0,-1} \zeta _3+52 H_{0,0} \zeta _3-\frac{3556 \zeta _5}{5}\biggr)
   +\epsilon ^6 \biggl(\frac{1391417 \pi ^6}{408240}+\frac{5}{18} \pi ^4 H_{-1,-1}+\frac{641}{90} \pi ^4 H_{-1,0}
\nn \\ &&  \hspace*{-150mm}   
   -\frac{1207}{90} \pi ^4 H_{0,-1}+\frac{3163}{180} \pi ^4 H_{0,0}-10 \pi ^2 H_{-1,-1,-1,-1}
   +\frac{50}{3} \pi ^2 H_{-1,-1,-1,0}+30 \pi ^2 H_{-1,-1,0,-1}
\nn \\ &&  \hspace*{-150mm}   
   -\frac{71}{3} \pi ^2 H_{-1,-1,0,0}+126 \pi ^2 H_{-1,0,-1,-1}-66 \pi ^2 H_{-1,0,-1,0}-98 \pi ^2 H_{-1,0,0,-1}+66 \pi ^2 H_{-1,0,0,0}
\nn \\ &&  \hspace*{-150mm}   
   -218 \pi ^2 H_{0,-1,-1,-1}+\frac{562}{3} \pi ^2 H_{0,-1,-1,0}+270 \pi ^2 H_{0,-1,0,-1}-\frac{527}{3} \pi ^2 H_{0,-1,0,0}
\nn \\ &&  \hspace*{-150mm}   
   +358 \pi ^2 H_{0,0,-1,-1}-\frac{926}{3} \pi ^2 H_{0,0,-1,0}-394 \pi ^2 H_{0,0,0,-1}+\frac{746}{3} \pi ^2 H_{0,0,0,0}
\nn \\ &&  \hspace*{-150mm}   
   -20 H_{-1,-1,-1,-1,0,0}+80 H_{-1,-1,-1,0,0,0}+60 H_{-1,-1,0,-1,0,0}-172 H_{-1,-1,0,0,0,0}
\nn \\ &&  \hspace*{-150mm}   
   +252 H_{-1,0,-1,-1,0,0}-144 H_{-1,0,-1,0,0,0}-196 H_{-1,0,0,-1,0,0}+296 H_{-1,0,0,0,0,0}
\nn \\ &&  \hspace*{-150mm}   
   -436 H_{0,-1,-1,-1,0,0}+688 H_{0,-1,-1,0,0,0}+540 H_{0,-1,0,-1,0,0}-940 H_{0,-1,0,0,0,0}
\nn \\ &&  \hspace*{-150mm}   
   +716 H_{0,0,-1,-1,0,0}-1136 H_{0,0,-1,0,0,0}-788 H_{0,0,0,-1,0,0}+1208 H_{0,0,0,0,0,0}+\frac{269}{3} \pi ^2 H_{-1} \zeta _3
\nn \\ &&  \hspace*{-150mm}   
   -\frac{1916}{9} \pi ^2 H_{0} \zeta _3-20 H_{-1,-1,-1} \zeta _3-20 H_{-1,-1,0} \zeta _3+252 H_{-1,0,-1} \zeta _3+32 H_{-1,0,0} \zeta _3
\nn \\ &&  \hspace*{-150mm}   
   -436 H_{0,-1,-1} \zeta _3+44 H_{0,-1,0} \zeta _3+716 H_{0,0,-1} \zeta _3-608 H_{0,0,0} \zeta _3+\frac{788 \zeta _3^2}{3}
\nn \\ &&  \hspace*{-150mm}   
   -516 H_{-1} \zeta _5+\frac{8432}{5} H_{0} \zeta _5\biggr)
+ \cO(\eps^7) \,.
\eea
\bea
f^{E}_{40}(x,\eps)=
-\frac{98}{9}+\frac{50}{3} \epsilon  H_{0}
   +\epsilon ^2 \biggl(\frac{755 \pi ^2}{54}-10 H_{0,0}\biggr)
   +\epsilon ^3 \biggl(28 \pi ^2 H_{-1}-\frac{635}{18} \pi ^2 H_{0}+56 H_{-1,0,0}
&&\nn \\ &&  \hspace*{-149mm}   
   -58 H_{0,0,0}+122 \zeta _3\biggr)
   +\epsilon ^4 \biggl(\frac{331 \pi ^4}{144}+84 \pi ^2 H_{-1,-1}-\frac{244}{3} \pi ^2 H_{-1,0}-92 \pi ^2 H_{0,-1}
   +\frac{463}{6} \pi ^2 H_{0,0}
\nn \\ &&  \hspace*{-149mm}   
   +168 H_{-1,-1,0,0}-320 H_{-1,0,0,0}-184 H_{0,-1,0,0}+310 H_{0,0,0,0}+168 H_{-1} \zeta _3-238 H_{0} \zeta _3\biggr)
\nn \\ &&  \hspace*{-149mm} 
 +\epsilon ^5 \biggl(\frac{197}{45} \pi ^4 H_{-1}+\frac{91}{80} \pi ^4 H_{0}+284 \pi ^2 H_{-1,-1,-1}
 -\frac{748}{3} \pi ^2 H_{-1,-1,0}-276 \pi ^2 H_{-1,0,-1}
\nn \\ &&  \hspace*{-149mm} 
 +\frac{554}{3} \pi ^2 H_{-1,0,0}-308 \pi ^2 H_{0,-1,-1}+\frac{724}{3} \pi ^2 H_{0,-1,0}+236 \pi ^2 H_{0,0,-1}-\frac{665}{6} \pi ^2 H_{0,0,0}
\nn \\ &&  \hspace*{-149mm} 
 +568 H_{-1,-1,-1,0,0}-928 H_{-1,-1,0,0,0}-552 H_{-1,0,-1,0,0}+1096 H_{-1,0,0,0,0}-616 H_{0,-1,-1,0,0}
\nn \\ &&  \hspace*{-149mm}   
 +832 H_{0,-1,0,0,0}+472 H_{0,0,-1,0,0}-826 H_{0,0,0,0,0}-\frac{839 \pi ^2 \zeta _3}{6}+568 H_{-1,-1} \zeta _3-392 H_{-1,0} \zeta _3
\nn \\ &&  \hspace*{-149mm}   
 -616 H_{0,-1} \zeta _3+370 H_{0,0} \zeta _3+\frac{17818 \zeta _5}{15}\biggr)
   +\epsilon ^6 \biggl(-\frac{393371 \pi ^6}{181440}+\frac{527}{45} \pi ^4 H_{-1,-1}+\frac{59}{45} \pi ^4 H_{-1,0}
\nn \\ &&  \hspace*{-149mm}     
   +\frac{319}{45} \pi ^4 H_{0,-1}-\frac{7927}{240} \pi ^4 H_{0,0}+948 \pi ^2 H_{-1,-1,-1,-1}-\frac{2452}{3} \pi ^2 H_{-1,-1,-1,0}
\nn \\ &&  \hspace*{-149mm}     
   -908 \pi ^2 H_{-1,-1,0,-1}+\frac{1630}{3} \pi ^2 H_{-1,-1,0,0}-988 \pi ^2 H_{-1,0,-1,-1}+\frac{2092}{3} \pi ^2 H_{-1,0,-1,0}
\nn \\ &&  \hspace*{-149mm}     
   +692 \pi ^2 H_{-1,0,0,-1}-220 \pi ^2 H_{-1,0,0,0}-1052 \pi ^2 H_{0,-1,-1,-1}+\frac{2380}{3} \pi ^2 H_{0,-1,-1,0}
\nn \\ &&  \hspace*{-149mm}     
   +740 \pi ^2 H_{0,-1,0,-1}-\frac{970}{3} \pi ^2 H_{0,-1,0,0}+836 \pi ^2 H_{0,0,-1,-1}-\frac{1444}{3} \pi ^2 H_{0,0,-1,0}
\nn \\ &&  \hspace*{-149mm}     
   -332 \pi ^2 H_{0,0,0,-1}-\frac{1009}{6} \pi ^2 H_{0,0,0,0}+1896 H_{-1,-1,-1,-1,0,0}-3008 H_{-1,-1,-1,0,0,0}
\nn \\ &&  \hspace*{-149mm}     
   -1816 H_{-1,-1,0,-1,0,0}+2968 H_{-1,-1,0,0,0,0}-1976 H_{-1,0,-1,-1,0,0}+2208 H_{-1,0,-1,0,0,0}
\nn \\ &&  \hspace*{-149mm}     
   +1384 H_{-1,0,0,-1,0,0}-2192 H_{-1,0,0,0,0,0}-2104 H_{0,-1,-1,-1,0,0}+2656 H_{0,-1,-1,0,0,0}
\nn \\ &&  \hspace*{-149mm}     
   +1480 H_{0,-1,0,-1,0,0}-1576 H_{0,-1,0,0,0,0}+1672 H_{0,0,-1,-1,0,0}-1216 H_{0,0,-1,0,0,0}
\nn \\ &&  \hspace*{-149mm}     
   -664 H_{0,0,0,-1,0,0}+118 H_{0,0,0,0,0,0}-518 \pi ^2 H_{-1} \zeta _3+\frac{2629}{6} \pi ^2 H_{0} \zeta _3+1896 H_{-1,-1,-1} \zeta _3
\nn \\ &&  \hspace*{-149mm}     
   -1272 H_{-1,-1,0} \zeta _3-1976 H_{-1,0,-1} \zeta _3+592 H_{-1,0,0} \zeta _3-2104 H_{0,-1,-1} \zeta _3+1800 H_{0,-1,0} \zeta _3
\nn \\ &&  \hspace*{-149mm}     
   +1672 H_{0,0,-1} \zeta _3-238 H_{0,0,0} \zeta _3-505 \zeta _3^2+1272 H_{-1} \zeta _5-2930 H_{0} \zeta _5\biggr)
+ \cO(\eps^7) \,.
\eea
\bea
f^{E}_{41}(x,\eps)=
-\frac{56}{9}+4 \epsilon  H_{0}
   +\epsilon ^2 \biggl(\frac{166 \pi ^2}{27}+4 H_{0,0}\biggr)
   +\epsilon ^3 \biggl(8 \pi ^2 H_{-1}-\frac{11}{3} \pi ^2 H_{0}+16 H_{-1,0,0}
&&\nn \\ &&  \hspace*{-140mm}     
   -12 H_{0,0,0}+\frac{200 \zeta _3}{3}\biggr)
   +\epsilon ^4 \biggl(-\frac{151 \pi ^4}{36}+12 \pi ^2 H_{-1,-1}-\frac{20}{3} \pi ^2 H_{-1,0}+20 \pi ^2 H_{0,-1}-21 \pi ^2 H_{0,0}
\nn \\ &&  \hspace*{-140mm}     
   +24 H_{-1,-1,0,0}-16 H_{-1,0,0,0}+40 H_{0,-1,0,0}-44 H_{0,0,0,0}+24 H_{-1} \zeta _3-44 H_{0} \zeta _3\biggr)
\nn \\ &&  \hspace*{-140mm}    
   +\epsilon ^5 \biggl(-\frac{248}{45} \pi ^4 H_{-1}+\frac{271}{24} \pi ^4 H_{0}-24 \pi ^2 H_{-1,-1,-1}
   +\frac{88}{3} \pi ^2 H_{-1,-1,0}+64 \pi ^2 H_{-1,0,-1}
\nn \\ &&  \hspace*{-140mm}       
   -\frac{140}{3} \pi ^2 H_{-1,0,0}+152 \pi ^2 H_{0,-1,-1}-\frac{344}{3} \pi ^2 H_{0,-1,0}-160 \pi ^2 H_{0,0,-1}+\frac{313}{3} \pi ^2 H_{0,0,0}
\nn \\ &&  \hspace*{-140mm}       
   -48 H_{-1,-1,-1,0,0}+128 H_{-1,-1,0,0,0}+128 H_{-1,0,-1,0,0}-224 H_{-1,0,0,0,0}+304 H_{0,-1,-1,0,0}
\nn \\ &&  \hspace*{-140mm}       
   -384 H_{0,-1,0,0,0}-320 H_{0,0,-1,0,0}+420 H_{0,0,0,0,0}-\frac{334 \pi ^2 \zeta _3}{3}-48 H_{-1,-1} \zeta _3-80 H_{-1,0} \zeta _3
\nn \\ &&  \hspace*{-140mm}       
   +304 H_{0,-1} \zeta _3-196 H_{0,0} \zeta _3+\frac{6856 \zeta _5}{15}\biggr)
   +\epsilon ^6 \biggl(-\frac{43585 \pi ^6}{27216}-\frac{197}{45} \pi ^4 H_{-1,-1}-\frac{341}{45} \pi ^4 H_{-1,0}
\nn \\ &&  \hspace*{-140mm}       
   +\frac{997}{45} \pi ^4 H_{0,-1}-\frac{3767}{360} \pi ^4 H_{0,0}-180 \pi ^2 H_{-1,-1,-1,-1}+\frac{700}{3} \pi ^2 H_{-1,-1,-1,0}
 \nn \\ &&  \hspace*{-140mm}      
   +316 \pi ^2 H_{-1,-1,0,-1}-\frac{694}{3} \pi ^2 H_{-1,-1,0,0}+364 \pi ^2 H_{-1,0,-1,-1}-332 \pi ^2 H_{-1,0,-1,0}
\nn \\ &&  \hspace*{-140mm}       
   -316 \pi ^2 H_{-1,0,0,-1}+\frac{416}{3} \pi ^2 H_{-1,0,0,0}+692 \pi ^2 H_{0,-1,-1,-1}-\frac{1724}{3} \pi ^2 H_{0,-1,-1,0}
\nn \\ &&  \hspace*{-140mm}       
   -636 \pi ^2 H_{0,-1,0,-1}+386 \pi ^2 H_{0,-1,0,0}-748 \pi ^2 H_{0,0,-1,-1}+\frac{1700}{3} \pi ^2 H_{0,0,-1,0}
\nn \\ &&  \hspace*{-140mm}       
   +540 \pi ^2 H_{0,0,0,-1}-\frac{643}{3} \pi ^2 H_{0,0,0,0}-360 H_{-1,-1,-1,-1,0,0}+1040 H_{-1,-1,-1,0,0,0}
\nn \\ &&  \hspace*{-140mm}       
   +632 H_{-1,-1,0,-1,0,0}-1320 H_{-1,-1,0,0,0,0}+728 H_{-1,0,-1,-1,0,0}-1264 H_{-1,0,-1,0,0,0}
\nn \\ &&  \hspace*{-140mm}       
   -632 H_{-1,0,0,-1,0,0}+1280 H_{-1,0,0,0,0,0}+1384 H_{0,-1,-1,-1,0,0}-2064 H_{0,-1,-1,0,0,0}
\nn \\ &&  \hspace*{-140mm}       
   -1272 H_{0,-1,0,-1,0,0}+1896 H_{0,-1,0,0,0,0}-1496 H_{0,0,-1,-1,0,0}+1904 H_{0,0,-1,0,0,0}
\nn \\ &&  \hspace*{-140mm}      
   +1080 H_{0,0,0,-1,0,0}-1532 H_{0,0,0,0,0,0}-\frac{530}{3} \pi ^2 H_{-1} \zeta _3
   +\frac{721}{3} \pi ^2 H_{0} \zeta _3-360 H_{-1,-1,-1} \zeta _3
\nn \\ &&  \hspace*{-140mm}       
   +136 H_{-1,-1,0} \zeta _3+728 H_{-1,0,-1} \zeta _3-584 H_{-1,0,0} \zeta _3+1384 H_{0,-1,-1} \zeta _3-904 H_{0,-1,0} \zeta _3
\nn \\ &&  \hspace*{-140mm}   
   -1496 H_{0,0,-1} \zeta _3+1076 H_{0,0,0} \zeta _3-\frac{1364 \zeta _3^2}{3}+984 H_{-1} \zeta _5-\frac{3892}{5} H_{0} \zeta _5\biggr)
+ \cO(\eps^7) \,.
\eea
 
\bibliographystyle{JHEP}

\bibliography{hss_arxiv}

\end{document}